%% file: DMA.tex
\documentclass[a4paper,fleqn,usenatbib]{mnras}

\usepackage{mathptmx}
\DeclareSymbolFont{matha}{OML}{txmi}{m}{it}
\DeclareMathSymbol{\varv}{\mathord}{matha}{118}

\usepackage[T1]{fontenc}
\usepackage{ae,aecompl}

\usepackage{graphicx}	
\usepackage{amsmath}	
\usepackage{amssymb}	

\usepackage[version=4]{mhchem}

\title[Dimethylamine]
      {Rotation-tunneling spectrum and astrochemical modeling of dimethylamine, CH$_3$NHCH$_3$, and searches for it in space}

\author[H. S. P. M{\"u}ller et al.]{
H.~S.~P. M{\"u}ller,$^{1}$\thanks{E-mail: hspm@ph1.uni-koeln.de (HSPM)}
R.~T. Garrod,$^{2}$ 
A. Belloche,$^{3}$ 
V.~M. Rivilla,$^{4}$ 
K.~M. Menten,$^{3}$ 
\newauthor{
I. Jim{\'e}nez-Serra,$^{4}$ 
J. Mart{\'i}n-Pintado,$^{4}$ 
F. Lewen$^{1}$ and S. Schlemmer$^{1}$}
\\
$^{1}$Astrophysik/I.~Physikalisches Institut, Universit{\"a}t zu K{\"o}ln,
      Z{\"u}lpicher Str. 77, 50937 K{\"o}ln, Germany\\
$^{2}$Departments of Chemistry and Astronomy, University of Virginia, Charlottesville, VA 22904, USA\\
$^{3}$Max-Planck-Institut f\"ur Radioastronomie, Auf dem H\"ugel 69, 53121 Bonn, Germany\\
$^{4}$Centro de Astrobiolog{\'i}a (CSIC, INTA), Ctra. de Ajalvir, km. 4, Torrej{\'o}n de Ardoz, 28850 Madrid, Spain
}

\date{Accepted 2023 May 18. Received 2023 May 18; in original form 2023 March 15}

\pubyear{2023}

\begin{document}
\label{firstpage}
\pagerange{\pageref{firstpage}--\pageref{lastpage}}
\maketitle

\hyphenation{For-schungs-ge-mein-schaft}
\hyphenation{Roh-de}

\begin{abstract}
Methylamine has been the only simple alkylamine detected in the interstellar medium for a long time. 
With the recent secure and tentative detections of vinylamine and ethylamine, respectively, dimethylamine has become a promising target for searches in space. 
Its rotational spectrum, however, has been known only up to 45~GHz until now. 
Here we investigate the rotation-tunneling spectrum of dimethylamine in selected regions between 76 and 1091~GHz using three different spectrometers in order to facilitate its detection in space. 
The quantum number range is extended to $J = 61$ and $K_a = 21$, yielding an extensive set of accurate spectroscopic parameters. 
To search for dimethylamine, we refer to the spectral line survey ReMoCA carried out with the Atacama Large Millimeter/submillimeter Array toward the high-mass star-forming region Sagittarius B2(N) 
and a spectral line survey of the molecular cloud G+0.693$-$0.027 employing the IRAM 30~m and Yebes 40~m radio telescopes. 
We report nondetections of dimethylamine toward the hot molecular cores Sgr~B2(N1S) and Sgr~B2(N2b) as well as G+0.693$-$0.027 which imply that dimethylamine is at least 14, 4.5 and 39 times less abundant than methylamine toward these sources, respectively.
The observational results are compared to computational results from a gas-grain astrochemical model. 
The modeled methylamine to dimethylamine ratios are compatible with the observational lower limits. 
However, the model produces too much ethylamine compared with methylamine which could mean that the already fairly low levels of dimethylamine in the models may also be too high.

\end{abstract}



\section{Introduction}
\label{intro}

Methylamine, CH$_3$NH$_2$, was among the earliest molecules to be discovered by radio-astronomical means; \citet{MeNH2_det1_1974} and \citet{MeNH2_det2_1974} reported detections toward Sagittarius (Sgr) B2 and Orion~A, 
although the detection toward Orion~A was dismissed a few years later \citep{MeNH2_not_OrionA_1984}.
Despite being a fairly small molecule, searches for methylamine toward other sources remained fruitless for quite some time. It was ultimately identified  
in the course of a spectral line survey of a $z \approx 0.89$ foreground galaxy located in front of the quasar PKS 1830$-$211 \citep{MeNH2_extragal_2011}. 
Surprisingly, CH$_3$NH$_2$ emission was detected toward the peculiar molecule-rich circumstellar environment of the ``red nova'' CK Vul, 
thought to be the remnant of a stellar merger \citep{CK_Vul_2017}, and imaged with the Atacama Large Millimeter/submillimeter Array (ALMA)\citep{CK_Vul_2020}. 
Recently, CH$_3$NH$_2$ was detected in the molecular cloud G+0.693$-$0.027 close to Sgr~B2(N) \citep{zeng2018}. 
Shortly thereafter, it was also found toward the high-mass star forming regions NGC6334I \citep{MeNH2_NGC6334I_2019} and G10.47+0.03 \citep{MeNH2_Ohishi_2019} and later toward several other high-mass protostars \citep{MeNH2_more_2022}.

\citet{zeng2021} detected vinylamine, C$_2$H$_3$NH$_2$, securely and ethylamine, C$_2$H$_5$NH$_2$, tentatively in a spectral line survey of G+0.693$-$0.027. 
Dimethylamine, CH$_3$NHCH$_3$, (DMA for short) is an isomer of ethylamine and related to methylamine, making it a viable candidate for searches in space. 
These relationships are similar to the relationships of dimethyl ether, CH$_3$OCH$_3$, 
with respect to ethanol, C$_2$H$_5$OH, and methanol, CH$_3$OH, all three species being well known interstellar molecules.

\citet{rot_gs_1968} investigated the rotation-tunneling spectra of several isotopologs of 
DMA in its ground vibrational state up to 45~GHz with $J \le 8$ and $K_a \le 3$. 
They determined hyperfine structure (HFS) parameters, dipole moment components, and 
structural parameters. Since they did not resolve any splitting caused by the internal 
rotation of the two equivalent methyl rotors, they studied the spectra of several 
isotopologs in their two fundamental torsional states \citep{rot_vt1_1971}.

Very recently, \citet{DMA_FTMW_2021} analyzed the ground state rotation-tunneling spectrum 
of the DMA main isotopic species applying microwave Fourier transform spectroscopy. 
They were able to resolve internal rotation splitting of the order of 0.2~MHz and improved the HFS parameters. 
But because of the limitations in frequency ($\le 32$~GHz) and the low 
rotational temperatures, their quantum number range is limited to $J \le 10$ and $K_a \le 1$.

Since the available data are insufficient to calculate accurate transition frequencies 
in the millimeter, let alone the submillimeter region, we carried out an extensive study 
of the ground state rotation-tunneling spectrum of the DMA main isotopolog in sections 
between 76 and 1091~GHz.

We use the results of this spectroscopic study to search for DMA in the interstellar medium, employing two spectral line surveys performed toward 
the Sgr~B2 molecular cloud complex: a survey obtained with ALMA toward hot molecular cores of the Sgr~B2(N) star forming region \citep[e.g.,][]{Belloche19} 
and a survey obtained with single-dish telescopes toward the Giant Molecular Cloud G+0.693$-$0.027, also located in the Sgr~B2 molecular cloud complex 
and that is experiencing a cloud-cloud collision \citep{zeng2018,zeng2020,rivilla2021a,rivilla2022c}.

We also present here the results of astrochemical models under generic hot-core conditions, using a new chemical network that includes DMA. 
Methylamine was included in the early interstellar grain-surface chemistry network of \citet{Allen77}; the radicals CH$_3$ and NH$_2$ could directly recombine to form methylamine, 
while related atoms and radicals could also react to form the various hydrogenation states CH$_\textrm{x}$NH$_\textrm{y}$, which could then be further hydrogenated all the way to CH$_3$NH$_2$. 
Similar schemes have been used in later gas-grain chemical networks, including those intended specifically for hot cores \citep[e.g.][]{Garrod08}. 
Although some of the species accessible in the \citet{Allen77} network (e.g.~CH$_2$NCH$_3$) came close in structure to DMA, the latter molecule does not appear to be present in any existing astrochemical networks up to now. 
Here, we adapt the hot-core model and chemical network used by \citet{Garrod22}, which already includes methylamine chemistry, to trace the production of DMA through a small selection of grain-surface/bulk-ice reaction pathways. 
The network also includes the usual mechanisms for grain-surface and gas-phase destruction of DMA and other associated species.

\section{Experimental details}
\label{exptl}

The investigation of the rotation-tunneling spectrum of DMA was carried out with three 
different spectrometers. All absorption cells were made of Pyrex glass, had an inner 
diameter of 10~cm, and were kept at room temperature. The cells were filled to a certain 
nominal pressure from an aqueous solution of DMA; after a few hours to about a day, 
the sample was pumped off and the cell refilled because of the pressure rise caused by small leakages.

We used two 7~m coupled glass cells in a double path arrangement for measurements in the 
76$-$124~GHz region, yielding an optical path length of 28~m. We employed a 5~m double path 
cell for the 159$-$375~GHz range. The cells of both spectrometers are equipped with 
Teflon windows. Additional information on these spectrometers is available in 
\citet{n-BuCN_rot_2012} and \citet{OSSO_rot_2015}, respectively.
The measurements between 793 and 1091~GHz were carried out in a 5~m long single path cell 
equipped with high-density polyethylene windows. Additional information on this spectrometer 
system is available in \citet{CH3SH_rot_2012}. All spectrometers utilize Virginia Diode, Inc. 
(VDI), frequency multipliers driven by Rohde \& Schwarz SMF~100A synthesizers as sources. 
Schottky diode detectors are used up to 375~GHz while a closed cycle liquid He-cooled 
InSb bolometer (QMC Instruments Ltd) is employed around 1~THz. Frequency modulation is applied 
to reduce baseline effects with demodulation at twice the modulation frequency. 
This causes absorption lines to appear approximately as second derivatives of a Gaussian.

We recorded mostly individual transitions in all frequency windows covering 5, 6, and 10~MHz 
around 100, 250, and 900~GHz, respectively at pressures of about 1, 2, and 3$-$5~Pa. 
Larger sections were also covered, in particular at higher frequencies. 
Uncertainties were evaluated mostly based on the symmetry of the line shape. 
The lines around 100~GHz were quite weak because of HFS, internal rotation splitting or intrinsically. 
Assigned uncertainties were between 5 and 30 kHz in this region. 
We applied uncertainties of 3 to 10~kHz for lines that were very symmetric or nearly so in the 159$-$375~GHz range and 15 to 50~kHz for moderately to less symmetric lines, 
lines closer to other lines or fairly weak lines. Similar uncertainties were achieved 
earlier, for example, in the case of 2-cyanobutane, which has a much richer rotational spectrum \citep{2-CAB_rot_2017}. 
Uncertainties for very good lines were 5$-$10~kHz, and larger 
uncertainties up to $\sim$300~kHz were assigned for example to weaker lines and lines 
close to other lines in the 793$-$1091~GHz region. Similar uncertainties at these frequencies 
were achieved for excited vibrational lines of CH$_3$CN \citep{MeCN_up2v4eq1_etc_2021} or 
for isotopic oxirane \citep{c-C2H4O_rot_2022,c-C2H3DO_rot_2023}.


\begin{figure}
\centering
\includegraphics[width=.75\columnwidth,angle=0]{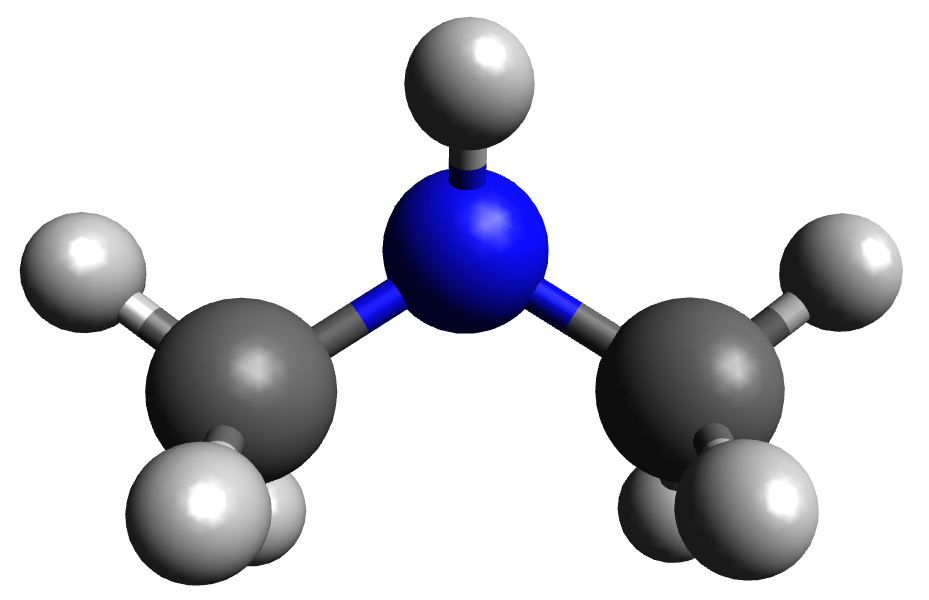}
\caption{Model of the dimethylamine molecule. Carbon atoms are symbolized by gray spheres, 
   hydrogen atoms are indicated by small, light gray spheres, and the nitrogen atom by a blue sphere.}
\label{mol-bild}
\end{figure}

\section{Spectroscopic properties of dimethylamine}
\label{rot_backgr}


\begin{figure}
\centering
\includegraphics[width=.65\columnwidth,angle=0]{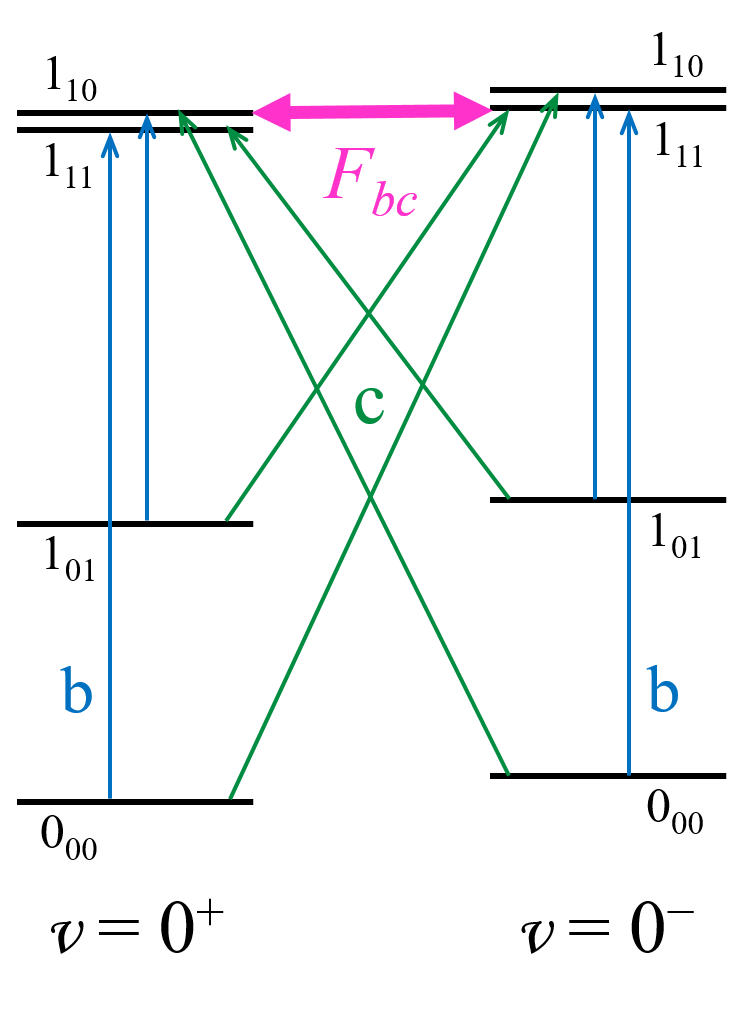}
\caption{Detail of the energy level structure of DMA. 
The stronger $c$-type transitions connect the two tunneling states while the weaker $b$-type transitions are within each 
tunneling state. 
Near-degeneracies of certain levels, here indicated by a magenta arrow, lead to perturbations; 
$F_{bc}$ is the lowest order parameter to treat these perturbations.}
\label{Interactions-bild}
\end{figure}


\begin{table}
\begin{center}
\caption{Values of $J$ of the strongest perturbations as indicated in Fig.~\ref{Interactions-bild} 
         pertaining to lower values of $K_a$ and number No. of different transition frequencies 
         in the final fit.}
\label{tab-interactions}
\begin{tabular}[h]{ccc}
\hline \hline
$K_a$ & $J$ & No. \\
\hline
 1 &   1   & 10$^a$ \\
 2 &   5   &  5     \\
 3 &  10   &  8     \\
 4 &  16   &  6     \\
 5 &  22   &  8     \\
 6 &  28   &  4     \\
 7 & 34/35 & 15     \\
 8 &  41   &  6     \\
 9 &  48   &  0$^b$ \\
10 &  55   &  0     \\
11 &  62   &  0     \\
\hline
\end{tabular}
\end{center}
$^a$ In addition, two transition frequencies from \citet{DMA_FTMW_2021} and seven from \citet{rot_gs_1968}.\\
$^b$ There are two transitions with $J = 47$ in the line list.
\end{table}


DMA is an asymmetric rotor with $\kappa = (2B - A - C)/(A - C) = -0.9140$ close to the prolate 
limit of $-$1. Fig.~\ref{mol-bild} shows that the molecule has $C_{\rm S}$ symmetry in a static 
picture; the H atom attached to the N atom can be above or below the CNC plane. 
However, the barrier to exchange of these two positions is relatively low, such that the H atom 
is able to tunnel between these equivalent positions. Rotation lifts the degeneracy and causes 
a symmetric tunneling state frequently labeled with $0^+$ and an antisymmetric tunneling state 
then labeled with $0^-$. The symmetry of the molecule in this dynamical picture is $C_{\rm 2v}$, 
as it is the symmetry of the transition state. The $b$-axis in this configuration is along the 
NH bond; the pure rotational transitions, those within each tunneling state, obey $b$-type 
selection rules. The $a$-axis is parallel to the line through the C atoms, and the $c$-axis is perpendicular to the CNC plane. 
The rotation-tunneling transitions, which connect the two tunneling states, obey $c$-type selection rules as the tunneling of the H atom 
takes place parallel to the $c$-axis. 
\citet{rot_gs_1968} determined the dipole moment components as 
$\mu _b = 0.295$~D and $\mu _c = 0.967$~D. Fig.~\ref{Interactions-bild} demonstrates possible transitions among the lowest energy levels of DMA.

Coriolis-type interaction occurs between the two tunneling states for levels with equal $J$ which differ 
in $K_a$ and $K_c$ by an even and odd number, respectively. 
The interaction is strongest (resonant) when the levels are close in energy. The most common interactions 
in a prolate rotor are those with $\Delta K_a = 0$ and $\Delta K_c = \pm1$. 
It is also shown in Fig.~\ref{Interactions-bild} that the first resonant interaction of 
this type occurs in DMA at $K_a = 1$ for $J = 1$. The figure demonstrates the general pattern 
of this type of resonance: the upper asymmetry level of the lower tunneling state interacts 
with the lower asymmetry level of the upper tunneling state. Since the asymmetry splitting 
decreases rapidly with increasing $K_a$ for a given $J$, this type of interaction is resonant 
at increasing $J$ for an increase in $K_a$. The usually one $J$ with strongest interaction 
for a given $K_a$ is listed in Table~\ref{tab-interactions} for $K_a \le 11$. The parameter 
$F_{bc}$ describes this interaction at lowest order.


\begin{figure}
\centering
\includegraphics[width=.65\columnwidth,angle=0]{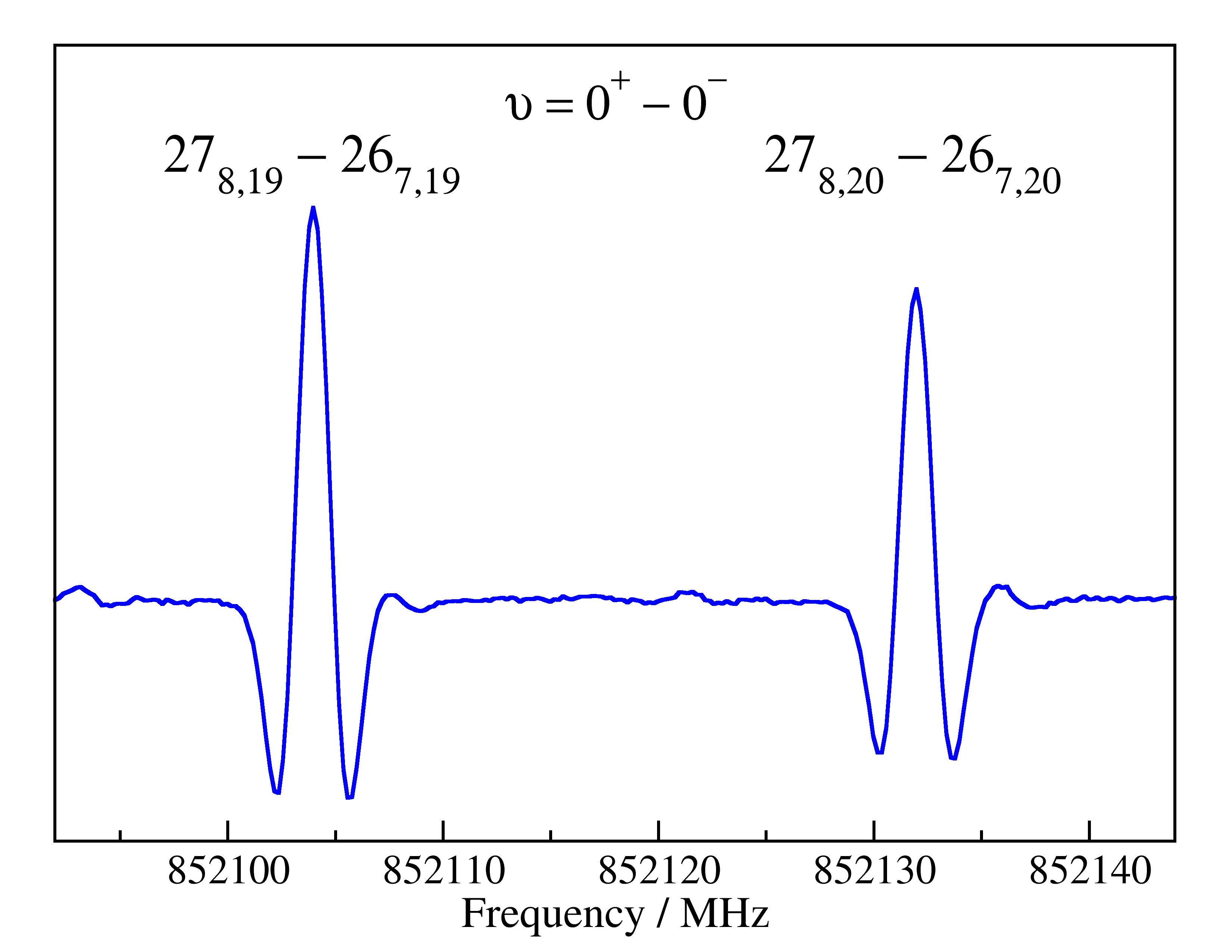}
\includegraphics[width=.65\columnwidth,angle=0]{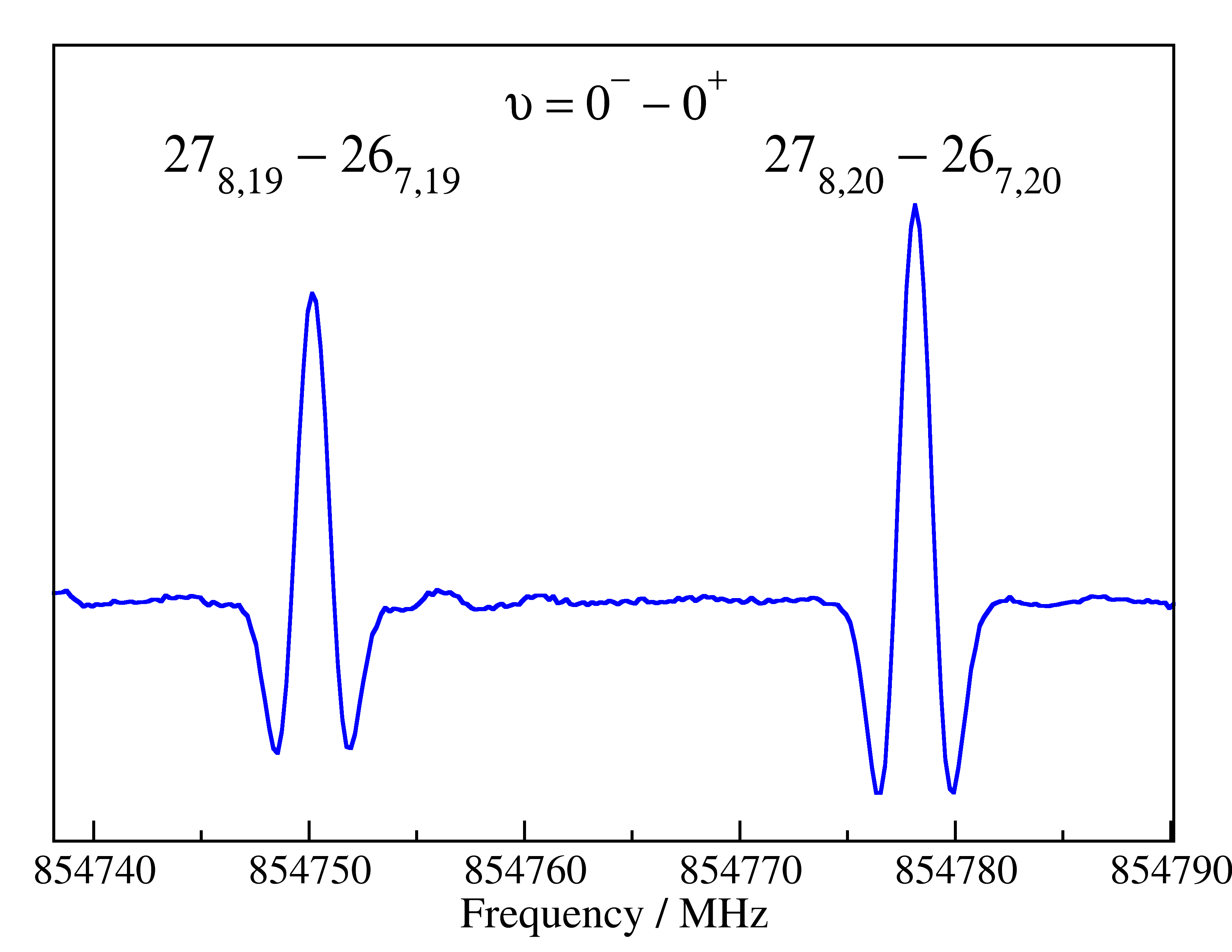}
\caption{Sections of the rotation-tunneling spectrum of DMA illustrating the spin-statistics 
   and the reversal upon exchange of tunneling states.}
\label{Spin-Stat-bild}
\end{figure}


The two equivalent methyl groups in DMA (Fig.~\ref{mol-bild}) lead to \textit{ortho} and \textit{para} spin-statistics 
with intensity ratios of 9~:~7. The \textit{ortho} and \textit{para} levels are described by $K_a + K_c$ being odd and even, 
respectively in $\upsilon = 0^+$ while it is opposite in $\upsilon = 0^-$, see Fig.~\ref{Spin-Stat-bild}.

\begin{figure}
\centering
\includegraphics[width=.75\columnwidth,angle=0]{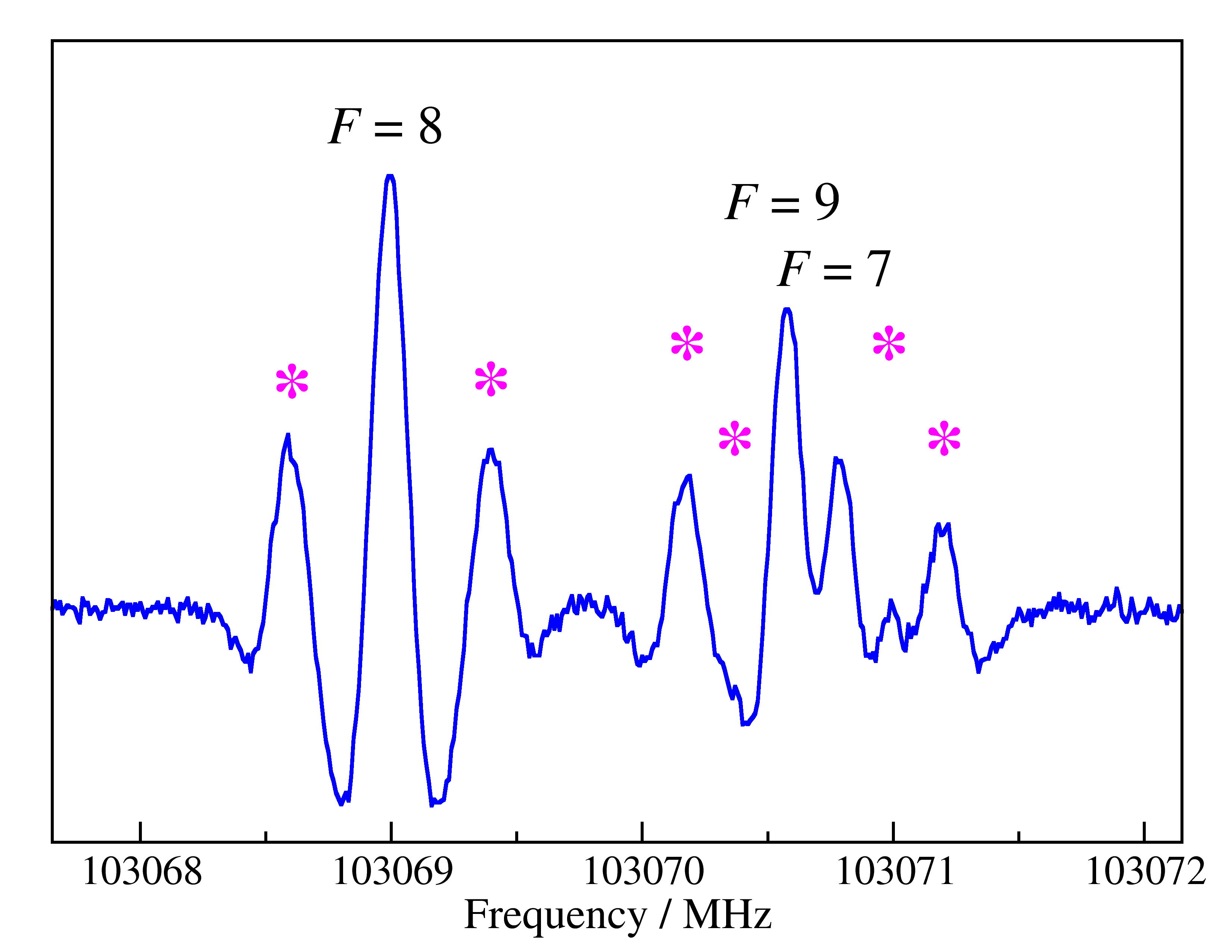}
\caption{The $\upsilon = 0^+ - 0^-$, $J_{K_a,K_c} = 8_{2,6} - 8_{1,8}$ transition of DMA. 
The $^{14}$N hyperfine splitting is resolved, but the $F = J \pm 1$ components are nearly blended whereas the $F = J$ component is isolated. 
Methyl internal rotation is also resolved, and the weaker internal rotation components are marked by magenta asterisks.}
\label{HFS-bild}
\end{figure}

The two methyl groups in DMA carry out hindered internal rotations, but this rotation is so 
strongly hindered that it was not resolved in the initial microwave study of the ground 
vibrational state \citep{rot_gs_1968}. Fig.~\ref{HFS-bild} shows weaker internal rotation 
components displaced by about 400~kHz to either side of the stronger central component.

The presence of the $^{14}$N nucleus ($I_{\rm N} = 1$) leads to HFS splitting. Each rotational 
level with $J \ge 1$ splits into three, and the strongest HFS components are those with 
$\Delta F = \Delta J$, where $F$ represents the combination of the rotational and nuclear spin 
angular momenta. Transitions with $\Delta F \ne \Delta J$ are also allowed as long as 
$\Delta F = 0, \pm1$ is fulfilled. These transitions are usually too weak to be observed 
except for transitions with low rotational quantum numbers or measured with very high signal-to-noise ratios (S/N). 
An example of HFS splitting is shown in Fig.~\ref{HFS-bild}. As one can see, the $F = J \pm 1$ 
components nearly coincide in frequency, and the frequency modulation causes a reduction 
of their intensities because the two components are not completely separated. 
In fact, the $F = J \pm 1$ components are frequently completely blended, whereas the $F = J$ component may be separated from these. 
This causes an asymmetric intensity ratio of about 2~:~1 for partly resolved HFS patterns.


\section{Spectroscopic results and determination of spectroscopic parameters}
\label{lab-results}

Fitting and calculation of the rotational spectrum of DMA was carried out with the SPFIT and SPCAT programs \citep{spfit_1991}. 
The two interacting tunneling states are commonly fit with a Hamiltonian that can be divided into a $2 \times 2$ matrix 
with the diagonal elements consisting of the usual Watson-type rotational Hamiltonians for $0^+$ and $0^-$ on the diagonal; 
the one for $0^-$ includes in addition the energy difference. The interaction Hamiltonian is off-diagonal. 
The treatment of the Coriolis interaction between the $0^+$ and $0^-$ states requires additional consideration. 
\citet{Coriolis_1970}, for example, discussed the need for two low order terms to treat the $c$-type Coriolis interaction between 
$\upsilon _1 = 1$ and $\upsilon _3 = 1$ of ozone: $iD_c J_c + F_{ab}(J_aJ_b + J_bJ_a)/2$. Additional rotational correction 
terms to either low order term may be needed if a large range of high quantum numbers is accessed, as in the example of ClClO$_2$ 
\citep{ClClO2_Coriolis_2002}. \citet{tunneling_2terms_1962} derived that both types of terms are allowed to treat the 
rotation-tunneling interaction in cyanamide, H$_2$NCN. \citet{tunneling_iD_1966} showed in the analysis of the ring-puckering in 
trimethylene sulfide, $c$-C$_3$H$_6$S, that at least under certain circumstances only one type of terms is required and proposed 
to use the odd order terms $iD_j$ with $j$ being $a$, $b$, or $c$, depending on the symmetry. \citet{rot_gs_1968} followed this 
recommendation in their analysis of the rotation-tunneling spectrum of DMA. 
\citet{Pickett-RAS_1972} proposed to use the reduced axis system 
in which the interaction Hamiltonian consists of appropriate axis-rotation terms $F_{ij}$ with $i$, $j$ being $a$, $b$, or $c$ and $i \ne j$. 
He showed that this method minimizes the effects of Coriolis coupling and may thus reduce the number of required interaction terms. 
The reduced axis system was the preferred method to treat tunneling-rotation and related spectra lately and was applied in all examples 
given below. 
The only non-zero interaction element in the case of DMA is $F_{bc}$ supplemented with its distortion corrections: 
$(F_{bc} + F_{bc,K}J_a^2 + F_{bc,J}J^2 + F_{2bc}(J_b^2 - J_c^2) + ...) \times (J_bJ_c + J_cJ_b)/2$. 
Two separate rotational Hamiltonians are traditionally employed, which is appropriate if the parameters in the two Hamiltonians differ considerably. 
This approach was applied, e.g., in the case of \textit{gauche}-propanal \citep{g-EtCHO_rot_2022}.

Noting that the two tunneling states $0^+$ and $0^-$ together comprise the ground vibrational state $\upsilon = 0$, 
it can be advantageous to rearrange the Hamiltonians and fit average spectroscopic parameters $X = (X(0^+) + X(0^-$))/2 
and differences $\Delta X = \{X(0^-) - X(0^+$)\}/2, as was done by 
\citet{aGg-eglyc_rot_2003} in their treatment of the lowest energy conformer of ethylene glycol, also known as ethanediol. 
\citet{aGg-eglyc_rot_2003} also pointed out that the differences in spectroscopic parameters can be interpreted 
as rotational corrections to the energy difference, 
e.g., $E_K$ is defined as $\Delta \{A - (B + C)/2\}$, $E_J$ as $\Delta \{(B + C)/2\}$, $E_2$ as $\Delta \{(B - C)/4\}$, and so forth.
We follow this interpretation in our current work. 
We should point out that in our definition of the differences, $0^-$ is higher in energy by $2E$ than $0^+$. 
The advantage of this formulation is that an average parameter or its difference can be employed 
individually in the fit independent of each other. A rotation-tunneling spectrum with a small 
energy difference may require fewer parameter differences than average parameters. On the other hand, 
the rotation-tunneling spectrum of a symmetric top molecule, such as NH$_3$, requires many differences 
in the fit to describe pure tunneling transitions. In addition, the purely axial average parameters 
(here $C$, $D_K$, etc.) are not determinable from regular ($\Delta K = 0$) transitions. 
Examples applying such rearranged Hamiltonians include the treatments of ethanethiol \citep{RSH_ROH_2016}, 
hydroxyacetonitrile \citep{HAN_rot_2017}, the low-energy conformer of ethanediol 
\citep{aGg-eglyc_rot_2020}, and isotopic cyanamide \citep{H2NCN-iso_rot_2019}; the difference 
between the number of average parameters and their differences is very small in the last example, 
which may, at least in part, be related to the large energy splitting between the two tunneling states.

The HFS splitting requires appropriate terms to be added to the Hamiltonian; these are the quadrupole 
coupling parameters $\chi _{ii}$ with $i = a$, $b$, $c$ on the diagonal and $\chi _{bc}$ off-diagonal. 
Since the quadrupole tensor is traceless, only two of the three $\chi _{ii}$ are independent.

The rotation-tunneling spectrum of DMA is relatively sparse on the level of the strongest transitions. 
In addition, several of the stronger series of transitions display HFS splitting that is neither well 
resolved, be it into the three strong components or be it into the $F = J$ and the overlapping 
$F = J \pm 1$ components, nor fully collapsed. Such patterns make it more difficult to determine 
rest frequencies very accurately. And finally, several series of interest are very weak such that 
their S/N may be insufficient in spectral recordings that cover large 
frequency regions in a reasonable amount of time. Therefore, we decided to search in most cases 
only for individual rotation-tunneling transitions.

An initial catalog file was created from the transition frequencies by \citet{rot_gs_1968} applying 
100~kHz as uncertainty for all lines as reported. We carried out exploratory measurements between 
76 and 124~GHz to evaluate the quality of the model. The transitions cover almost exclusively low 
or very low values in $J$ ($1 \le J \le 11$), notably the $R$-branch ($\Delta J = +1$) transitions 
with $K_a = 1 - 0$, $0 - 1$, $2 - 1$, and $1 - 2$ as well as $Q$-branch ($\Delta J = 0$) transitions 
with $K_a = 2 - 1$. Most of these transitions displayed fully (Fig.~\ref{HFS-bild}) or partially 
resolved HFS splitting. Internal rotation splitting affected sometimes the positions of the strongest 
central component; the uncertainty of the corresponding line was increased somewhat as long as 
the effect appeared to be small, otherwise, the line was not used in the fit.

In the course of the investigation, three of the initial transition frequencies \citep{rot_gs_1968} 
showed large residuals that were eventually interpreted as typographical errors as two residuals were 
very close to 1.0~MHz and one very close to 2.0~MHz.


\begin{figure*}
\centering
\includegraphics[width=1.9\columnwidth,angle=0]{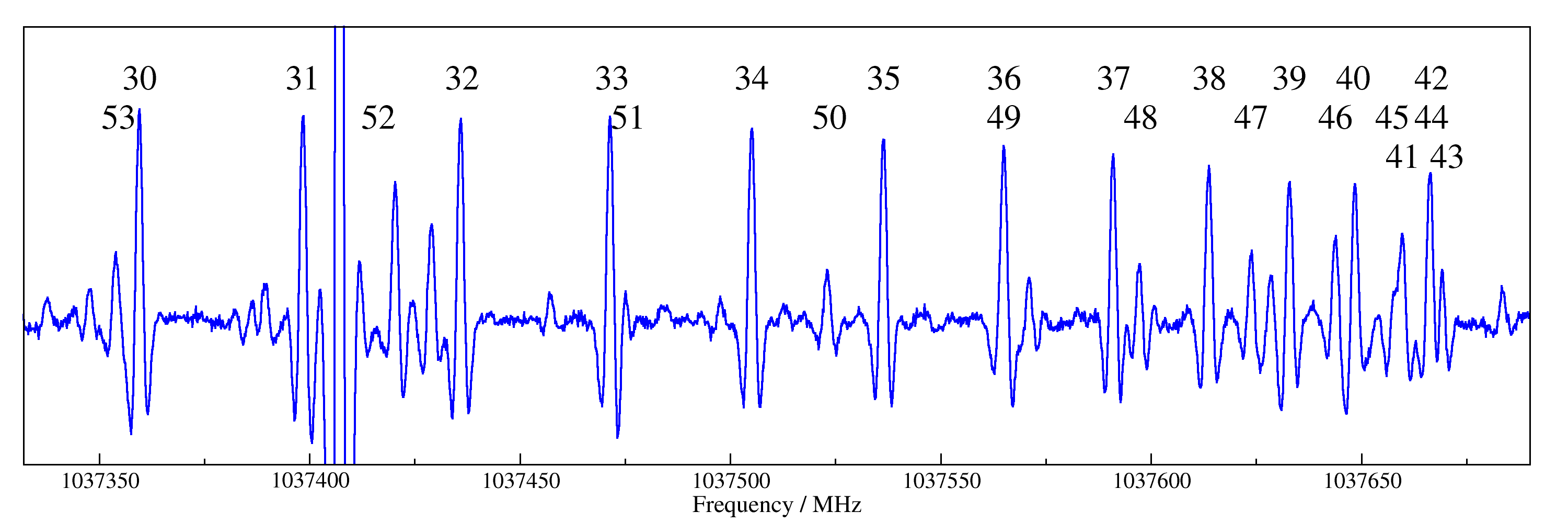}
\caption{Section of the dimethylamine terahertz spectrum displaying part of a $Q$-branch, i.e., $\Delta J = 0$, 
and the $J$ values are indicated; $K_a = 21 - 20$ and $\upsilon = 0^- - 0^+$.}
\label{rQ20-bild}
\end{figure*}


The majority of the measurements were made between 159 and 375~GHz, fairly extensive recordings 
were also carried out between 793 and 1091~GHz. We focused on transitions that had a splitting 
between the $F = J$ and the overlapping $F = J \pm 1$ components of at least one line width 
up to 254~GHz or on transitions having HFS splitting well within half of the estimated line width 
throughout the entire frequency region. 
We tried to record as many members of specific series as possible with limitations caused by 
the intrinsic weakness of a series, the reduced sensitivity of the spectrometer in certain regions, 
especially at the edges, or by accidental blending. We covered several different series of transitions 
pertaining to adjacent $K_a$ for low values of $K_a$. The strong $^rR$-branch transitions were 
covered up to $J = K_a = 16 - 15$ around 1067~GHz. All $K_a$ values were covered up to $K_a = 21 - 20$, 
which was accessed through both $Q$-branches, with $0^+$ or with $0^-$ representing the lower state 
level; part of the former one is shown in Fig.~\ref{rQ20-bild}. The highest $J$ level that we accessed 
had $J = 61$. The highest $J$ value with partially resolved HFS was 44 for the $K_a = 4 - 3$ $Q$-branch 
transitions within $0^+$ and $0^-$ near 252565~MHz; we only used the overlapping $F = J \pm 1$ components 
because the uncertainties of the $F = J$ components were much larger. The $F = J$ components were also 
kept in the line list for several $K_a = 6 - 5$ $Q$-branch transitions between $0^+$ and $0^-$ with 
38 being the highest $J$ value near 163889~MHz. 


\begin{table}
\begin{center}
\caption{Spectroscopic parameters (MHz) of dimethylamine.}
\label{tab-spec-parameters}
\renewcommand{\arraystretch}{1.10}
\begin{tabular}[t]{lr@{}l}
\hline \hline
Parameter & \multicolumn{2}{c}{Value}  \\
\hline
$A$                      &   34242&.442246~(208) \\    
$B$                      &    9334&.296360~(77)  \\   
$C$                      &    8215&.651275~(72)  \\   
$D_{K}   \times 10^{3}$  &     255&.2442~(33)    \\ 
$D_{JK}  \times 10^{3}$  &   $-$32&.49219~(66)   \\  
$D_{J}   \times 10^{3}$  &       8&.273664~(98)  \\
$d_{1}   \times 10^{3}$  &    $-$1&.614217~(34)  \\
$d_{2}   \times 10^{3}$  &       0&.014938~(16)  \\
$H_{K}   \times 10^{9}$  &    9104&.5~(220)      \\
$H_{KJ}  \times 10^{9}$  & $-$2063&.2~(41)       \\
$H_{JK}  \times 10^{9}$  &      66&.83~(47)      \\
$H_{J}   \times 10^{9}$  &       6&.592~(51)     \\
$h_{1}   \times 10^{9}$  &       3&.8166~(103)   \\
$h_{2}   \times 10^{9}$  &       0&.6967~( 61)   \\
$h_{3}   \times 10^{9}$  &       0&.0386~(13)    \\
$L_{K}   \times 10^{12}$ & $-$1691&.~( 64)       \\
$L_{KKJ} \times 10^{12}$ &  $-$734&.2~(117)      \\
$L_{JK}  \times 10^{12}$ &   $-$29&.92~(174)     \\
$L_{JJK} \times 10^{12}$ &    $-$2&.255~(114)    \\
$L_{J}   \times 10^{12}$ &       0&.1061~(96)    \\
$P_{K}   \times 10^{12}$ &       0&.521~(64)     \\
$P_{KKJ} \times 10^{12}$ &       0&.6835~(137)   \\
$P_{KJ}  \times 10^{12}$ &       0&.0345~(28)    \\
$F_{bc}$                 &      10&.98176(137)   \\  
$F_{bc,K} \times 10^{3}$ &    $-$6&.31~(27)      \\
$F_{bc,J} \times 10^{3}$ &    $-$0&.2747~(30)    \\
$2E$                     &    1322&.95355~(21)   \\  
$E_{K}   \times 10^{3}$  &  $-$288&.982~(22)     \\
$E_{J}   \times 10^{3}$  &      22&.1617~(23)    \\ 
$E_{KK}  \times 10^{6}$  &      70&.931~(160)    \\
$E_{JK}  \times 10^{6}$  &   $-$10&.808~(22)     \\
$E_{JJ}  \times 10^{6}$  &       0&.3817~(26)    \\
$E_{KKK} \times 10^{9}$  &   $-$12&.05~(33)      \\
$E_{JKK} \times 10^{9}$  &       2&.736~(87)     \\
$E_{JJK} \times 10^{9}$  &    $-$0&.2006~(123)   \\
$E_{JJJ} \times 10^{12}$ &       5&.56~(100)     \\
$E_{2} \times 10^{6}$    &     916&.10~(242)     \\
$E_{2,J} \times 10^{9}$  &      37&.46~(169)     \\
$\chi _{aa}$             &       2&.99972~(95)   \\ 
$\chi _{bb}$$^a$         &       0&.94217~(176)  \\ 
$\chi _{cc}$             &    $-$3&.94189~(146)  \\   
$\chi _{bc}$             &    $-$2&.959~(74)     \\
\hline
\end{tabular}
\end{center}
Watson's S reduction was used in the representation $I^r$. Numbers in parentheses are one standard deviation in units of the least significant figures.\\
$^a$ Derived parameter. 
\end{table}


We took care to describe the rotation-tunneling interaction in DMA well. To this end, we tried 
to cover as many near-degenerate levels as available. We accessed several transitions for each $K_a$ 
between 1 and 8, as can be seen in Table~\ref{tab-interactions}; we reached only one $J$ below 
the strongest interaction for $K_a = 9$. The interaction transfers intensity from $c$-type transitions 
to some of the weaker $b$-type transitions. This is best described with $F_{bc}$ being positive when 
$\mu _b$ and $\mu _c$ are chosen to be positive. The parameter $\chi_{bc}$ affects in particular 
the line positions in the transitions involving the $J = K_a = 1$ levels. 
Its value needed to be negative once $F_{bc}$ was chosen to be positive.

The determination of the spectroscopic parameters was carried out in the usual way. 
We tested after each round of assignments if one or more spectroscopic parameters would 
improve the quality of the fit by amounts that warranted keeping the respective parameter 
in the fit. Care was taken to try only parameters that are reasonable with respect to those 
already used in the fit. If at least one parameter improved the quality of the fit sufficiently, 
we chose the one that improved the quality the most and searched for additional parameters. 
Correlation among the parameters may cause a parameter that looked well determined, 
i.e., with a magnitude at least several times its uncertainty, to be not well determined. 
It was omitted from the fit, often only temporarily, if its omission did not cause a 
large deterioration of the fit.

Some of our spectral recordings resolved internal rotation splitting, as can be seen in Fig.~\ref{HFS-bild}. 
DMA is isoelectronic to dimethyl ether; the internal rotation patterns for this molecule were described in detail 
for example by \citet{DME_rot_2009}. Briefly, the nine equivalent internal rotation positions caused by the two 
equivalent methyl rotors in DMA and dimethyl ether lead to four distinct internal rotation substates 
$AA$, $AE$, $EA$, and $EE$. The lines pertaining to $AE$ and $EA$ are frequently blended, 
in particular in dimethyl ether and even more so in DMA with their small internal rotation splittings. 
The combined $AE$ and $EA$ peak has then the same intensity as the $AA$ peak, and the two peaks occur 
to either side of the central $EE$ peak displaced by essentially the same amount in the absence of 
strong torsion-rotation interaction. Information on the internal rotation splitting in DMA is 
provided in the microwave study on torsionally excited DMA \citep{rot_vt1_1971}. 
We did not use the internal rotation splitting of ground state DMA in almost all of our fits 
and only used the central, strong internal rotation peak ($EE$).

The SPFIT program is capable of fitting spectra with small internal rotation splitting \citep{propane_rot_2006,DME_rot_2009}, 
but attempts to introduce internal rotation into the treatment of the rotation-tunneling spectrum of DMA failed. 
The transition frequencies calculated with internal rotation splitting differed from those calculated without splitting by a few to several tens of kilohertz, 
often well outside the experimental uncertainties, and we were unable to reduce this difference. 
The reason is that the internal rotation splitting affects the tunneling splitting differently for different values of $K_a$. 
Therefore, we also used only the frequencies of the central internal rotation components from 
the Fourier transform microwave measurements \citep{DMA_FTMW_2021}. 
Four of the $c$-type HFS components and most of the few $b$-type HFS components showed larger residuals 
of around 10~kHz or more than the majority of the data, so the $b$-type transitions 
and the four $c$-type HFS components were omitted from the final fit. 
The residuals are a result of the density of lines for $b$-type transitions with low values of $J$ and for some of the $c$-type transitions. 
The Fourier transform microwave data did not require any 
additional parameters in the fit, 
such as nuclear spin-rotation parameters, but improved some of the lower order parameters substantially, in particular $\chi _{aa}$ and $\chi _{cc}$ as well as the rotational parameters 
and $\chi _{bc}$; effects were smaller for most other parameters. 
The resulting spectroscopic parameters are listed in Table~\ref{tab-spec-parameters}.

The 48 transition frequencies from \citet{rot_gs_1968} were fit to 103~kHz on average, acceptably close 
to the assigned 100 kHz uncertainty. The more recent microwave data from \citet{DMA_FTMW_2021} were fit 
to 1.39~kHz on average. We had chosen 1.5~kHz as final uncertainties for these 88 transition frequencies 
after we recognized that the value of 1.0~kHz, initially assumed by us, was too optimistic. Our 1294 transitions 
were reproduced to an rms error, also known as weighted or normalized rms, of 0.952. 
As we employed a substantial range of different uncertainties, the rms value of 28.7~kHz is dominated by the data with larger uncertainties. 
The number of different frequencies is 943 because of overlapping HFS components, unresolved asymmetry splitting or accidental blending.

The fit file and calculations of the rotational spectrum of DMA with and without HFS have been deposited as supplementary material to this paper together with an explanatory file. 
These files as well as additional files are also available in the Cologne Database for Molecule Spectroscopy \citep[CDMS\footnote{https://cdms.astro.uni-koeln.de/};][]{CDMS_2005,CDMS_2016}. 
The partition function $Q_{\rm rot}$ was calculated for selected temperatures beyond the standard temperatures in the CDMS by summation over the ground state energies up to $J = 130$ and $K_a = 60$. 
Spin-statistics and the spin-weight from the $^{14}$N nucleus were taken into account.
In the warm ISM ($T \gtrsim 100$~K), excited vibrational states are populated non-negligibly. 
We evaluated the vibrational partition function for a posteriori correction in the harmonic approximation. 
\citet{DMA_IR_Ra_1973} and \citet{DMA_FIR-Ra_1977} reported on gas phase infrared and Raman measurements. 
These reports are, however, not always consistent with each other and are incomplete as not all of the 24 fundamental vibrations were identified. 
Explanations for the incompleteness and inconsistency are that some fundamentals are very weak and may be blended with potentially stronger overtone and combination bands. 
Moreover, some fundamentals occur in a very narrow energy range. 
We carried out quantum-chemical calculations at the Regionales Rechenzentrum der Universit{\"a}t zu K{\"o}ln (RRZK) using the commercially available program Gaussian~09 \citep{Gaussian09E} in order to calculate the harmonic and anharmonic fundamentals of DMA. 
We employed the B3LYP hybrid density functional \citep{Becke_1993,LYP_1988} for that matter together with the correlation consistent basis set augmented with diffuse basis functions commonly denoted aug-cc-pVTZ \citep{cc-pVXZ_1989}.
There are 17 fundamentals below $\sim$1500~cm$^{-1}$; the highest ones correspond to a Boltzmann factor of $\sim$10$^{-3}$ at 300~K; the remaining fundamentals are above 2700~cm$^{-1}$. 
Wherever fundamentals are isolated and not too weak, there is usually good to reasonable agreement between experimental and calculated values. 
\citet{DMA_IR_Ra_1973} and \citet{DMA_FIR-Ra_1977} reported also vibrations around $\sim$1450~cm$^{-1}$, but fewer than 7, and some of their values may correspond to overtones or combination bands. 
The experimental and quantum-chemically evaluated fundamentals are presented in Table~\ref{vibs}, and 
the resulting values for $Q_{\rm rot}$ and $Q_{\rm vib}$ are summarized in Table~\ref{Q-values}.


\begin{table}
\begin{center}
\caption{Mode numbers$^a$ and  experimental fundamental vibrations (cm$^{-1}$) of DMA 
  along with harmonic (Harm.) and anharmonic (Anharm.) values from a B3LYP/aug-cc-pVTZ 
  quantum-chemical calculation employed for the calculation 
  of the vibrational partition function.}
\label{vibs}
\begin{tabular}[h]{ccccc}
\hline 
Mode No. & (b) & (c) & Harm. & Anharm. \\
\hline 
24 &  ---  &  219.4 &  224.8 &  243.1 \\
13 &  230? &  256.3 &  254.2 &  259.0 \\
12 &  384  &  383   &  380.5 &  375.4 \\
11 &  735  &   ?    &  755   &  685   \\
10 &  928  &  929   &  935   &  913   \\
23 &  ---  &  ---   & 1034   & 1026   \\
22 &  ---  &  ---   & 1100   & 1078   \\
21 & 1145? & 1123   & 1165   & 1132   \\
9  & 1145? & 1159   & 1184   & 1148   \\
8  & 1240  & 1240   & 1265   & 1234   \\
20 &       &        & 1442   & 1412   \\
7  &       &        & 1467   & 1438   \\
19 &       &        & 1475   & 1438   \\
18 &       &        & 1487   & 1460   \\
6  &       &        & 1495   & 1449   \\
5  &       &        & 1516   & 1475   \\
17 &       &        & 1516   & 1478   \\
\hline 
\end{tabular}
\end{center}
$^a$ In the static $C_{\rm S}$ symmetry; modes 1 to 13 are in a', 14 to 24 in a''. 
Values are give only for vibrations up to 1500~cm$^{-1}$, see text.\\ 
$^a$ Experimental data from gas phase IR and Raman measurements \citep{DMA_IR_Ra_1973}.\\
$^b$ Experimental data from gas phase IR and Raman measurements \citep{DMA_FIR-Ra_1977}.
\end{table}


\begin{table}
\begin{center}
\caption{Rotational$^a$ and vibrational partition function, 
  $Q_{\rm rot}$ and $Q_{\rm vib}$, of DMA at selected temperatures (K).}
\label{Q-values}
\begin{tabular}[h]{r@{}lr@{}lc}
\hline 
\multicolumn{2}{c}{$T$} & \multicolumn{2}{c}{$Q_{\rm rot}$} & $Q_{\rm vib}$ \\
\hline 
 300&.    & 822487&.8817 & 2.7773 \\
 250&.    & 625513&.5474 & 2.0932 \\
 225&.    & 534004&.2559 & 1.8313 \\
 200&.    & 447470&.9752 & 1.6117 \\
 180&.    & 382025&.6185 & 1.4627 \\
 170&.    & 350623&.5753 & 1.3964 \\
 160&.    & 320135&.2065 & 1.3353 \\
 150&.    & 290588&.3240 & 1.2793 \\
 140&.    & 262013&.4910 & 1.2284 \\
 125&.    & 221048&.8001 & 1.1615 \\
 100&.    & 158173&.3256 & 1.0757 \\
  75&.    & 102752&.3393 & 1.0232 \\
  50&.    &  55958&.0584 & 1.0025 \\
  37&.5   &  36366&.3518 & 1.0003 \\
  18&.75  &  12889&.3154 & 1.0000 \\
   9&.375 &   4580&.6180 & 1.0000 \\
   5&.0   &   1800&.4981 & 1.0000 \\
   2&.725 &    736&.4721 & 1.0000 \\
\hline 
\end{tabular}
\end{center}
$^a$ We employ a reduced spin-weight of 9~:~7 for \textit{ortho} and \textit{para} states, respectively, 
instead of the full spin-weight of 72~:~56 to circumvent limitations for the upper state degeneracies in the catalog files.
\end{table}

\section{Discussion of the spectroscopic parameters}
\label{spec-discussion}

The spectroscopic parameters of DMA are prototypical for an asymmetric top molecule close to the prolate limit, 
as the series of parameters describing the asymmetry (the $d$'s and $h$'s) converge faster than that of the 
purely $J$-dependent parameters ($D_J$, $H_J$, $L_J$), and that, in turn, converges faster than the series of 
purely $K$-dependent parameters ($D_K$, $H_K$, $L_K$, $P_K$). 
In addition, the magnitudes of the diagonal distortion parameters within a given order decrease considerably from the purely $K$-dependent parameters to the purely $J$-dependent 
parameters with the minor exception of $P_K$ and $P_{KKJ}$. A strong decrease is also seen for $d_1$ and $d_2$ 
as well as for $h_1$ to $h_3$. The distortion corrections to the energy behave in a similar way, and their number 
is smaller than that of the regular distortion parameters, commensurate with the moderate energy difference 
between $0^+$ and $0^-$.

The value of $F_{bc}$ is quite small, and only two distortion corrections are required to fit this fairly 
large data set well. The HFS parameters are very well determined, including the interaction parameter $\chi _{bc}$. 
It is not surprising that the $^{14}$N nuclear spin-rotation parameters could not be determined, as the magnetic 
moment of $^{14}$N is quite small. Inclusion of resolved internal rotation components other than the central 
$EE$ components into a fit should not only enable the direct determination of a probably small number of 
internal rotation parameters, but may also reduce the uncertainties of some of the lower order parameters slightly.

Our present spectroscopic parameters are sufficiently accurate for all kinds of astronomical observations. 
The neglect of the internal rotation does not pose a limitation in sources such as Sgr~B2(N) or G+0.693$-$0.027. 
Moreover, it should affect our current parameter values negligibly. 
Internal rotation may need to be considered around 100~GHz for sources with line widths less than 2~km\,s$^{-1}$. 
The internal rotation splitting is likely too small to be resolved at much higher frequencies. 
Conversely, the DMA lines are probably too weak at lower frequencies in the warm ISM (around 100 to 200~K).

\section{Search for dimethylamine toward G+0.693--0.027}
\label{obs_Gplus}

\subsection{Observations}
\label{obs-det_Gplus}

We searched for DMA toward the G+0.693$-$0.027 molecular cloud, located in the Sgr~B2 Giant Molecular Cloud of the Galactic center region. 
This source exhibits an extremely rich molecular complexity, and numerous molecular species have been detected for the first time toward it 
(see e.g. \citealt{rivilla2019b,rivilla2020b,bizzocchi2020,rivilla2021a,rivilla2021b,rivilla2022a,rivilla2022b,rodriguez-almeida2021a,rodriguez-almeida2021b,zeng2021,jimenez-serra2022}).
We used a sensitive unbiased spectral survey performed with the Yebes 40m (Guadalajara, Spain) and the IRAM~30m (Granada, Spain) telescopes. 
The position switching observations were centered at $\alpha$(J2000.0)=$\,$17$^{\rm h}$47$^{\rm m}$22$^{\rm s}$, $\delta$(J2000.0)=$\,-$28$^{\circ}$21$^{\prime}$27$^{\prime\prime}$. 
The Yebes 40m observations cover a spectral range from 31.0~GHz to 50.4~GHz, while
the IRAM 30m observations cover the spectral ranges 71.77$-$116.72~GHz, 124.8–175.5~GHz, and 199.8$-$238.3~GHz.
The line intensity of the spectra was measured in units of $T_{\mathrm{A}}^{\ast}$ as the molecular emission toward G+0.693 is extended 
over the beam (\citealt{requena-torres_organic_2006,requena-torres_largest_2008,zeng2018,zeng2020}).
The noise of the spectra (in $T_{A}^{*}$) depends on the frequency range, and varies from 1 to 10~mK. 
The spectra were smoothed to velocity resolutions of 1.0$-$2.6~km~s$^{-1}$, depending on the frequency. 
For more detailed information of the observational survey we refer to \citet{rivilla2022c}.


\begin{table}
\begin{center}
\caption{Column densities and abundances of DMA and related species obtained toward G+0.693.}
\label{table-G0693}
\begin{tabular}{cccc}
\hline 
Molecule & $N$ & $\chi$$^a$ & $\frac{N_{\rm ref}}{N}$$^b$ \\
         & (cm$^{-2}$) & & \\
\hline 
CH$_3$OH$^c$     & 1.5$\times$10$^{16}$       & 1.1$\times$10$^{-7}$        & 1 \\
CH$_3$OCH$_3$    & 1.1$\times$10$^{14}$       & 8.1$\times$10$^{-10}$       & 136 \\
CH$_3$NH$_2$$^d$ & 3.0$\times$10$^{15}$       & 2.2$\times$10$^{-8}$        & 5 \\
CH$_3$NHCH$_3$   & $\leq$7.6$\times$10$^{13}$ & $\leq$5.6$\times$10$^{-10}$ & $\geq 197$ \\ 
\hline
\end{tabular} 
\end{center}
$^a$ Abundances are derived using a H$_2$ column density of $N$(H$_2$)=1.35$\times$10$^{23}$$\,$cm$^{-2}$ \citep{martin_tracing_2008}.\\
$^b$ $N_{\rm ref} = N$(CH$_3$OH).\\
$^c$ Extracted from \citet{jimenez-serra2022}.\\
$^d$ From \citet{zeng2018}.
\end{table}

\subsection{Nondetection of dimethylamine toward G+0.693--0.027}
\label{obs-res_Gplus}

\begin{figure}
\centering
\includegraphics[width=\columnwidth,angle=0]{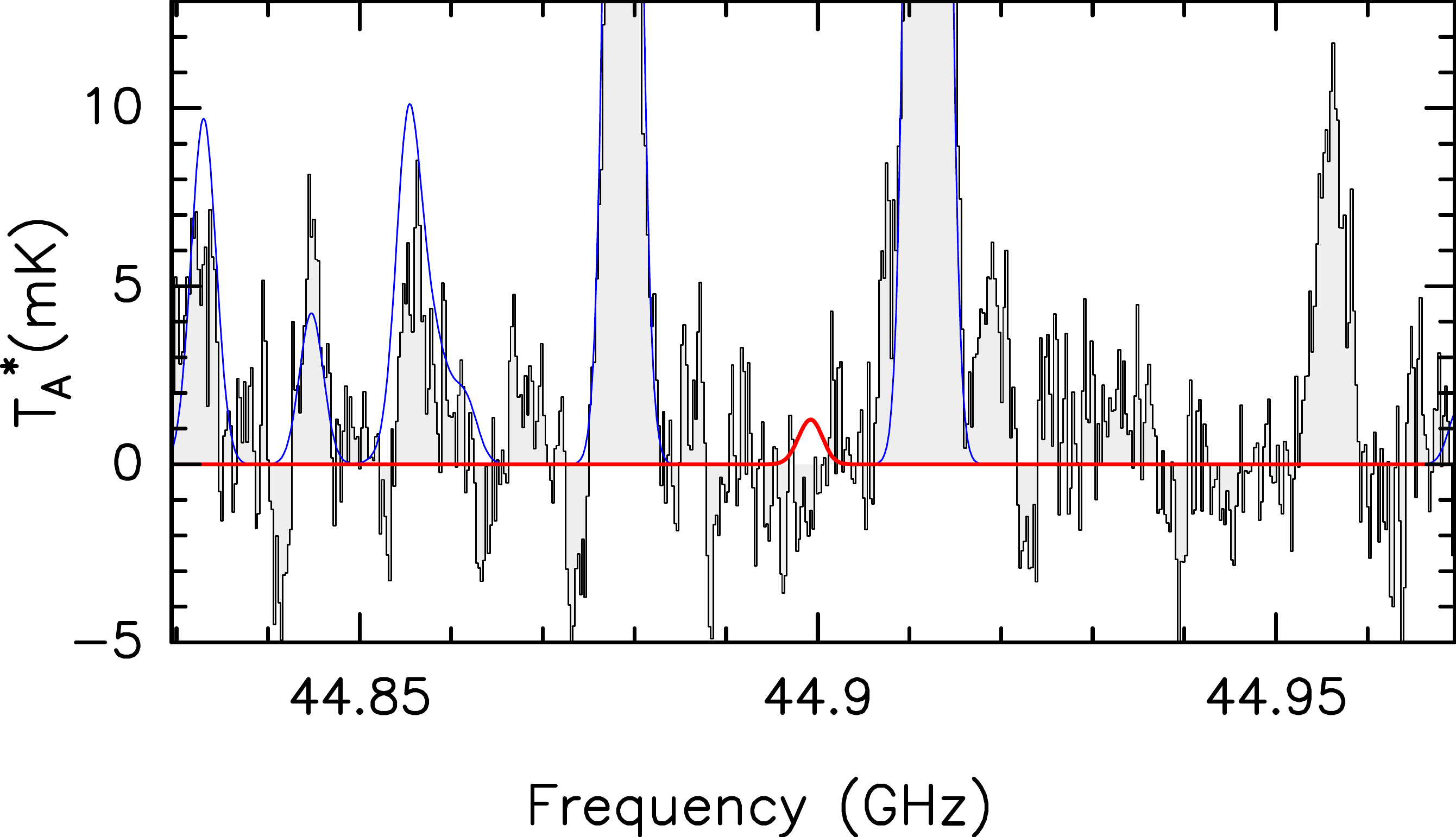}
\caption{Simulation of the $\upsilon = 0^- - 0^+$ $1_{1,0} - 0_{0,0}$ transition of DMA near 44.9~GHz with unresolved $F=$ 0$-$1, 2$-$1 and 1$-$1 
hyperfine components used to estimate the upper limit to its column density toward G+0.693 (see text). 
The black line and gray histogram show the observed spectrum, the red curve is the LTE synthetic model of DMA using 
the 3$\sigma$ column density upper limit derived using the parameters outlined in the text. 
The blue curve indicates the combined emission of all the molecules identified toward the source so far.}
\label{fig-G0693}
\end{figure}

We implemented the spectroscopic entry of DMA presented in this work into the MADCUBA package{\footnote{Madrid Data Cube Analysis is a software developed at the Center of Astrobiology (CAB) in Madrid and built using the Image Processing and Analysis 
in Java (ImageJ) infrastructure; http://cab.inta-csic.es/madcuba/}} (version 28/10/2022; \citealt{martin2019}).
Using the SLIM (Spectral Line Identification and Modeling) tool of MADCUBA, we generated a synthetic spectrum of DMA under the assumption of local thermodynamic equilibrium (LTE) and compared it with the observed spectra. 
The molecule is not detected in the observational survey. Most of the DMA transitions appear heavily blended with brighter transitions from other molecules already identified in the survey.
To derive the upper limit to the column density, we have used the brightest emission according to the simulated LTE model that appears unblended, 
which are three unresolved HFS components of the $\upsilon = 0^- - 0^+$ $1_{1,0} - 0_{0,0}$ transition with $F=$ 0$-$1, 2$-$1 and 1$-$1 that fall in the Q-band survey performed with the Yebes telescope and that are shown in Fig.~\ref{fig-G0693}. 
For the LTE simulated spectrum, we have used the excitation temperature derived for \ce{CH3NH2} by \citet{zeng2018}, 16~K, which is similar to those derived for \ce{C2H3NH2} and \ce{C2H5NH2} (18 and 12~K, respectively; \citealt{zeng2021}). 
We also used the velocity and linewidth found for the latter species: $V_{\rm LSR}=67$~km~s$^{-1}$ and $FWHM=18$~km~s$^{-1}$; the beam size at 44.9~GHz is 39''. 
Unfortunately, the predicted emission just appears in a spectral region which shows a local minimum in the data with respect to the adjusted broad band baseline. 
Considering the noise of the data, we obtained a 3$\sigma$ upper limit in integrated area of $N<$ 7.6$\times$10$^{13}$~cm$^{-2}$, which corresponds to an abundance compared to H$_2$ of $<$ 5.6$\times$10$^{-10}$, 
using $N$(H$_{2}$)=1.35$\times$10$^{23}$~cm$^{-2}$ (\citealt{martin_tracing_2008}). 
This upper limit as well as column densities of methylamine, methanol and dimethyl ether are summarized in Table~\ref{table-G0693} together with derived quantities. 
The upper limit to the column density of DMA could be affected by the local baseline by less than a factor of two when the local baseline uncertainty is considered. 
The rms noise level in the Yebes spectrum near 44.9~GHz is 1.6~mK in a 1.5~km~s$^{-1}$ channel. 
Fortunately, the spectral region of these transitions seems to be clean from contamination, opening the possibility of a detection in future more sensitive Q band searches.


\section{Search for dimethylamine toward Sgr~B2(N)}
\label{obs_SgrB2}

\subsection{Observations}
\label{obs-det_SgrB2}

We also used the imaging spectral line survey Reexploring Molecular Complexity
with ALMA (ReMoCA) that targeted the high-mass star forming protocluster 
Sgr~B2(N) with ALMA to search for dimethylamine in the interstellar medium. 
The data reduction and the method of analysis of this survey were described in 
\citet{Belloche19}. We summarize the main features of the survey below. 
The phase center is located at the equatorial position 
($\alpha, \delta$)$_{\rm J2000}$= 
($17^{\rm h}47^{\rm m}19{\fs}87, -28^\circ22'16{\farcs}0$). This position is 
half-way between the two hot molecular cores Sgr~B2(N1) and Sgr~B2(N2). The
survey covers the frequency range from 84.1~GHz to 114.4~GHz at a spectral 
resolution of 488~kHz (1.7 to 1.3~km~s$^{-1}$). This frequency coverage was
obtained with five different tunings of the receivers. The observations 
achieved a sensitivity per spectral channel ranging between 
0.35~mJy~beam$^{-1}$ and 1.1~mJy~beam$^{-1}$ (rms) depending on the setup, with 
a median value of 0.8~mJy~beam$^{-1}$. The angular resolution (HPBW) varies 
between $\sim$0.3$\arcsec$ and $\sim$0.8$\arcsec$ with a median value of 
0.6$\arcsec$. This corresponds to $\sim$4900~au at the distance of Sgr~B2 
\citep[8.2~kpc,][]{Reid19}. We further improved the process that separates 
the line and continuum emission by adding two reference positions to the pool 
of positions that were used to find the frequency ranges that contain 
absorption features \citep[for more details about the separation of line and 
continuum emission, see][]{Belloche19}.

For this work we analyzed the spectra of two positions. The first 
one, Sgr~B2(N1S), is located at ($\alpha, \delta$)$_{\rm J2000}$= 
($17^{\rm h}47^{\rm m}19{\fs}87$, $-28^\circ22\arcmin19{\farcs}5$). It is 
offset by about 1$\arcsec$ to the south of the main hot core Sgr~B2(N1). This
position was chosen by \citet{Belloche19} for its lower continuum opacity 
compared to the peak of the hot core. The second position is the position 
called Sgr~B2(N2b) by \citet{Belloche22}. It is located in the secondary hot 
core Sgr~B2(N2) at ($\alpha, \delta$)$_{\rm J2000}$= 
($17^{\rm h}47^{\rm m}19{\fs}83, -28^\circ22'13{\farcs}6$). This position was 
chosen as a compromise between getting narrow line widths to reduce the level
of spectral confusion and keeping a high enough H$_2$ column density to detect 
less abundant molecules.

Like in our previous ReMoCA studies \citep[e.g.,][]{Belloche19,Belloche22}, we compared the observed spectra to synthetic spectra computed 
under the assumption of LTE with the astronomical software Weeds \citep[][]{Maret11}. 
This assumption is justified by the high densities of the regions where hot-core emission is detected in Sgr~B2(N) 
\citep[$>1 \times 10^{7}$~cm$^{-3}$, see][]{Bonfand19}. The calculations take
into account the finite angular resolution of the observations and the optical 
depth of the rotational transitions. For each position, we derived by eye a 
best-fit synthetic spectrum for each molecule separately, and then added 
together the contributions of all identified molecules. Each species was 
modeled with a set of five parameters: size of the emitting region 
($\theta_{\rm s}$), column density ($N$), temperature ($T_{\rm rot}$), linewidth 
($\Delta V$), and velocity offset ($V_{\rm off}$) with respect to the assumed 
systemic velocity of the source, $V_{\rm sys}=62.0$~km~s$^{-1}$ for Sgr~B2(N1S) 
and $V_{\rm sys}=74.2$~km~s$^{-1}$ for Sgr~B2(N2b). The linewidth and velocity 
offset are obtained directly from the well detected and not contaminated 
lines. The emission of complex organic molecules is extended over several 
arcseconds around Sgr~B2(N1) \citep[see][]{Busch22}. For the LTE modeling, we 
assumed like in \citet{Belloche19} an emission size of 2$\arcsec$, which is 
much larger than the beam, meaning that the derived column densities do not 
depend on the exact value of this size parameter. In the case of Sgr~B2(N2b), 
the size of the emission of a given molecule was estimated from integrated 
intensity maps of transitions of this given molecule that were found to be 
relatively free of contamination from other species.

\subsection{Search for DMA toward Sgr~B2(N1S) and Sgr~B2(N2b)}
\label{obs-res_SgrB2}

\begin{figure}
\centerline{\resizebox{1.0\hsize}{!}{\includegraphics[angle=0]{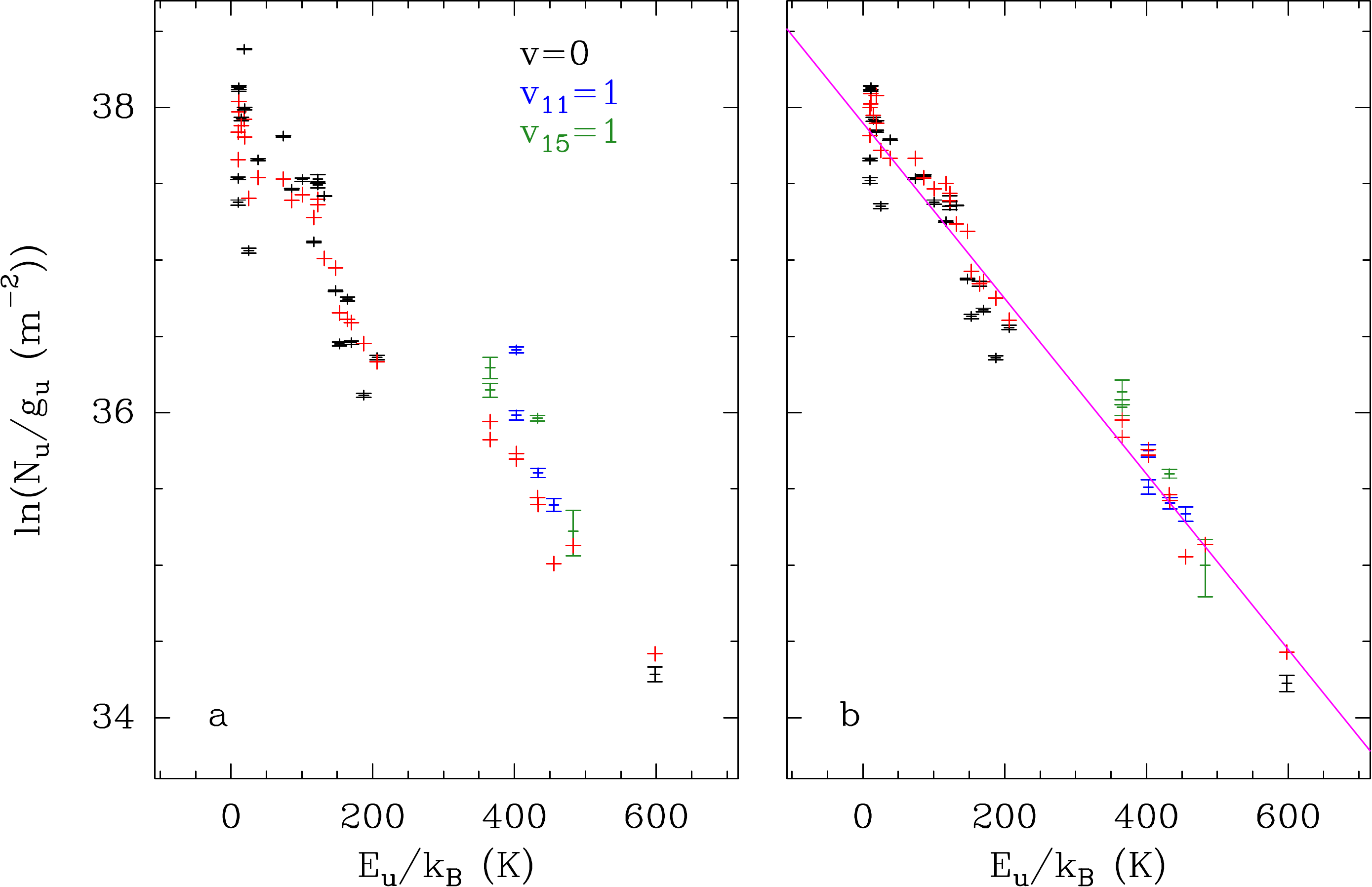}}}
\caption{Population diagram of CH$_3$OCH$_3$ toward Sgr~B2(N1S). The
observed data points are shown in black, blue, or green while the synthetic
populations are shown in red. No correction is applied in panel {\bf a}.
In panel {\bf b}, the optical depth correction has been applied to both the
observed and synthetic populations and the contamination by all other
species included in the full model has been subtracted from the observed
data points. The purple line is a linear fit to the observed populations (in
linear-logarithmic space).
}
\label{f:popdiag_ch3och3_n1s}
\end{figure}

\begin{figure}
\centerline{\resizebox{1.0\hsize}{!}{\includegraphics[angle=0]{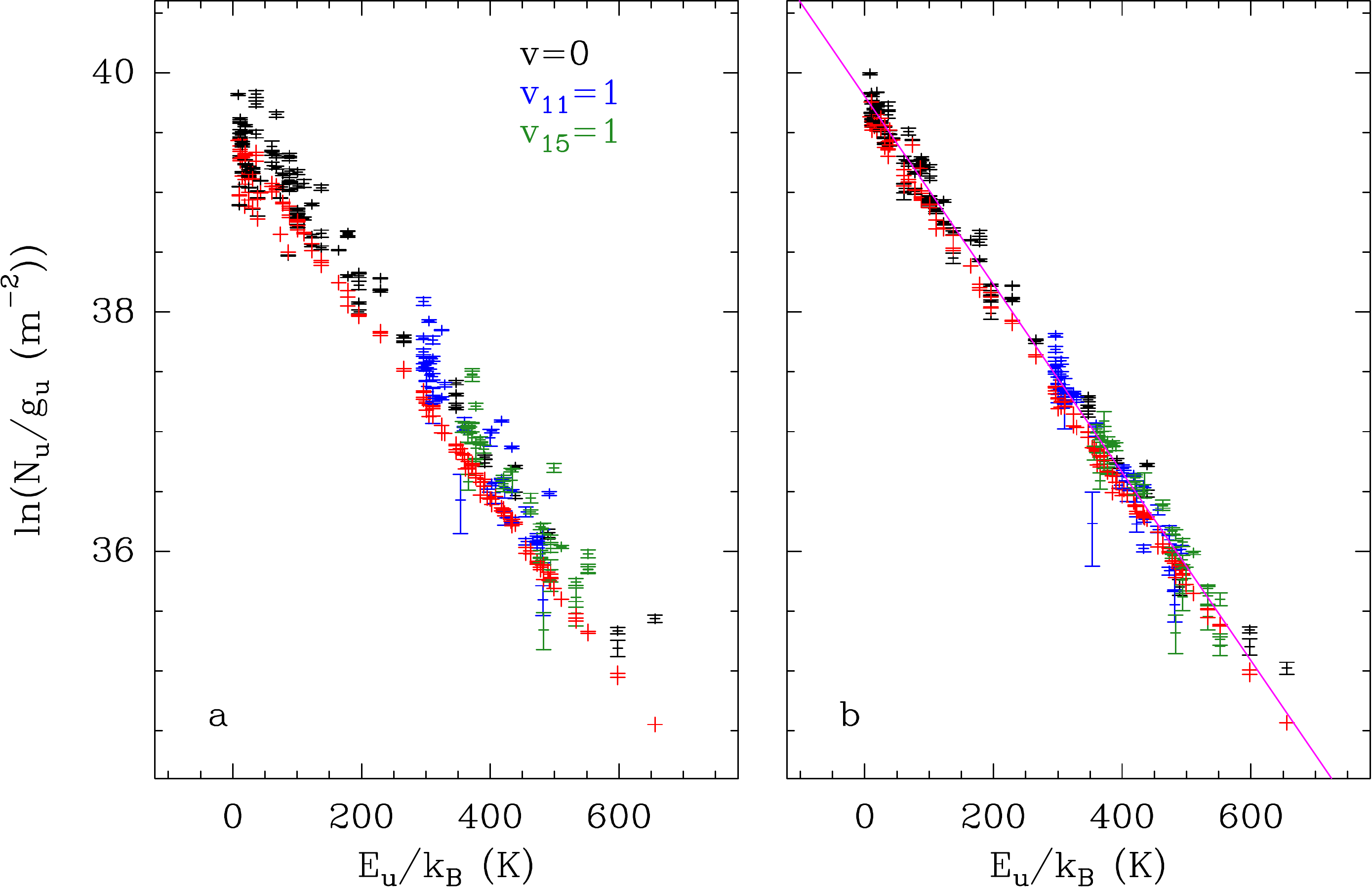}}}
\caption{Same as Fig.~\ref{f:popdiag_ch3och3_n1s}, but for Sgr~B2(N2b).
}
\label{f:popdiag_ch3och3_n2b}
\end{figure}

\input{tab_ch3nhch3_popfit_n1s.tex}

\input{tab_ch3nhch3_popfit_n2b.tex}

Before searching for dimethylamine, CH$_3$NHCH$_3$, toward Sgr~B2(N1S) and Sgr~B2(N2b), we modeled the rotational emission of the similar molecule dimethyl ether, CH$_3$OCH$_3$. 
We used the spectroscopic calculations available in the CDMS \citep{CDMS_2005}
for the vibrational ground state (version 1 of entry 46514), which are based
on \citet{DME_rot_2009} and references therein. 
For the vibrationally excited states $\upsilon_{11}=1$ and $\upsilon_{15}=1$, we used like in \citet{Belloche13}
predictions provided by C. Endres. Dimethyl ether is well detected toward both 
Sgr~B2(N1S) and Sgr~B2(N2b), with dozens of lines in its vibrational ground 
state easily identified (see Figs.~\ref{f:remoca_ch3och3_ve0_n1s} and 
\ref{f:remoca_ch3och3_ve0_n2b}). Two dozen lines in each of its 
vibrationally excited states $\upsilon_{11}=1$ and $\upsilon_{15}=1$  are also 
clearly detected toward Sgr~B2(N2b) 
(Figs.~\ref{f:remoca_ch3och3_v11e1_n2b}--\ref{f:remoca_ch3och3_v15e1_n2b}).
Because of the broader line widths that increase the level of spectral 
confusion (5~km~s$^{-1}$ versus 3.5~km~s$^{-1}$ toward Sgr~B2(N2b)), only a few 
lines from within $\upsilon_{11}=1$ and $\upsilon_{15}=1$ are 
sufficiently free of contamination to be identified toward Sgr~B2(N1S) 
(Figs.~\ref{f:remoca_ch3och3_v11e1_n1s}--\ref{f:remoca_ch3och3_v15e1_n1s}).
As explained in Sect.~\ref{obs-det_SgrB2}, we assumed an emission size of 
2$\arcsec$ to model the emission of dimethyl ether toward Sgr~B2(N1S). In the
case of Sgr~B2(N2b), integrated intensity maps of lines of dimethyl ether that
are free of contamination suggest an emission size on the order of 
0.8$\arcsec$. Figures~\ref{f:popdiag_ch3och3_n1s} and 
\ref{f:popdiag_ch3och3_n2b} show population diagrams of dimethyl ether for
Sgr~B2(N1S) and Sgr~B2(N2b), respectively. A fit to these population diagrams
yields rotational temperatures of $\sim$170~K and $\sim$130~K, respectively
(see Tables~\ref{t:popfit_n1s} and \ref{t:popfit_n2b}). Assuming these
temperatures, we adjusted synthetic LTE spectra to the observed spectra and
obtained the column densities reported in Tables~\ref{t:coldens_n1s} and 
\ref{t:coldens_n2b} for dimethyl ether. We also list in these tables the 
parameters that we previously obtained from the ReMoCA survey for methanol
 toward both positions \citep[][]{Motiyenko20,Belloche22}, as well as the
parameters that we recently obtained for methylamine, CH$_3$NH$_2$, toward 
Sgr~B2(N1S) \citep[][]{Gyawali23}, using the spectroscopic predictions 
available in the Lille Spectroscopic Database (version 2021.08.hfs of entry 
31802), which are based on \citet{Motiyenko14}.

\input{tab_ch3nhch3_weedsmodel_n1s.tex}

\input{tab_ch3nhch3_weedsmodel_n2b.tex}

Methylamine turns out to be more difficult to identify toward Sgr~B2(N2b) 
than toward Sgr~B2(N1S). Methylamine has a similar rotational temperature as 
methanol toward Sgr~B2(N1S), which led us to assume the same rotational 
temperature and emission size as methanol toward Sgr~B2(N2b). The best-fit 
synthetic LTE spectrum of methylamine using these parameters is shown in 
Fig.~\ref{f:remoca_ch3nh2_ve0_n2b}. Most transitions of this molecule are 
contaminated by emission from other species. Only two transitions are 
relatively free of contamination (at 84\,306 and 112\,273~MHz). As a result, 
we only claim a tentative detection of methylamine toward Sgr~B2(N2b). The 
parameters of its best-fit LTE model are given in Table~\ref{t:coldens_n2b}.

\begin{figure*}
\centerline{\resizebox{0.85\hsize}{!}{\includegraphics[angle=0]{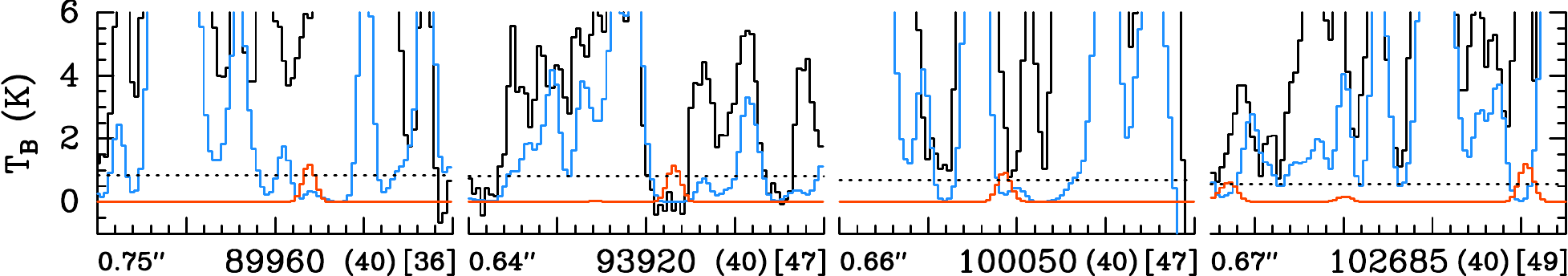}}}
\caption{Selection of rotational transitions of dimethylamine CH$_3$NHCH$_3$ 
covered by the ReMoCA survey. The LTE synthetic spectrum of CH$_3$NHCH$_3$
used to derive the upper limit on its column density toward Sgr~B2(N1S) is 
displayed in red and overlaid on the observed spectrum shown in black. The blue 
synthetic spectrum contains the contributions of all molecules identified in 
our survey so far, but not the species shown in red. The 
values written below each panel correspond from left to right to the half-power 
beam width, the central frequency in MHz, the width in MHz of each panel in 
parentheses, and the continuum level in K of the baseline-subtracted spectra 
in brackets. The y-axis is labeled in brightness temperature units (K). The 
dotted line indicates the $3\sigma$ noise level.}
\label{f:remoca_ch3nhch3_n1s}
\end{figure*}

\begin{figure*}
\centerline{\resizebox{0.85\hsize}{!}{\includegraphics[angle=0]{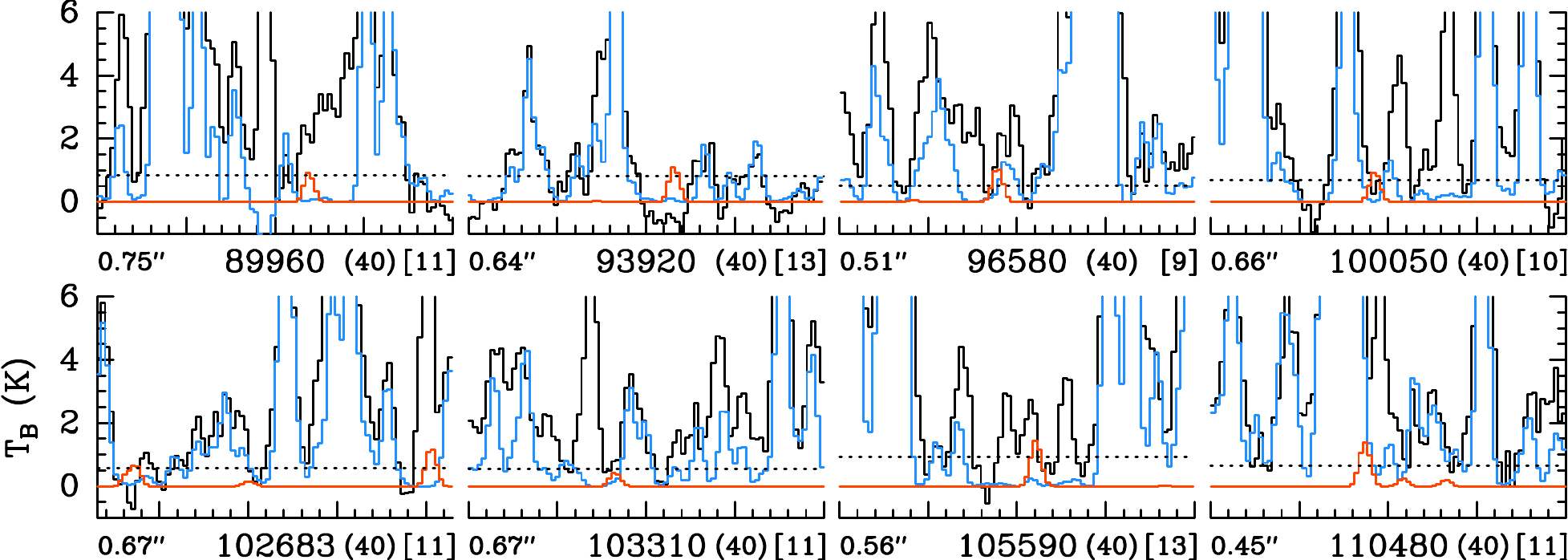}}}
\caption{Same as Fig.~\ref{f:remoca_ch3nhch3_n1s}, but for Sgr~B2(N2b).}
\label{f:remoca_ch3nhch3_n2b}
\end{figure*}

To search for dimethylamine toward each position using the spectroscopic predictions 
obtained in Sect.~\ref{lab-results}, we computed synthetic LTE spectra assuming 
the same emission size and rotational temperature as those derived for dimethyl ether. 
We did not find any evidence for dimethylamine in either source. 
The upper limit to the column density of DMA was evaluated for each source by varying the column density by hand 
and comparing the synthetic and observed spectra by eye, using the 3~$\sigma$ level 
(dashed lines in Figs.~\ref{f:remoca_ch3nhch3_n1s} and \ref{f:remoca_ch3nhch3_n2b}) as a guide, 
but also accounting for the blends with other species and the uncertainties on the baseline level.
The synthetic spectra used to estimate the upper limit to its column density are shown in 
Figs.~\ref{f:remoca_ch3nhch3_n1s} and \ref{f:remoca_ch3nhch3_n2b} for Sgr~B2(N1S) and Sgr~B2(N2b), respectively. 
These upper limits are reported in Tables~\ref{t:coldens_n1s} and \ref{t:coldens_n2b}, respectively.

\section{Astrochemical modeling}
\label{astro-model}

To explore the possible chemistry of DMA and its likely yield in the interstellar medium, we adapt the astrochemical models presented by \citet{Garrod22} 
to include DMA and a selection of related species. 
The physical treatment and chemical framework of the models correspond to the {\tt final} model of \citet{Garrod22}, which uses a three-phase 
(gas, grain/ice-surface and bulk-ice) treatment of the coupled gas and grain chemistry of hot cores. 
The grain chemistry framework \citep[explained in detail by][]{Jin20} includes nondiffusive mechanisms for surface and bulk-ice reactions, 
which allows the production of complex organic molecules to occur on the grains even at very low temperatures. 
The physical modeling uses a two-stage treatment, in which relatively diffuse gas ($n_{\mathrm{H}}=3000$~cm$^{-3}$) first collapses to high density 
($2 \times 10^{8}$~cm$^{-3}$); as the visual extinction increases from 3 to 500~mag., the dust cools from $\sim$14.6~K to 8~K, 
while the gas temperature is held steady at 10~K during the collapse as in past models. 
Much of the grain-surface ice builds up toward the end of the collapse stage. During the subsequent warm-up stage, this material is heated 
such that the gas and dust temperatures rise monotonically to a maximum of 400~K. 
Following various past models that have used the same physical treatment, the warm-up occurs over one of three characteristic timescales, 
labeled {\em fast} ($5 \times 10^{4}$~yr), {\em medium} ($2 \times 10^{5}$~yr), and {\em slow} ($1 \times 10^{6}$~yr), where the timescale technically corresponds 
to the time required to reach a representative hot-core temperature of 200~K. 
Desorption of the ice mantles into the gas phase takes place mainly in the 100--200~K temperature range.

The chemical network is based on that of \citet{Belloche22}, which included propanol chemistry and which is itself derived from the \citet{Garrod22} network. 
The new chemistry for DMA is based around production on grain surfaces or in the bulk ices, through the radical recombination reaction
\begin{eqnarray}
\ce{CH3} + \ce{CH3NH} \rightarrow \ce{CH3NHCH3}&& \label{reac1}
\end{eqnarray}
and the two-stage reaction process,
\begin{subequations}
\label{reac2}
\begin{eqnarray}
\ce{CH2} + \ce{CH3NH} \rightarrow \ce{CH2NHCH3}&& \label{reac2a} \\ 
\ce{H} + \ce{CH2NHCH3} \rightarrow \ce{CH3NHCH3} && \label{reac2b}
\end{eqnarray}
\end{subequations}
The intermediate radical CH$_3$NH may be produced by the hydrogenation of other species of the form CH$_\textrm{x}$NH$_\textrm{y}$, or through the addition of other radicals leading to such species. 
Crucially, the CH$_3$NH radical may additionally be formed through the photodissociation of methylamine, while CH$_2$ and CH$_3$ can be formed via repetitive H-addition to atomic carbon, or by the photodissociation of methane or other species containing a methyl group.

\begin{table*}
\begin{center}
\caption{Selected grain-surface chemical reactions and photoprocesses relevant to DMA chemistry that are included in the network. 
Activation energy barriers ($E_{\mathrm{A}}$) and rates, where applicable, are also indicated. ``$h\nu$ [CR]'' indicates photodissociation by a cosmic-ray induced photon. 
``$h\nu$ [ext]'' indicates photodissociation by an external photon from the interstellar radiation field. Rates and activation energies are based on similar reactions 
already present in the network, unless otherwise marked. $\zeta$ is the cosmic-ray ionization rate adopted in the particular model.}
\label{tab-reac}
\begin{tabular}[h]{lllllllll}
\hline
Reaction  &&&&&&&  $E_{\mathrm{A}}$ (K) & Rate (s$^{-1}$) \\
 \hline 
\ce{CH3}    &+& \ce{CH3NH}     &$\rightarrow$&  \ce{CH3NHCH3} & &             & 0         &--\\
\ce{CH2}    &+& \ce{CH3NH}     &$\rightarrow$&  \ce{CH2NHCH3} & &             & 0         &--\\
\ce{H}      &+& \ce{CH2NHCH3}  &$\rightarrow$&  \ce{CH3NHCH3} & &             & 0         &--\\
\ce{CH2OH}  &+& \ce{CH2NHCH3}  &$\rightarrow$&  \ce{CH3NHCH3} &+& \ce{H2CO}   & 0         &--\\
\ce{CH3O}   &+& \ce{CH2NHCH3}  &$\rightarrow$&  \ce{CH3NHCH3} &+& \ce{H2CO}   & 0         &--\\
\ce{COOH}   &+& \ce{CH2NHCH3}  &$\rightarrow$&  \ce{CH3NHCH3} &+& \ce{CO2}    & 0         &--\\
\ce{HCO}    &+& \ce{CH2NHCH3}  &$\rightarrow$&  \ce{CH3NHCH3} &+& \ce{CO}     & 0         &--\\
\ce{H}      &+& \ce{CH3NHCH3}  &$\rightarrow$&  \ce{CH2NHCH3} &+& \ce{H2}     & 1600~$^a$ &--\\
\ce{CH3O}   &+& \ce{CH3NHCH3}  &$\rightarrow$&  \ce{CH2NHCH3} &+& \ce{CH3OH}  & 2020~$^b$ &--\\
\ce{CN}     &+& \ce{CH3NHCH3}  &$\rightarrow$&  \ce{CH2NHCH3} &+& \ce{HCN}    &  100      &--\\
\ce{NH2}    &+& \ce{CH3NHCH3}  &$\rightarrow$&  \ce{CH2NHCH3} &+& \ce{NH3}    & 2710~$^b$ &--\\
\ce{OH}     &+& \ce{CH3NHCH3}  &$\rightarrow$&  \ce{CH2NHCH3} &+& \ce{H2O}    & 1000      &--\\

\ce{CH3NHCH3}&+& $h\nu$ [CR]   &$\rightarrow$&  \ce{CH3NH}    &+& \ce{CH3}    &--& 10,000 $\zeta$ \\
\ce{CH3NHCH3}&+& $h\nu$ [CR]   &$\rightarrow$&  \ce{CH2NHCH3} &+& \ce{H}      &--& 500    $\zeta$ \\

\ce{CH3NHCH2}&+& $h\nu$ [CR]   &$\rightarrow$&  \ce{CH2NH}    &+& \ce{CH3}    &--& 10,000 $\zeta$ \\
\ce{CH3NHCH2}&+& $h\nu$ [CR]   &$\rightarrow$&  \ce{CH3NH}    &+& \ce{CH2}    &--& 10,000 $\zeta$ \\

\ce{CH3NHCH3}&+& $h\nu$ [ext]  &$\rightarrow$&  \ce{CH3NH}    &+& \ce{CH3}    &--& $1.00 \times 10^{-9}  \, \exp(-1.7 \, A_{\mathrm{V}})$ \\
\ce{CH3NHCH3}&+& $h\nu$ [ext]  &$\rightarrow$&  \ce{CH2NHCH3} &+& \ce{H}      &--& $1.00 \times 10^{-9}  \, \exp(-1.7 \, A_{\mathrm{V}})$ \\

\ce{CH3NHCH2}&+& $h\nu$ [ext]  &$\rightarrow$&  \ce{CH2NH}    &+& \ce{CH3}    &--& $1.00 \times 10^{-9}  \, \exp(-1.7 \, A_{\mathrm{V}})$ \\
\ce{CH3NHCH2}&+& $h\nu$ [ext]  &$\rightarrow$&  \ce{CH3NH}    &+& \ce{CH2}    &--& $1.00 \times 10^{-9}  \, \exp(-1.7 \, A_{\mathrm{V}})$ \\
\\

\multicolumn{3}{l}{Pre-existing in the network:}\\
\ce{H}      &+& \ce{CH2NH}     &$\rightarrow$&  \ce{CH3NH}    & &             & 605~$^c$ &--\\
\ce{H}      &+& \ce{CH2NH}     &$\rightarrow$&  \ce{CH2NH2}   & &             & 605~$^c$ &--\\

\ce{CH4}    &+& $h\nu$ [CR]    &$\rightarrow$&  \ce{CH3}      &+& \ce{H}      &--& 1170   $\zeta$ \\
\ce{CH4}    &+& $h\nu$ [CR]    &$\rightarrow$&  \ce{CH2}      &+& \ce{H2}     &--& 1170   $\zeta$ \\

\ce{CH3NH2} &+& $h\nu$ [CR]    &$\rightarrow$&  \ce{CH3NH}    &+& \ce{H}      &--& 1000   $\zeta$ \\
\ce{CH3NH2} &+& $h\nu$ [CR]    &$\rightarrow$&  \ce{CH2NH2}   &+& \ce{H}      &--& 500    $\zeta$ \\
\ce{CH3NH2} &+& $h\nu$ [CR]    &$\rightarrow$&  \ce{CH3}      &+& \ce{NH2}    &--& 10,000 $\zeta$ \\

\ce{CH4}    &+& $h\nu$ [ext]   &$\rightarrow$&  \ce{CH3}      &+& \ce{H}      &--& $3.20 \times 10^{-10} \, \exp(-2.2 \, A_{\mathrm{V}})$ \\
\ce{CH4}    &+& $h\nu$ [ext]   &$\rightarrow$&  \ce{CH2}      &+& \ce{H2}     &--& $4.80 \times 10^{-10} \, \exp(-2.2 \, A_{\mathrm{V}})$ \\

\ce{CH3NH2} &+& $h\nu$ [ext]   &$\rightarrow$&  \ce{CH3NH}    &+& \ce{H}      &--& $5.00 \times 10^{-10} \, \exp(-1.7 \, A_{\mathrm{V}})$ \\
\ce{CH3NH2} &+& $h\nu$ [ext]   &$\rightarrow$&  \ce{CH2NH2}   &+& \ce{H}      &--& $1.00 \times 10^{-9}  \, \exp(-1.7 \, A_{\mathrm{V}})$ \\
\ce{CH3NH2} &+& $h\nu$ [ext]   &$\rightarrow$&  \ce{CH3}      &+& \ce{NH2}    &--& $1.00 \times 10^{-9}  \, \exp(-1.7 \, A_{\mathrm{V}})$ \\
\hline 
\end{tabular}
\end{center}
$^a$ \citet{Shang19}.
$^b$ Evans-Polanyi interpolation.\\
$^c$ Based on hydrogenation of \ce{C2H4}; \citet{Michael05}.
\end{table*}

\begin{figure}
\centering
\centerline{\resizebox{1.0\hsize}{!}{\includegraphics[angle=0]{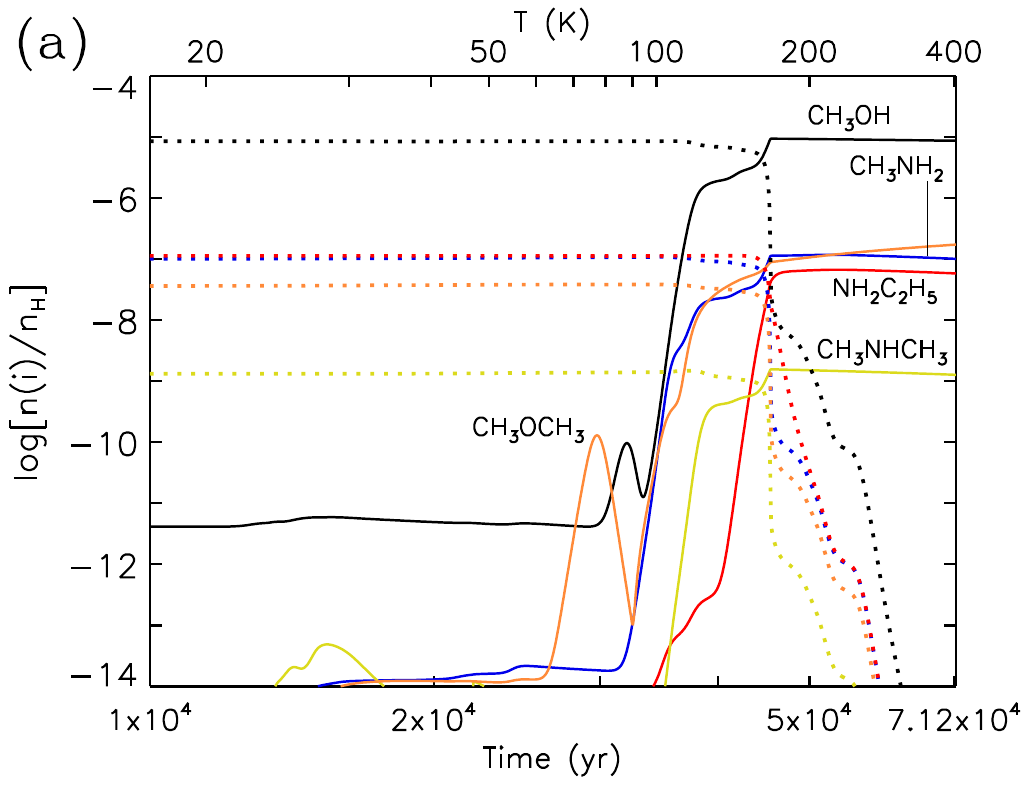}}}
\centerline{\resizebox{1.0\hsize}{!}{\includegraphics[angle=0]{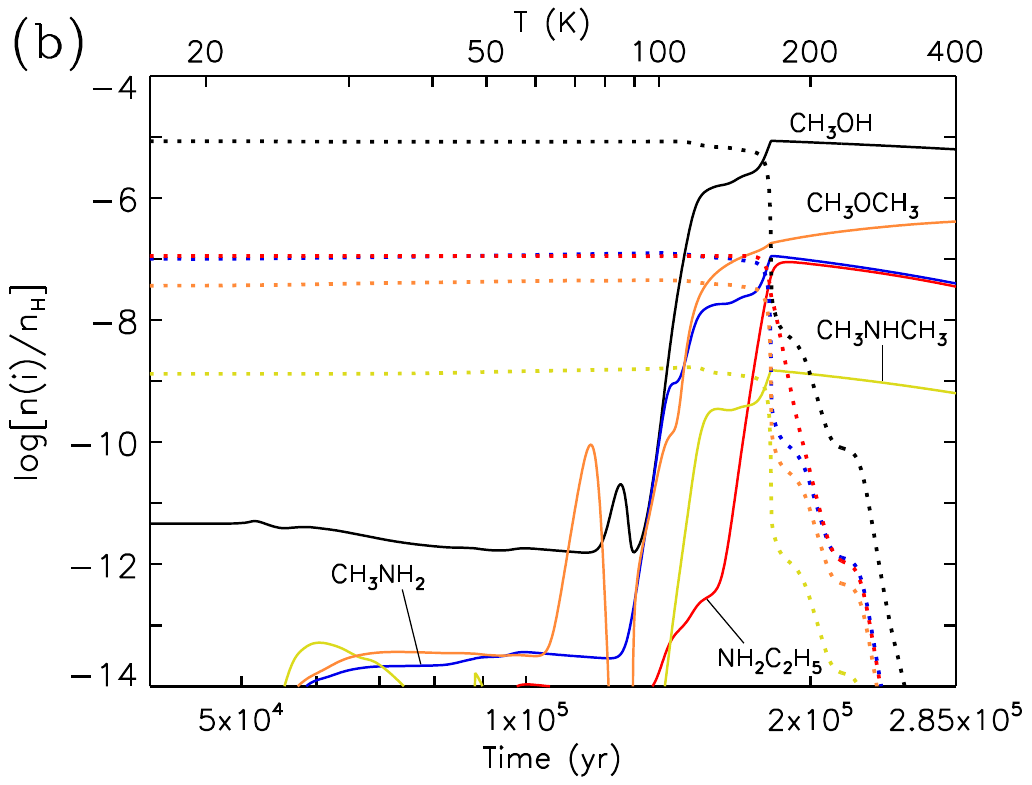}}}
\centerline{\resizebox{1.0\hsize}{!}{\includegraphics[angle=0]{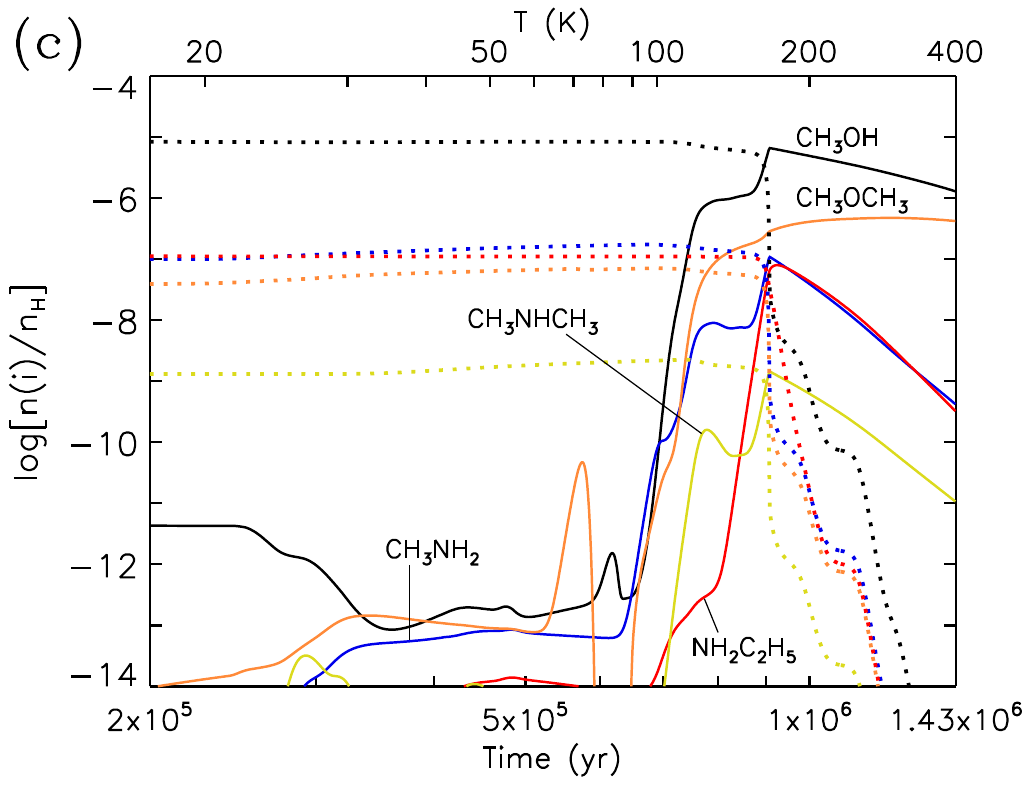}}}
\caption{Abundances with respect to total H, plotted logarithmically against time, for DMA and selected other molecules during the warm-up stage. 
Dotted lines of the same colours indicate the same species on the dust grains. Panels show results using the (a) {\em fast}, (b) {\em medium} and (c) {\em slow} warm-up models.}
\label{f:chem1}
\end{figure}

Each of the reaction sequences (\ref{reac1}) and (\ref{reac2}) may occur either on the grain/ice surfaces or within the bulk ice. 
On surfaces, reactions may be driven by diffusion or by nondiffusive encounters resulting from prior reactions (labeled ``three-body'' reactions, 3-B) 
or photodissociation events (labeled ``photodissociation-induced'' reactions, PDI). 
At the low temperatures (i.e. $<$20~K) achieved during the collapse stage and early in the warm-up stage, surface reactions such as 
(\ref{reac1}) and (\ref{reac2a}) occur mainly through nondiffusive processes, due to the immobility of the reactants, while reaction (\ref{reac2b}) is driven by the diffusion of mobile atomic H. 
At higher temperatures, surface diffusion of radicals also contributes somewhat to the production of large molecules. 
Within the bulk ice, diffusive motions are restricted to H and H$_2$ only, which can occur through quantum tunneling as well as thermal diffusion. 
Thus, in the bulk ice, reactions between radicals (i.e. not H) may only occur through the nondiffusive 3-B and PDI mechanisms.

Based on the above chemistry, two new neutral species are added to the network: DMA and its radical precursor \ce{CH2NHCH3}. 
Following past grain-surface chemistry treatments \citep[e.g.][]{Garrod13}, both of these species may also be destroyed by reactions with radicals; 
DMA may undergo H-abstraction by e.g. the OH radical, forming water and \ce{CH2NHCH3}, while \ce{CH2NHCH3} may abstract H from a selection of other radical species such as HCO, 
to form DMA and the stable molecule CO. DMA and \ce{CH2NHCH3} can also be photodissociated on the grains or in the gas phase.

Grain-surface/ice reactions included in the network that are relevant to DMA are listed in Table \ref{tab-reac}. 
Subject to the caveats outlined above, each may occur as the result of diffusive meetings, the 3-B and PDI processes, or by the Eley-Rideal (E-R) mechanism (surface only). 
Adsorption onto the grains, as well as thermal and nonthermal desorption mechanisms (photo- and chemical desorption) are treated as per past models. 
The surface binding energies of DMA and \ce{CH2NHCH3} used in the model are 5856~K and 5731~K, respectively, based on extrapolation from other species/functional groups 
in the network, as per previous models \citep{Garrod08,N-Me-Formamide_2017,Belloche19}. Molecular desorption driven by the direct impingement of cosmic rays 
on dust grains is not included in the present model, although the photodesorption mechanism includes a contribution from the cosmic ray-induced photon field.

The main destruction mechanisms for most complex organic molecules in the network are gas-phase ion-molecule reactions that take place following the desorption of the ice mantles into the gas. These are dominated by proton transfer from ions H$_3^+$, HCO$^+$ and H$_3$O$^+$; the resulting protonated molecule (\ce{CH3NH2CH3+}, in the case of DMA) may then recombine with electrons, fragmenting the molecule. A description of the procedures for the construction of this basic gas-phase chemistry may be found in \citet{Garrod08} and \citet{Garrod13}.

Following \citet{Taquet16}, \citet{Garrod22} added to their reaction network a number of gas-phase proton-transfer reactions between ammonia and various protonated complex organics, producing NH$_4^+$ while returning the original complex molecule unscathed. 
The inclusion of this mechanism was found to provide a competitive alternative to the  destructive electronic recombination of protonated complex organics, thus enhancing their post-desorption gas-phase lifetimes. 
However, the proton affinity of DMA is greater than that of ammonia \citep[929.5 versus 853.6~kJ~mol$^{-1}$][]{Hunter98}, meaning that protonated DMA cannot transfer its proton in this way; such a mechanism is therefore not included in the present chemical network. 
The possible importance of the converse process of proton transfer from NH$_4^+$ to certain complex organics is explored by \cite{Garrod23} ({\em submitted}).

The new chemical network already incorporates the chemistry of a large selection of simple and complex molecules, including methylamine, methanol, dimethyl ether and ethylamine; hot-core chemical model results for each of these are presented with those of DMA in order to compare with observations toward Sgr~B2(N).

In order to test the effects on molecular ratios of the elevated cosmic ray ionization rates expected toward the Galactic center \citep{goto2014}, the hot-core models are run using a selection of logarithmically spaced rates ranging from $\zeta = \zeta_{0}$ to $10^{1.5}~\zeta_{0}$, where $\zeta_{0} = 1.3 \times 10^{-17}$~s$^{-1}$.

For comparison with the gas-phase abundances observed toward G+0.693$-$0.027, the most appropriate modeling approach would include an explicit treatment of the passage of a shock \citep[e.g.,][]{requena-torres_organic_2006,zeng2018,rivilla2022b}, including the grain-heating and sputtering processes, as well as post-desorption gas-phase chemistry. 
Here, to allow a simpler comparison between the modeled abundances in the ices and the post-shock gas-phase abundances, we run collapse models similar to the first stage of the hot-core models, that would correspond to the pre-shock evolution of the dust and gas in G+0.693$-$0.027. 
The shock models of \citet{rivilla2022b} suggest cosmic-ray ionization rates in this molecular cloud ranging from 100$-$1000 $\zeta_{0}$; we run collapse models with $\zeta$ using the two extremes of this range. 
To better represent the pre-shock gas density of G+0.693$-$0.027, the final gas density at the end of the collapse is reduced to $n_{\mathrm{H}} = 2 \times 10^4$~cm$^{-3}$, as inferred by \citet{zeng2020}.

\subsection{Chemical model results}

Figure~\ref{f:chem1} shows fractional abundances for DMA and other molecules during the warm-up stage, using the three typical warm-up timescales, under conditions of $\zeta = \zeta_{0}$. 
Each of the molecules shown inherits a substantial solid-phase abundance (indicated by dotted lines) from the cold collapse stage. 
The material remains on the grains until the ice begins to desorb strongly into the gas phase at temperatures greater than 100~K. 
Much of this release is driven by the thermal desorption of water, which comprises much of the ice abundance; it begins to be lost substantially at temperatures around 114~K, as noted by \citet{Garrod22}, and this continues until most water has been released by around 164~K. 
This desorption model is supported by observational results from \citet{Busch22}. They found in a study of Sgr~B2(N1) that COMs formed on grains desorb thermally from the grain surface at $\sim$100~K, concomintantly with water.  
Because the model treats the ice as a distinct surface layer with a separate bulk ice beneath, the loss of water helps to release other species within the same time/temperature range. 
As a result, DMA, along with methanol and methylamine, reach their peak gas-phase abundances at around the temperature when water desorption reaches its maximum. 
These peak abundances and corresponding temperatures are shown in Table~\ref{tab-abuns}. The behavior of other molecules not shown in the figures are well represented by the extensive results presented by \citet{Belloche22} and \citet{Garrod22}.

Post-desorption gas-phase abundances are seen to fall more strongly as a function of temperature in the longer warm-up timescales runs, due to the longer period spent in the gas phase by those molecules. 
In all cases but for dimethyl ether, the peak gas-phase values of the molecules shown are seen to track fairly closely with the peak ice-mantle abundances of the same species. 
Thus, the relative abundances in the gas are more or less preserved from the much earlier and colder period when those molecules were formed on the grains. 
Dimethyl ether, as noted in various past modeling papers, has a strong gas-phase production mechanism based on the reaction of methanol with protonated methanol; its abundance therefore rises in the gas phase following desorption of the existing dimethyl ether in the ices.

Production of DMA in the models occurs during two main periods; the first occurs during the cold collapse stage and is thus a shared feature of all of the subsequent warm-up models. 
Reaction~(\ref{reac1}) occurs at very early times in the model, driven by a combination of PDI and 3-B processes within the bulk ice, in the proportion $\sim$2 : 1, respectively. 
The initial visual extinction of around 3~mag. is low enough to allow external UV photons to dissociate methane (CH$_4$) and methylamine, producing CH$_3$ and CH$_3$NH and allowing one or other radical to be produced, on occasion, in the presence of the other, leading to immediate nondiffusive reaction. 
Since this process does not involve diffusion, most of the reactions leading to DMA production occur in the thicker bulk-ice layer, rather than on the surface. At this early stage, the ice is up to a few tens of monolayers in thickness in total. 
The 3-B reaction mechanism in the bulk ice occurs around the same time, when H atoms released in the bulk ice by other photodissociation processes diffuse to find the stable molecule methanimine, CH$_2$NH, with which it reacts to form CH$_3$NH; the latter radical may then react with any contiguous CH$_3$ in the bulk ice.

Production of the related species ethylamine (C$_2$H$_5$NH$_2$) occurs in a similar way, through the reaction of radicals \ce{CH3} and \ce{CH2NH2}; however, \ce{CH2NH2} is formed mainly through H-atom abstraction from \ce{CH3NH2} by atomic H in the bulk ice. 
This abstraction reaction is substantially faster than the comparable photodissociation process by which it might form, because of the availability of diffusive atomic H in the bulk ice caused by the photodissociation of numerous different ice species, including water. 
Abstraction from the methyl group on methylamine to form \ce{CH2NH2} is expected to be strongly dominant over the alternative, \ce{CH3NH}, at very low temperatures \citep[see][who calculated branching down to 200~K]{Kerkeni07}. 
The availability of this effective mechanism for producing \ce{CH2NH2} therefore leads to much greater production of ethylamine versus DMA in the model.

\begin{figure*}
\centerline{\resizebox{0.95\hsize}{!}{\includegraphics[angle=0]{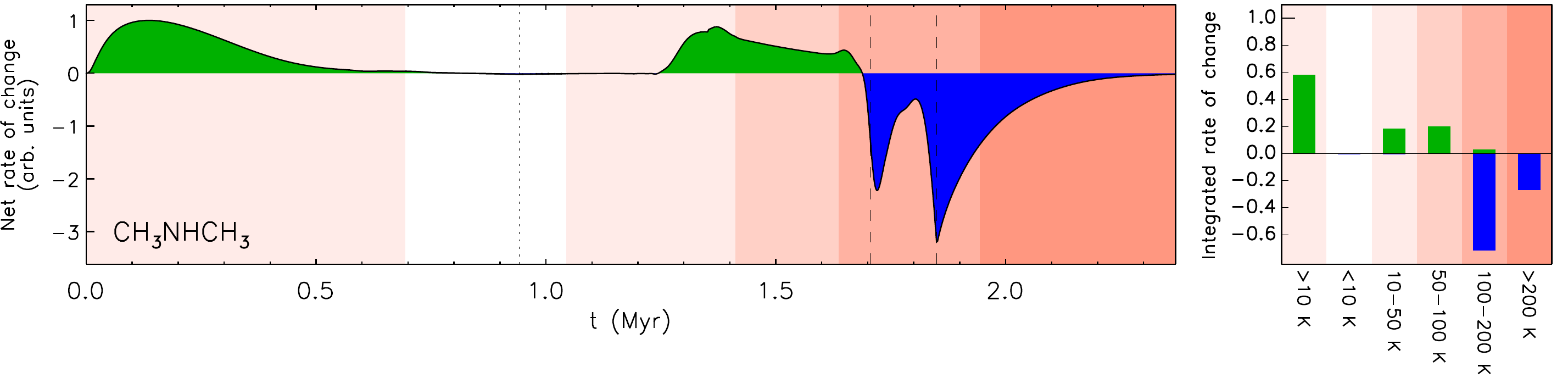}}}
\caption{{\bf {\em Left panel:}} Net rate of change (arbitrary units) in the abundances of DMA, summed over all chemical phases {\bf (i.e. gas + surface + bulk-ice)}, during the collapse and warm-up stages. 
  Data correspond to the {\em slow} warm-up timescale; the vertical dotted line indicates the start of the warm-up phase. 
  Net gain is shown in green, net loss in blue; rates are normalized to the peak value.
  Vertical dashed lines indicate the onset and end-point of water desorption. 
  Background colors indicate the dust temperature regime; from left to right these are: $>$10~K, $<$10~K, 10--50~K, 50--100~K, 100--200~K, 200--400~K. 
  The initial dust temperature is $\sim$14.7~K. {\bf {\em Right panel:}} Net rates of change integrated over each temperature range. 
  Positive (formation) and negative (destruction) rates are integrated independently; both are normalized to the total integrated formation rate.}
\label{f:roc}
\end{figure*}

The second stage of DMA production occurs during the warm-up stage, and is most prevalent in the {\em slow} warm-up model, due to the longer period available for ice photochemistry. 
In this case, the DMA formation occurs through cosmic ray-induced photodissociation of methylamine and methane. The left panel of Fig.~\ref{f:roc} shows the total rate of production of DMA in all phases (surface, ice-mantle, and gas phase) as a function of time through the full collapse stage followed by the {\em slow} warm-up stage. The first and second stages of DMA production are seen clearly in green. 
The second production stage becomes significant beginning around 30~K, and continues to produce DMA up until the point of desorption, which begins to occur strongly between the two vertical dashed lines that indicate the beginning and end of water desorption. 
The onset of DMA production around 30~K is related to the falling abundance of atomic H in the ice as temperatures increase, reducing its ability to recombine with simple radicals before they can form more complex species. 
The right panel of Fig.~\ref{f:roc} shows the production and destruction of DMA in several key temperature regimes. The early production mechanism for DMA contributes around 60\% of total production. 
The blue sections of the left and right panels indicate where DMA is destroyed in the gas phase, mainly through proton-transfer from H$_3$O$^+$, followed by dissociative electronic recombination.

Table~\ref{tab-abuns} also shows peak gas-phase abundances and associated temperatures for three additional sets of model-runs, using $\zeta = 10^{0.5}~\zeta_{0}$, $10~\zeta_{0}$ and $10^{1.5}~\zeta_{0}$. 
Among the nitrogen-bearing species, the peak abundances fall somewhat with increased cosmic ray ionization rate, with the effect more pronounced with longer warm-up timescales. 
The effect is caused by more rapid gas-phase destruction via ion-molecule reactions. The two highest $\zeta$ values, when combined with the longest warm-up timescale, produce a more extreme degree of destruction for all the molecules. 
Outside of those two cases, the oxygen-bearing species methanol and dimethyl ether are more robust to changes in $\zeta$, generally varying by less than a factor 2 between the various models. 
Although the peak values are less affected, however, both increased warm-up timescale and $\zeta$ value lead to more rapid destruction in the gas phase for all species following attainment of the post-desorption peak.

\begin{figure}
\centering
\centerline{\resizebox{1.0\hsize}{!}{\includegraphics[angle=0]{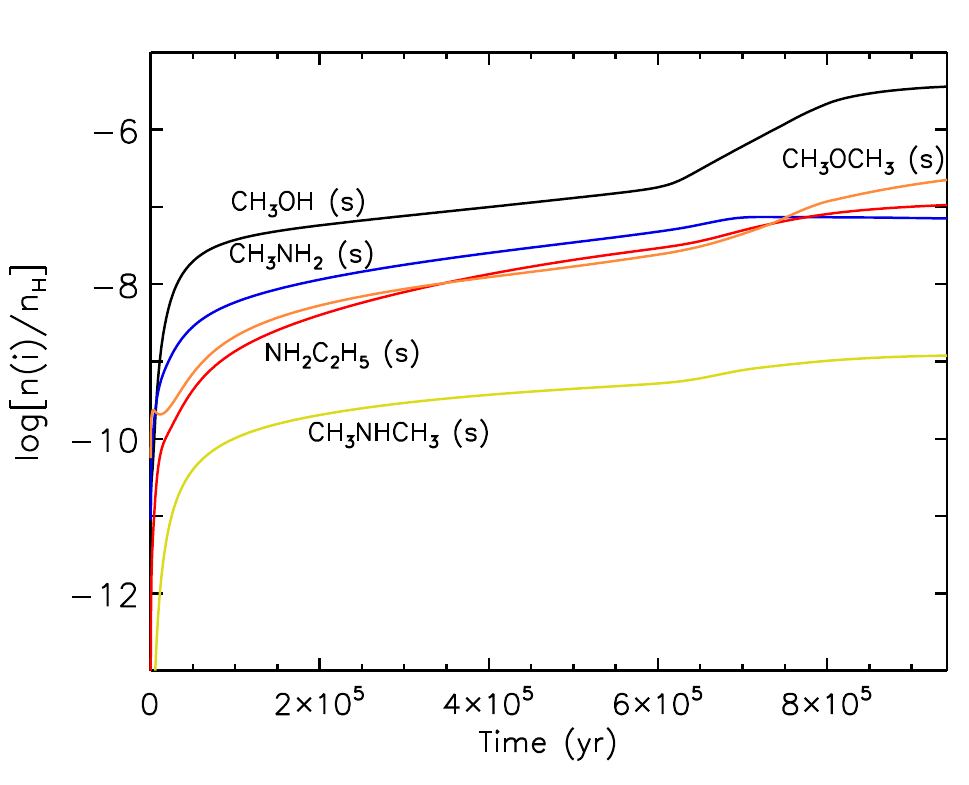}}}
\caption{Abundances in the ice mantle with respect to total H, plotted logarithmically against time, for DMA and selected other grain-surface/ice molecules for the low gas-density collapse model with $\zeta$ = 100~$\zeta_{0}$.}
\label{f:chem2}
\end{figure}

Fig.~\ref{f:chem2} shows results from the low-density collapse model with $\zeta$ = 100~$\zeta_{0}$, which is intended to represent the pre-shock behavior of G+0.693; time-dependent solid-phase abundances are shown for the same five species as in Fig.~\ref{f:chem1}. 
During the collapse, the increases in gas density and decreases in dust temperature become more substantial after a time of around 0.5~Myr, which manifests as a steeper rise in methanol abundance in particular. 
DMA abundance is seen never to exceed that of methylamine, while ethylamine slightly exceeds the latter toward the end of the model run.

The final solid-phase abundances of these molecules are shown in Table~\ref{tab-abuns2}, where the results for the $\zeta$ = 1000~$\zeta_{0}$ model are also shown. 
The abundance of methanol is notably affected in the highest $\zeta$ model, while the abundance of dimethyl ether is lower by a factor $\sim$4. 
The abundances of the N-bearing species shown are each lower by a factor $\sim$2 in the $\zeta$ = 1000~$\zeta_{0}$ model.

\begin{table*}
\begin{center}
\caption{Peak gas-phase chemical abundances for selected molecules during the warm-up stage. The temperature at which the peak value is achieved is also indicated. Notation $a(-b)$ indicates $a \times 10^{-b}$.}
\label{tab-abuns}
\begin{tabular}[h]{lllllllllll}
\hline
 &&& \multicolumn{2}{c}{{\em Fast}} && \multicolumn{2}{c}{{\em Medium}} && \multicolumn{2}{c}{{\em Slow}} \\
\cline{4-5} \cline{7-8} \cline{10-11} \\
Cosmic ray ionization rate & Molecule && $n[i]/n_{\mathrm{H}}$ & $T$ (K) && $n[i]/n_{\mathrm{H}}$ & $T$ (K) && $n[i]/n_{\mathrm{H}}$ & $T$ (K) \\
\hline

$\zeta=\zeta_0$ 
& CH$_3$OH         && 9.4(-6) & 168 && 8.7(-6) & 167 && 6.6(-6) & 166  \\
& CH$_3$OCH$_3$    && 1.7(-7) & 400 && 4.1(-7) & 398 && 4.7(-7) & 292  \\
& CH$_3$NH$_2$     && 1.2(-7) & 218 && 1.1(-7) & 167 && 1.1(-7) & 166  \\
& CH$_3$NHCH$_3$   && 1.6(-9) & 167 && 1.5(-9) & 167 && 1.4(-9) & 166  \\
& NH$_2$C$_2$H$_5$ && 6.7(-8) & 227 && 9.0(-8) & 180 && 8.0(-8) & 172  \\
\\

$\zeta=3.16 \times \zeta_0$
& CH$_3$OH         && 1.4(-5) & 166 && 1.1(-5) & 165 && 5.8(-6)  & 164  \\
& CH$_3$OCH$_3$    && 8.0(-7) & 400 && 1.2(-6) & 334 && 5.0(-7)  & 187  \\
& CH$_3$NH$_2$     && 8.2(-8) & 166 && 7.4(-8) & 165 && 4.8(-8)  & 164  \\
& CH$_3$NHCH$_3$   && 1.5(-9) & 166 && 1.3(-9) & 165 && 8.2(-10) & 164  \\
& NH$_2$C$_2$H$_5$ && 4.0(-8) & 184 && 4.9(-8) & 174 && 2.6(-8)  & 167  \\
\\

$\zeta=10 \times \zeta_0$
& CH$_3$OH         && 1.5(-5) & 165 && 7.9(-6)  & 163 && 1.1(-7)  & 118  \\
& CH$_3$OCH$_3$    && 1.8(-6) & 335 && 9.6(-7)  & 189 && 3.6(-9)  & 118  \\
& CH$_3$NH$_2$     && 5.5(-8) & 165 && 3.3(-8)  & 163 && 5.9(-10) & 117  \\
& CH$_3$NHCH$_3$   && 1.1(-9) & 165 && 6.4(-10) & 163 && 1.6(-11) & 118  \\
& NH$_2$C$_2$H$_5$ && 2.7(-8) & 175 && 1.9(-8)  & 169 && 1.3(-10) & 155  \\
\\

$\zeta=31.6 \times \zeta_0$
& CH$_3$OH         && 9.7(-6)  & 168 && 2.5(-6)  & 163 && 2.5(-9)  & 112  \\
& CH$_3$OCH$_3$    && 8.7(-7)  & 205 && 1.9(-7)  & 165 && 3.2(-10) & 112  \\
& CH$_3$NH$_2$     && 3.4(-8)  & 168 && 1.1(-8)  & 163 && 9.1(-11) & 112  \\
& CH$_3$NHCH$_3$   && 6.0(-10) & 168 && 2.0(-10) & 163 && 1.7(-12) & 113  \\
& NH$_2$C$_2$H$_5$ && 1.8(-8)  & 171 && 5.3(-9)  & 163 && 5.4(-12) & 149  \\
\hline
\end{tabular}
\end{center}
\end{table*}

\begin{table*}
\begin{center}
\caption{Molecular abundances based on final ice-mantle chemical abundances for selected molecules during the low-density collapse models.}
\label{tab-abuns2}
\begin{tabular}[h]{lllrr}
\hline
Cosmic ray ionization rate & Molecule (ice mantle) && $n[i]/n_{\mathrm{H}}$ \\
\hline

$\zeta=100 \times \zeta_0$
& CH$_3$OH         && 3.5(-6)   \\
& CH$_3$OCH$_3$    && 2.2(-7)   \\
& CH$_3$NH$_2$     && 7.1(-8)   \\
& CH$_3$NHCH$_3$   && 1.2(-9)   \\
& NH$_2$C$_2$H$_5$ && 1.1(-7)   \\
\\

$\zeta=1000 \times \zeta_0$
& CH$_3$OH         && 2.6(-7)   \\
& CH$_3$OCH$_3$    && 6.0(-8)   \\
& CH$_3$NH$_2$     && 3.8(-8)   \\
& CH$_3$NHCH$_3$   && 6.8(-10)   \\
& NH$_2$C$_2$H$_5$ && 5.7(-8)   \\
\hline
\end{tabular}
\end{center}
\end{table*}

\begin{table*}
\begin{center}
\caption{Molecular ratios based on peak gas-phase chemical abundances for selected molecules during the warm-up stage.}
\label{tab-ratios}
\begin{tabular}[h]{lllrrr}
\hline
Cosmic ray ionization rate & Molecular ratio (gas phase) && {\em Fast} & {\em Medium} & {\em Slow} \\
\hline

$\zeta=\zeta_0$ 
& CH$_3$OH : CH$_3$OCH$_3$      &&   55  &   21  &   14 \\
& CH$_3$NH$_2$ : CH$_3$NHCH$_3$ &&   75  &   73  &   79 \\
& CH$_3$OH : CH$_3$NHCH$_3$     && 6030  & 5740  & 4590 \\
\\

$\zeta=3.16 \times \zeta_0$
& CH$_3$OH : CH$_3$OCH$_3$      && 18    &  9    &   12 \\
& CH$_3$NH$_2$ : CH$_3$NHCH$_3$ && 55    & 57    &   59 \\
& CH$_3$OH : CH$_3$NHCH$_3$     && 9350  & 8850  & 7090 \\
\\

$\zeta=10 \times \zeta_0$
& CH$_3$OH : CH$_3$OCH$_3$      &&  8      &  8       &   31 \\
& CH$_3$NH$_2$ : CH$_3$NHCH$_3$ && 50      & 52       &   37 \\
& CH$_3$OH : CH$_3$NHCH$_3$     && 13100   & 12400    & 6680 \\
\\

$\zeta=31.6 \times \zeta_0$
& CH$_3$OH : CH$_3$OCH$_3$      && 11       & 13       &    8 \\
& CH$_3$NH$_2$ : CH$_3$NHCH$_3$ && 57       & 55       &   54 \\
& CH$_3$OH : CH$_3$NHCH$_3$     && 16200    & 12700    & 1410 \\
\hline
\end{tabular}
\end{center}
\end{table*}
\begin{table*}
\begin{center}
\caption{Molecular ratios based on final ice-mantle chemical abundances for selected molecules during the low-density collapse models.}
\label{tab-ice-ratios2}
\begin{tabular}[h]{lllrr}
\hline
Cosmic ray ionization rate & Molecular ratio (ice mantle) && Value \\
\hline

$\zeta=100 \times \zeta_0$
& CH$_3$OH : CH$_3$OCH$_3$      &&  16   \\
& CH$_3$NH$_2$ : CH$_3$NHCH$_3$ &&  60   \\
& CH$_3$OH : CH$_3$NHCH$_3$     &&  2900 \\
\\

$\zeta=1000 \times \zeta_0$
& CH$_3$OH : CH$_3$OCH$_3$      &&   4  \\
& CH$_3$NH$_2$ : CH$_3$NHCH$_3$ &&  56  \\
& CH$_3$OH : CH$_3$NHCH$_3$     &&  379 \\
\hline
\end{tabular}
\end{center}
\end{table*}

\section{Discussion of the astrochemical results}
\label{astro-discussion}

\subsection{G+0.693}

The column density upper limit derived for DMA is 7.6$\times$10$^{13}$~cm$^{-2}$, slightly higher than the column density derived for \ce{C2H3NH2} and \ce{C2H5NH2}, which are (4.5$\pm$0.6)$\times$10$^{13}$~cm$^{-2}$ and (2.5$\pm$0.7)$\times$10$^{13}$~cm$^{-2}$, respectively (\citealt{zeng2021}). 
The abundance ratio compared to methylamine, whose abundance is 2.2$\times$10$^{-8}$ (\citealt{zeng2018}), is \ce{CH3NH2}/DMA $>$39. 
We can compare this ratio with other $-$H and $-$CH$_3$ pairs already detected towards G+0.693, such as methanol (\ce{CH3OH}) and dimethyl ether (\ce{CH3OCH3}). 
The \ce{CH3OH} abundance is  1.1$\times$10$^{-7}$ (\citealt{jimenez-serra2022}), and that of \ce{CH3OCH3} is 8.1$\times$10$^{-10}$ (Rivilla, private communication), which yields a ratio \ce{CH3OH}/\ce{CH3OCH3} $\sim$135. 
This value is consistent with the lower limit found for the \ce{CH3NH2}/DMA ratio ($>$39), see also Table~\ref{table-G0693}. 
This  might suggest, if the $-$H/$-$CH$_3$ ratio also holds for amines, that the detection of DMA would still need a significant improvement of the sensitivity of the current observations.

\subsection{Sgr~B2(N)}

Table~\ref{t:coldens_n1s} shows that dimethyl ether is 14 times less abundant
than methanol in Sgr~B2(N1S) and dimethylamine is at least 14 times less
abundant than methylamine toward this position. This limit is not constraining
enough to infer whether or not the formation routes that possibly relate 
methanol to dimethyl ether and methylamine to dimethylamine operate 
differently. Toward Sgr~B2(N2b), dimethyl ether is 26 times less abundant than methanol 
(Table~\ref{t:coldens_n2b}). Because the abundance ratio of methanol to 
methylamine is much higher in Sgr~B2(N2b) than in Sgr~B2(N1S) (300 versus 
15), the upper limit obtained for dimethylamine is much less constraining 
toward Sgr~B2(N2b): we can only say that dimethylamine is at least 4.5 times 
less abundant than methylamine toward this position.

\subsection{Comparison of chemical models with observations}

As with the observational data themselves, molecular abundance ratios may be the most useful means of comparison between models and observations. 
Table~\ref{tab-ratios} shows chemical model ratios between peak gas-phase abundance values for CH$_3$OH : CH$_3$OCH$_3$, CH$_3$NH$_2$ : CH$_3$NHCH$_3$ and CH$_3$OH : CH$_3$NHCH$_3$. 
The first of these, comparing with the ratios obtained for Sgr B2(N1S) and Sgr B2(N2b) of 14 and 26, respectively, provides an acceptable match with the observed values within the range of model outcomes, which range from 8 to 55. 
The modeled ratio CH$_3$NH$_2$ : CH$_3$NHCH$_3$, however, is rather more extreme, ranging from 37 to 75, with most values exceeding 50. 
Lower limits on this ratio for Sgr~B2(N1S) and Sgr~B2(N2b) are 14 and 4.5; thus the models predict abundances for DMA that are substantially below its observational upper limits in Sgr~B2(N). 
The modeled ratio of methanol to DMA shows a similar picture, although the ratio for the high-$\zeta$/{\em slow} warm-up model of 1410 is close to the observed lower limit for Sgr B2(N2b) of 1360. 
Either way, the chemical models are consistent with the non-detection of DMA toward Sgr~B2(N).

Table~\ref{tab-ice-ratios2} shows molecular ratios based on the {\em ice-mantle} abundances achieved for each molecule at the end of the low-density, high-$\zeta$ collapse models. 
We compare these ratios with those obtained from the gas-phase abundances observed toward G+0.693$-$0.027, on the assumption that the gas-phase abundances may retain the underlying ratios of the originating ice composition. 

In comparison with the observed CH$_3$OH : CH$_3$OCH$_3$ ratio of 135, the model ratios are too low, producing values of 16 and 4 for the $\zeta$ = 100~$\zeta_{0}$ and 1000~$\zeta_{0}$ models, respectively. 
At cosmic-ray ionization rates that are this high, the gas-phase survival of CO is substantially affected by ion-molecule reactions involving H$_3^+$ and He$^+$, such that the amount of CO available to produce grain-surface methanol, via hydrogenation by atomic H, is diminished. 
Furthermore, under these high-$\zeta$, low-density conditions, in which photodissociation of ice-mantle species is quite rapid in comparison to the rate 
at which gas-phase CO and other species are deposited onto the grains, bulk-ice photochemistry involving CH$_3$OH dominates dimethyl ether production on the grains, increasing its ratio relative to methanol.

Meanwhile the CH$_3$NH$_2$ : CH$_3$NHCH$_3$ ratios achieved in the models still slightly exceed the observational lower limit of 39, ranging from 56 to 60. 
These results are thus consistent with the non-detection of DMA toward G+0.693$-$0.027, although they also suggest that a detection may be rather easier to achieve toward this source than toward Sgr B2(N).

We note, however, that the ice-mantle abundance ratio for CH$_3$NH$_2$ : C$_2$H$_5$NH$_2$ is close to unity in both the hot-core models and the lower-density, high-$\zeta$ models. 
This compares unfavorably with the observational ratio of $\sim$120 toward G+0.693$-$0.027, suggesting that the models substantially over-produce ethylamine. 
Furthermore, toward Sgr B2(N1S), this ratio has been found to be $>$6 \citep{Margules22}. 
Due to the similarities in the chemistry between ethylamine and DMA, this could indicate that DMA is also over-produced, rendering a possible future detection of this molecule even more challenging.

The rates used to simulate the production of DMA and related species are only poorly constrained, largely by comparison between photodissociation rates of similar molecules (under gas-phase conditions). 
There may therefore be a substantial amount of possible variation in model outcomes. 
The strongest mechanism for DMA production in the hot-core models also occurs early in the physical evolution, when visual extinction is low. 
It remains to be seen whether such a low initial visual extinction is appropriate to all regions of a hot core, thus DMA abundances could again be lower than the models predict.

There remains also the possibility that DMA could be formed in the gas phase, possibly through a similar mechanism to that which is so effective for the production of protonated dimethyl ether, \ce{CH3OH2+} + \ce{CH3OH} $\rightarrow$ \ce{CH3OHCH3+} + \ce{H2O}, followed by dissociative recombination. 
However, the authors are unaware of any experimental studies \citep{Anicich03} of such a reaction between any combination of \ce{CH3OH2+}, \ce{CH3OH}, \ce{CH3NH3+} or \ce{CH3NH2} that would lead to protonated DMA. 
Furthermore, the gas-phase reaction producing dimethyl ether is especially effective due to the exceedingly high gas-phase abundances achieved by both methanol and protonated methanol (relative to other protonated complex organics), due to the presence of methanol in great abundance in the ice mantles. 
Even with a large reaction rate coefficient, the absolute reaction rate for an alternative process involving methylamine might not be great enough to compete with production of DMA on grains, limited as that may be.

While the DMA abundance produced by the low-density, high-$\zeta$ models is consistent with the non-detection, the behavior of dimethyl ether with respect to methanol is not consistent with their observed ratio toward G+0.693$-$0.027. 
However, we emphasize that a fair comparison between models and observations should include the substantial influence of the shock chemistry and shock-induced grain-mantle release, which the present models do not do.

\section{Conclusion and outlook}
\label{conclusion}

We have investigated the rotation-tunneling spectrum of DMA extensively in the millimeter and submillimeter region. 
This yielded very accurate spectroscopic parameters which are well-suited for almost all searches for this molecule in space, 
except possibly in cases in which the methyl internal rotation needs to be taken into account.

Dimethylamine was not detected toward G+0.693$-$0.027. 
The lower limit to the \ce{CH3NH2}/DMA ratio of $>$39 is constraining. 
But if the \ce{CH3NH2}/DMA ratio is the same as the \ce{CH3OH}/\ce{CH3OCH3} ratio of $\sim$135 an unambiguous identification of DMA in this source will require a considerable improvement in sensitivity.

We report nondetections of dimethylamine with ALMA toward Sgr~B2(N1S) and Sgr~B2(N2b). 
The nondetections imply that dimethylamine is at least 14 
and 4.5 times less abundant than methylamine toward these sources, 
respectively, while dimethyl ether is 14 and 26 times less abundant than methanol, respectively.

Dimethylamine was included in astrochemical kinetic modeling calculations, assuming grain-surface and ice-mantle formation mechanisms. The calculated \ce{CH3NH2}/DMA ratios are compatible with our observational non-detections. 
DMA in the models is formed mainly through an early mechanism that relies on photoprocessing of the young dust-grain ices by external UV under low-extinction conditions. 
Further processing of the ices by cosmic ray-induced UV photons allows DMA to be formed at higher temperatures. 
The models overproduce \ce{C2H5NH2} with respect to \ce{CH3NH2}, suggesting that the abundance of DMA in the interstellar medium could be substantially lower than the predicted values if both DMA and ethylamine are similarly overproduced by the models. 
The calculated ratios of peak molecular abundances remain fairly stable for the range of cosmic-ray ionization rates tested here.


\section*{Acknowledgements}

We are grateful to H.~V.~L. Nguyen for communicating results of the Fourier transform microwave investigation 
prior to publication. We also thank C.~P. Endres for providing SPFIT files of the ground state rotational spectrum 
of dimethyl ether. We acknowledge support by the Deutsche Forschungsgemeinschaft via the collaborative research 
center SFB~956 (project ID 184018867) project B3 as well as the Ger{\"a}tezentrum SCHL~341/15-1 
(``Cologne Center for Terahertz Spectroscopy''). 
RTG thanks the Astronomy \& Astrophysics program of the National Science Foundation (grant No. AST 19-06489) for funding the chemical modeling presented here.
V.M.R. has received support from the project RYC2020-029387-I funded by MCIN/AEI /10.13039/501100011033.
I.J-.S and J.M.-P. acknowledge funding from grant No. PID2019-105552RB-C41 awarded by the Spanish Ministry of Science and Innovation/State Agency of Research MCIN/AEI/10.13039/501100011033.
Our research benefited from NASA's Astrophysics Data System (ADS).
This paper makes use of the following ALMA data: 
ADS/JAO.ALMA\#2016.1.00074.S. 
ALMA is a partnership of ESO (representing its member states), NSF (USA), and 
NINS (Japan), together with NRC (Canada), NSC and ASIAA (Taiwan), and KASI 
(Republic of Korea), in cooperation with the Republic of Chile. The Joint ALMA 
Observatory is operated by ESO, AUI/NRAO, and NAOJ. The interferometric data 
are available in the ALMA archive at https://almascience.eso.org/aq/.

\section*{Data Availability}
The spectroscopic line lists and associated files are available as supplementary material through the journal and in the data section of the 
CDMS\footnote{https://cdms.astro.uni-koeln.de/classic/predictions/daten/DMA/}. 
The underlying original spectral recordings will be shared on reasonable request to the corresponding author. 
The radio astronomical data on Sgr~B2(N) are available through the 
ALMA archive\footnote{https://almascience.eso.org/aq/}.


\bibliographystyle{mnras}
\bibliography{DMA,rivilla} 


\begin{appendix}
\label{appendix}

\section{Additional figures from the ReMoCA survey}
\label{a:remoca}

Figures~\ref{f:remoca_ch3och3_ve0_n1s}--\ref{f:remoca_ch3och3_v15e1_n1s} show
transitions of CH$_3$OCH$_3$ $\upsilon = 0$, CH$_3$OCH$_3$ $\upsilon_{11} = 1$, and
CH$_3$OCH$_3$ $\upsilon_{15} = 1$, respectively, that are covered by the ReMoCA 
survey and significantly contribute to the signal detected toward Sgr~B2(N1S). 
Figures~\ref{f:remoca_ch3och3_ve0_n2b}--\ref{f:remoca_ch3nh2_ve0_n2b} show
transitions of CH$_3$OCH$_3$ $\upsilon = 0$, CH$_3$OCH$_3$ $\upsilon_{11} = 1$,
CH$_3$OCH$_3$ $\upsilon_{15} = 1$, and CH$_3$NH$_2$ $\upsilon=0$, respectively, that 
are covered by the ReMoCA survey and significantly contribute to the signal 
detected toward Sgr~B2(N2b). Transitions of a given molecule that are too 
heavily blended with much stronger emission from other molecules and therefore 
cannot contribute to the identification of this molecule are not shown in 
these figures.

\begin{figure*}
\centerline{\resizebox{0.85\hsize}{!}{\includegraphics[angle=0]{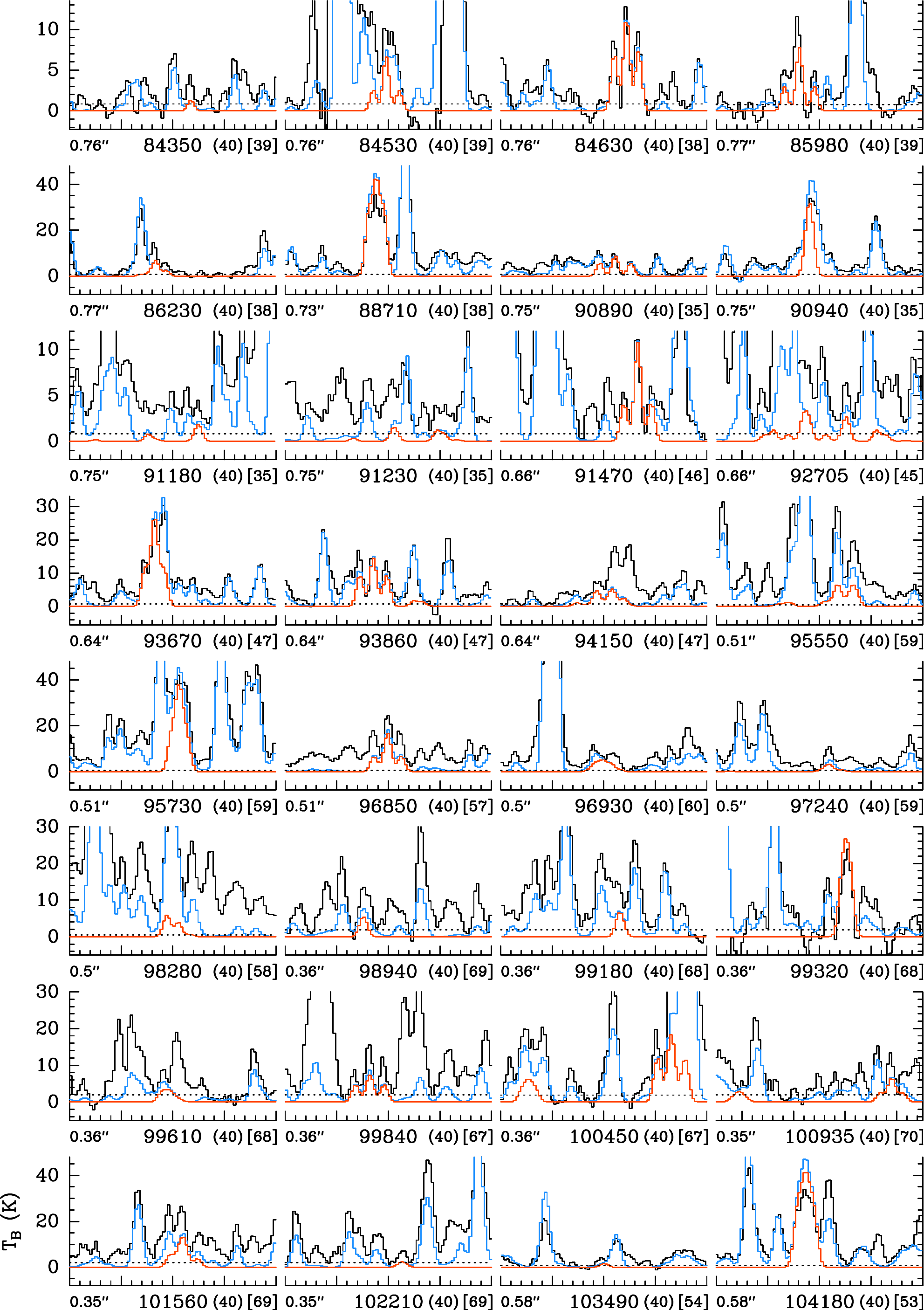}}}
\caption{Selection of rotational transitions of dimethyl ether CH$_3$OCH$_3$ in 
its vibrational ground state covered by the ReMoCA survey. The LTE 
synthetic spectrum of CH$_3$OCH$_3$, $\upsilon = 0$ is displayed in red and 
overlaid on the observed spectrum of Sgr~B2(N1S) shown in black. The blue 
synthetic spectrum contains the contributions of all molecules identified in 
our survey so far, including the contribution of the species shown in red. The 
values written below each panel correspond from left to right to the half-power 
beam width, the central frequency in MHz, the width in MHz of each panel in 
parentheses, and the continuum level in K of the baseline-subtracted spectra 
in brackets. The y-axis is labeled in brightness temperature units (K). The 
dotted line indicates the $3\sigma$ noise level.}
\label{f:remoca_ch3och3_ve0_n1s}
\end{figure*}

\begin{figure*}
\addtocounter{figure}{-1}
\centerline{\resizebox{0.88\hsize}{!}{\includegraphics[angle=0]{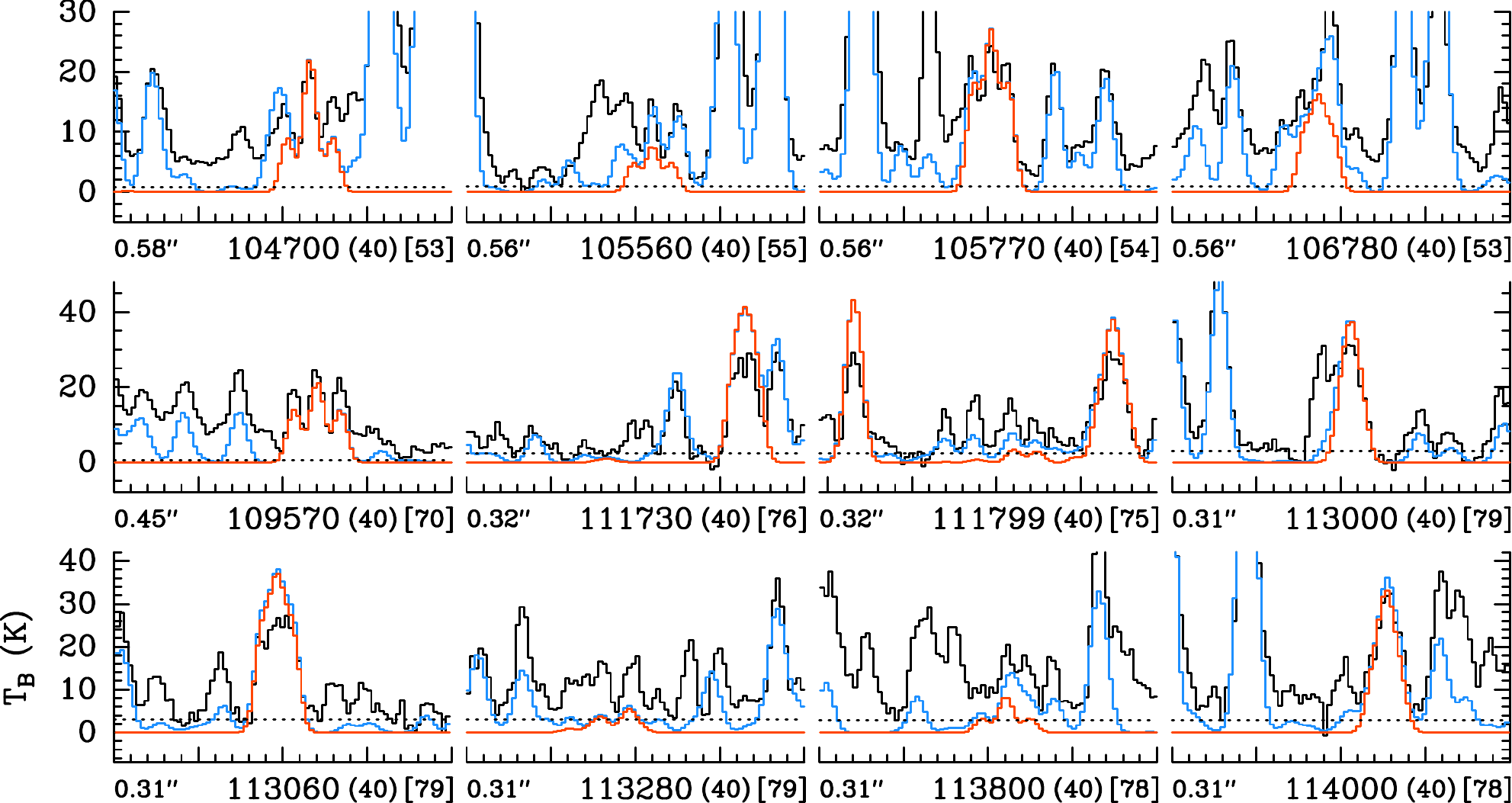}}}
\caption{continued.}
\end{figure*}

\begin{figure*}
\centerline{\resizebox{0.88\hsize}{!}{\includegraphics[angle=0]{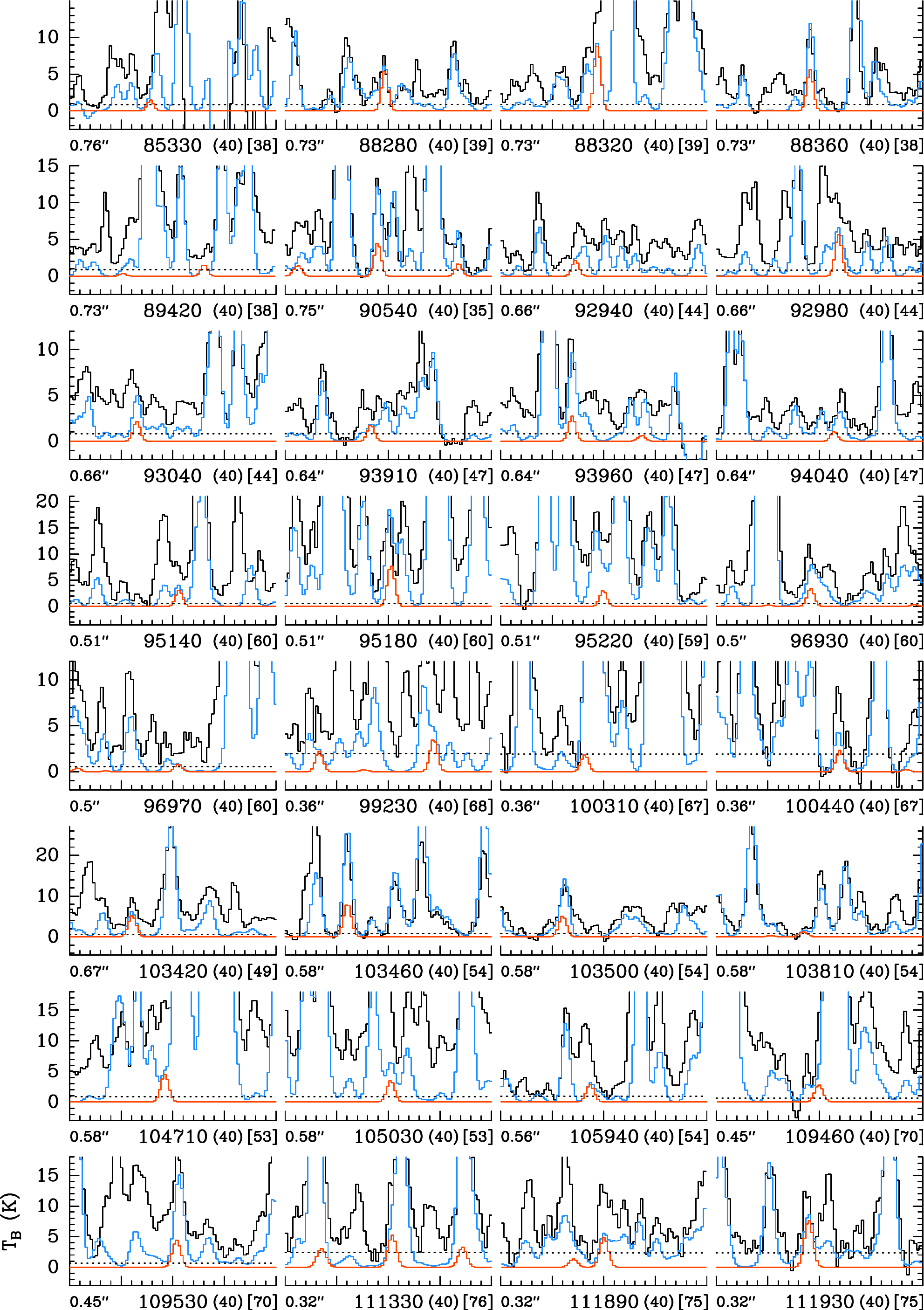}}}
\caption{Same as Fig.~\ref{f:remoca_ch3och3_ve0_n1s}, but for CH$_3$OCH$_3$, $\upsilon_{11} = 1$.
}
\label{f:remoca_ch3och3_v11e1_n1s}
\end{figure*}

\begin{figure*}
\addtocounter{figure}{-1}
\centerline{\resizebox{0.88\hsize}{!}{\includegraphics[angle=0]{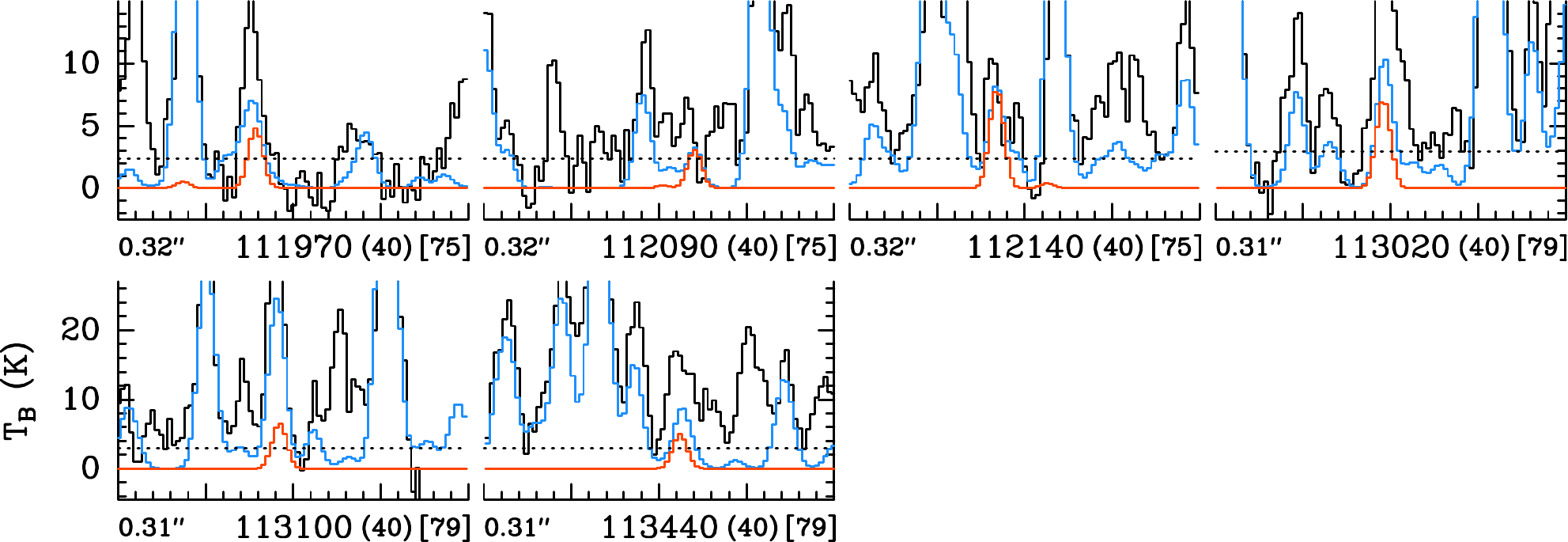}}}
\caption{continued.}
\end{figure*}

\begin{figure*}
\centerline{\resizebox{0.88\hsize}{!}{\includegraphics[angle=0]{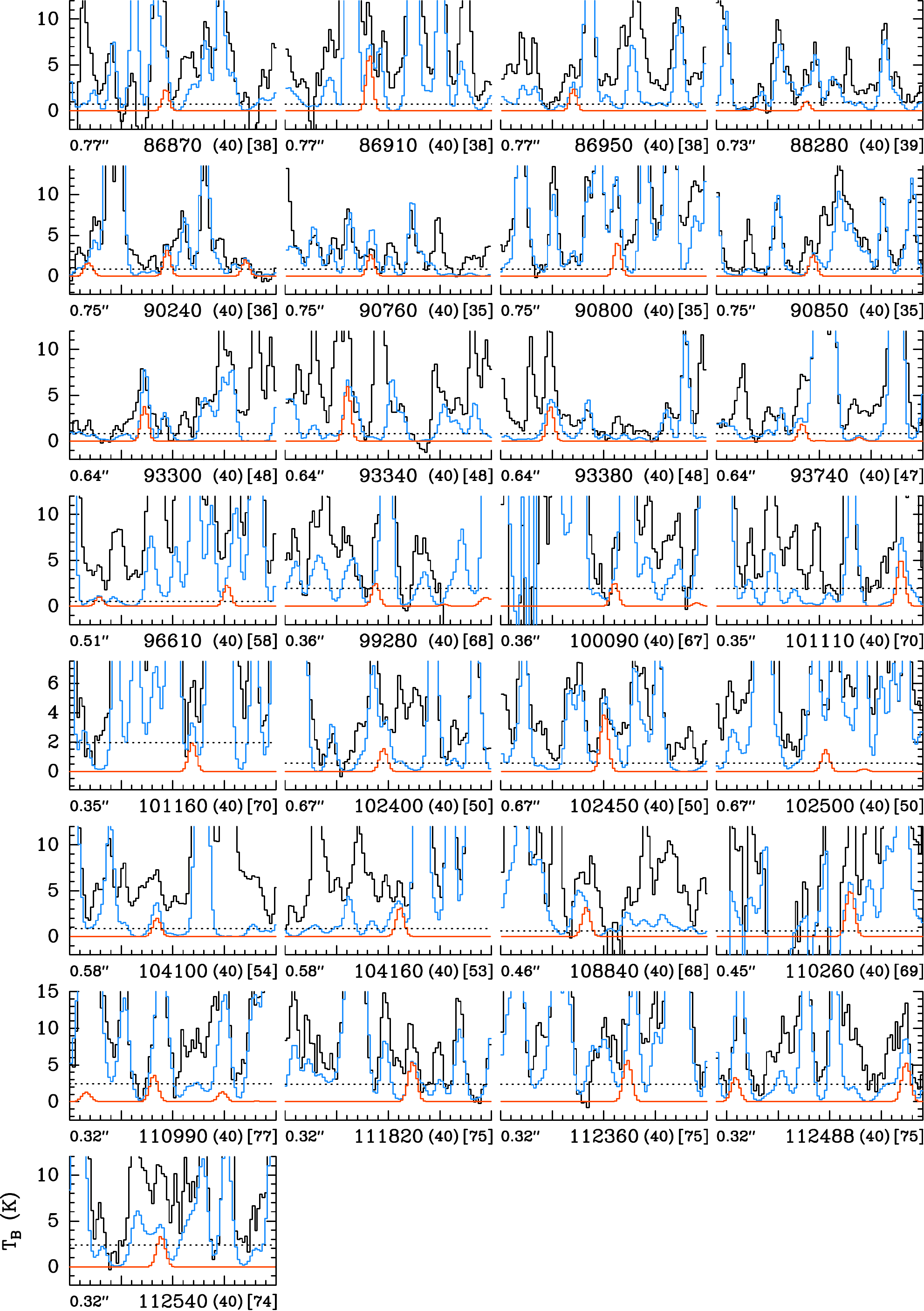}}}
\caption{Same as Fig.~\ref{f:remoca_ch3och3_ve0_n1s}, but for CH$_3$OCH$_3$, $\upsilon_{15} = 1$.
}
\label{f:remoca_ch3och3_v15e1_n1s}
\end{figure*}

\begin{figure*}
\centerline{\resizebox{0.85\hsize}{!}{\includegraphics[angle=0]{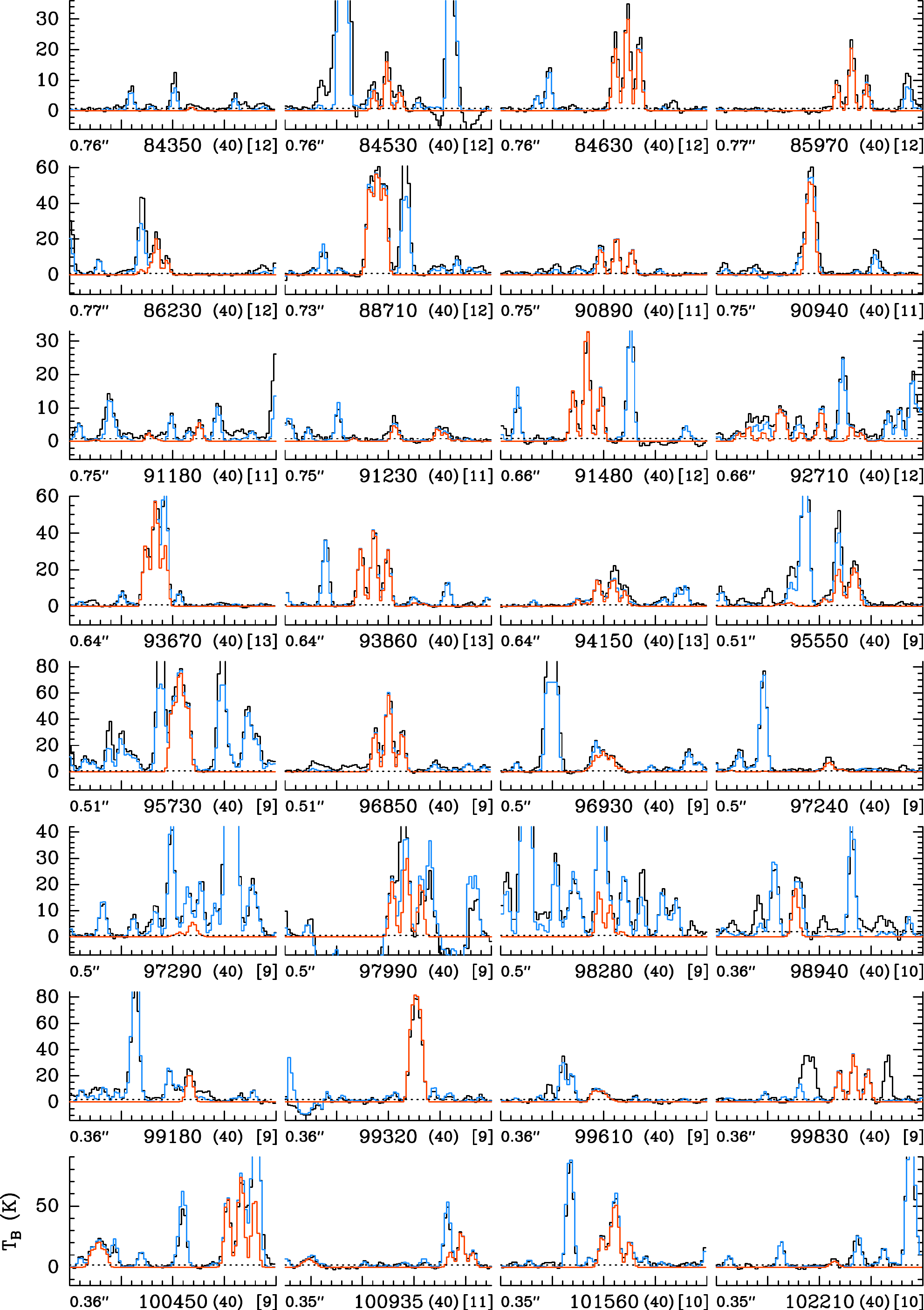}}}
\caption{Selection of rotational transitions of dimethyl ether CH$_3$OCH$_3$ in 
its vibrational ground state covered by the ReMoCA survey. The LTE 
synthetic spectrum of CH$_3$OCH$_3$, $\upsilon = 0$ is displayed in red and 
overlaid on the observed spectrum of Sgr~B2(N2b) shown in black. The blue 
synthetic spectrum contains the contributions of all molecules identified in 
our survey so far, including the contribution of the species shown in red. The 
values written below each panel correspond from left to right to the half-power 
beam width, the central frequency in MHz, the width in MHz of each panel in 
parentheses, and the continuum level in K of the baseline-subtracted spectra 
in brackets. The y-axis is labeled in brightness temperature units (K). The 
dotted line indicates the $3\sigma$ noise level.}
\label{f:remoca_ch3och3_ve0_n2b}
\end{figure*}

\begin{figure*}
\addtocounter{figure}{-1}
\centerline{\resizebox{0.88\hsize}{!}{\includegraphics[angle=0]{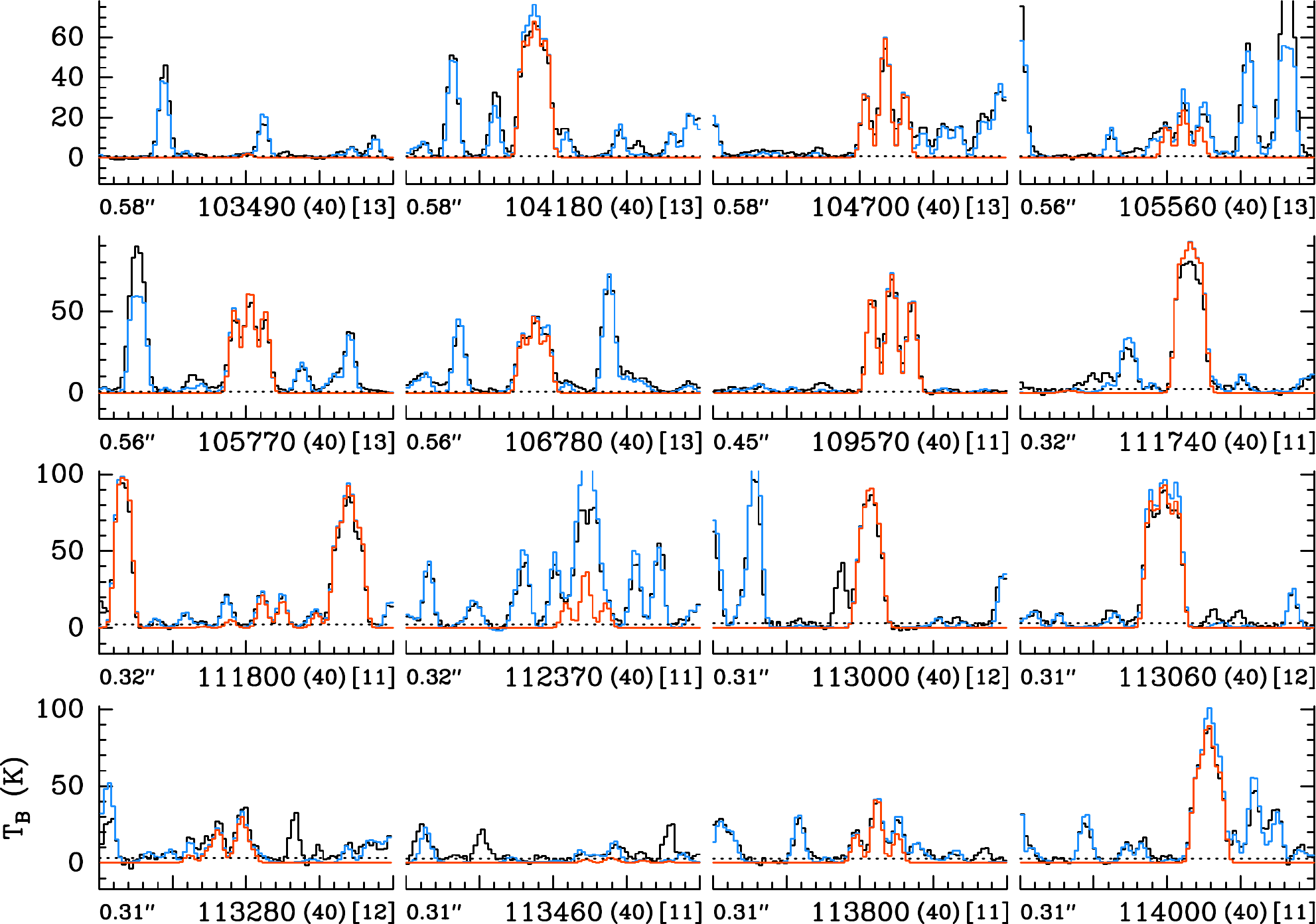}}}
\caption{continued.}
\end{figure*}

\begin{figure*}
\centerline{\resizebox{0.88\hsize}{!}{\includegraphics[angle=0]{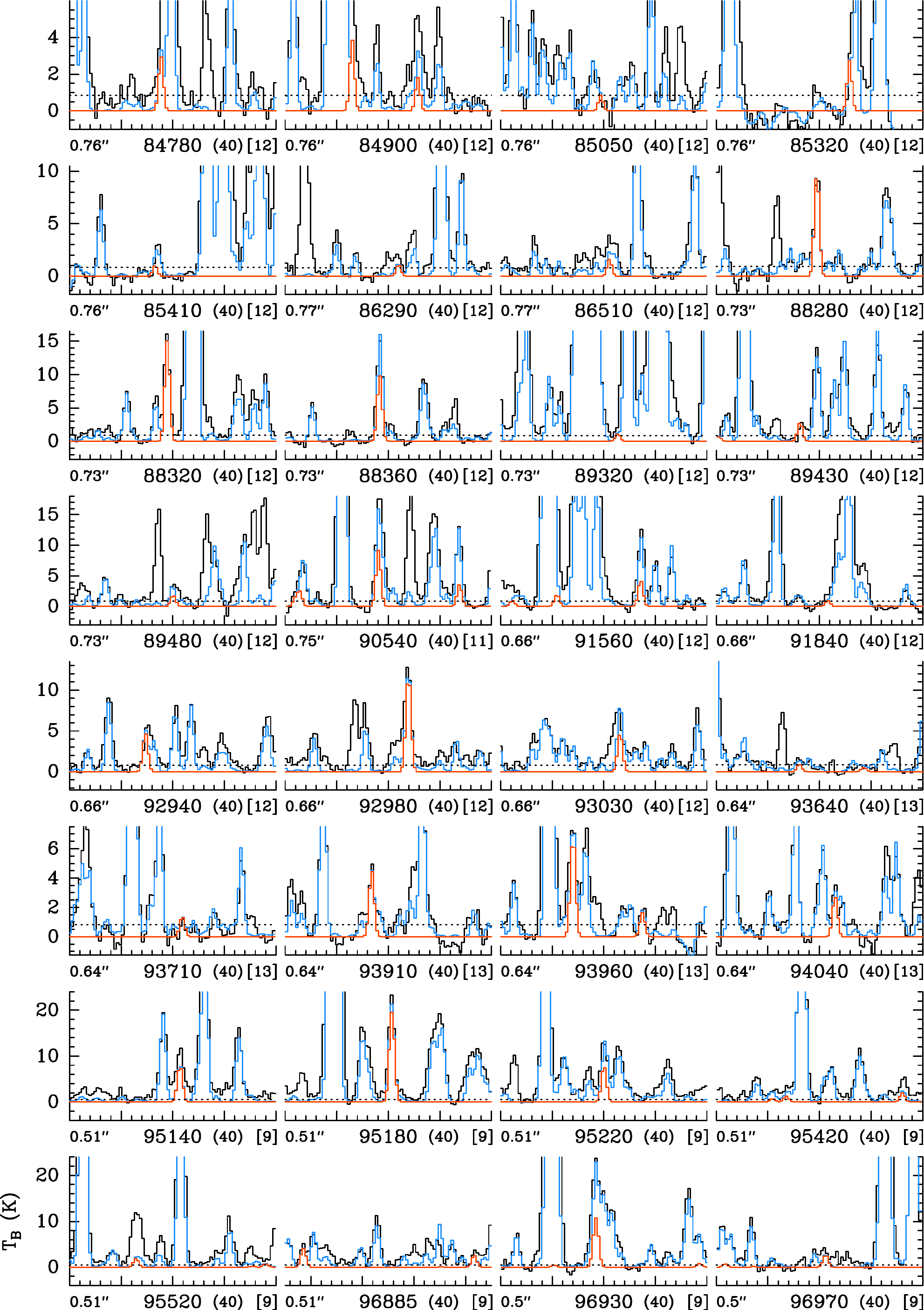}}}
\caption{Same as Fig.~\ref{f:remoca_ch3och3_ve0_n2b}, but for CH$_3$OCH$_3$, $\upsilon_{11} = 1$.
}
\label{f:remoca_ch3och3_v11e1_n2b}
\end{figure*}

\begin{figure*}
\addtocounter{figure}{-1}
\centerline{\resizebox{0.88\hsize}{!}{\includegraphics[angle=0]{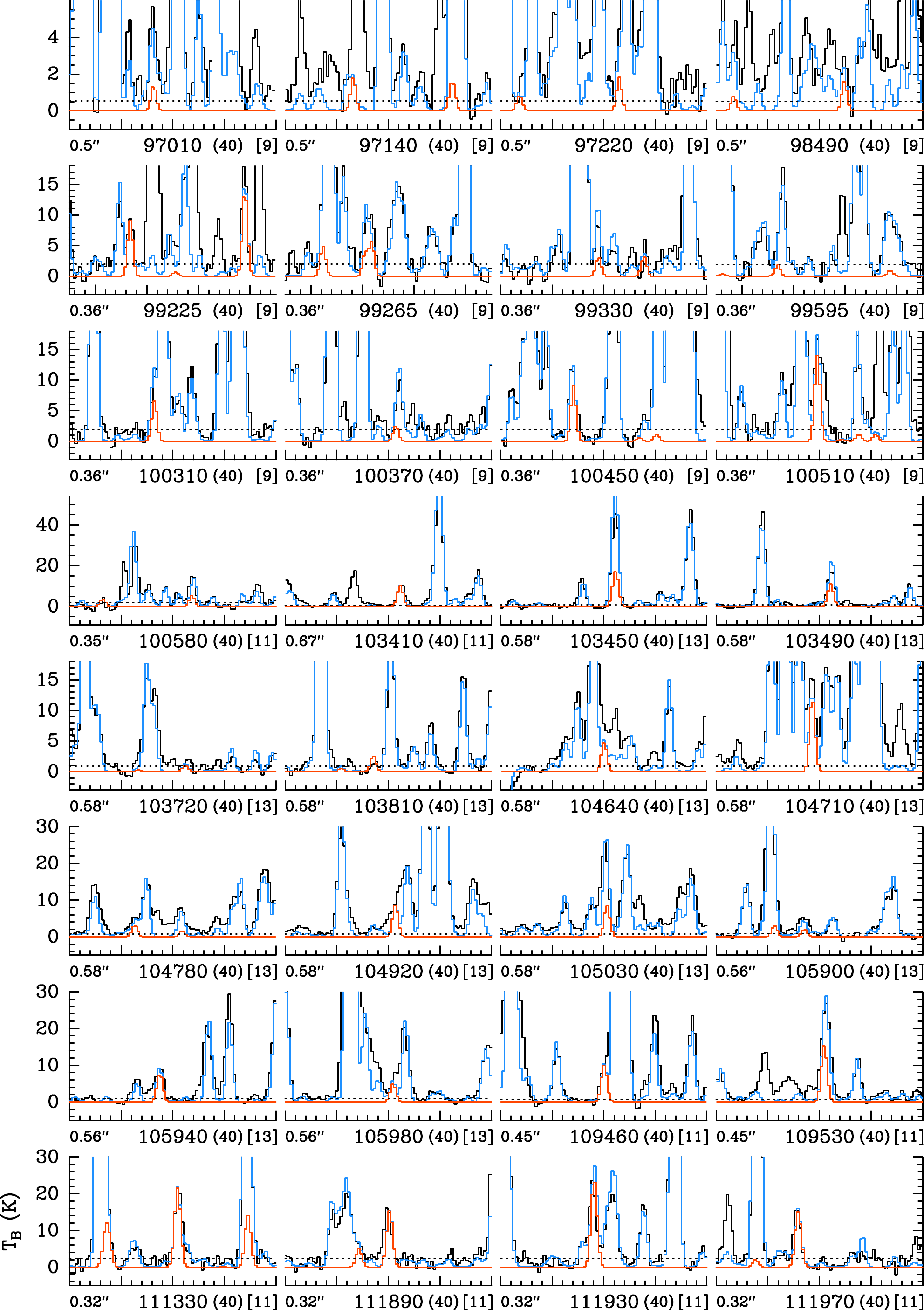}}}
\caption{continued.}
\end{figure*}

\begin{figure*}
\addtocounter{figure}{-1}
\centerline{\resizebox{0.88\hsize}{!}{\includegraphics[angle=0]{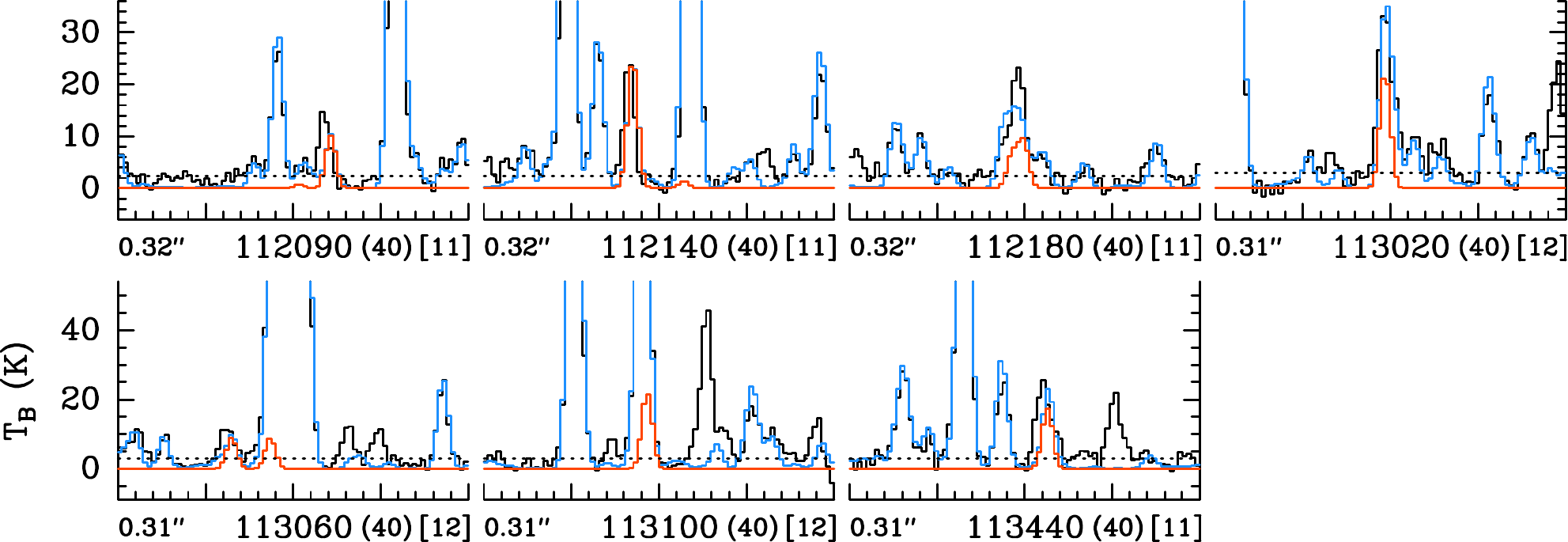}}}
\caption{continued.}
\end{figure*}

\begin{figure*}
\centerline{\resizebox{0.88\hsize}{!}{\includegraphics[angle=0]{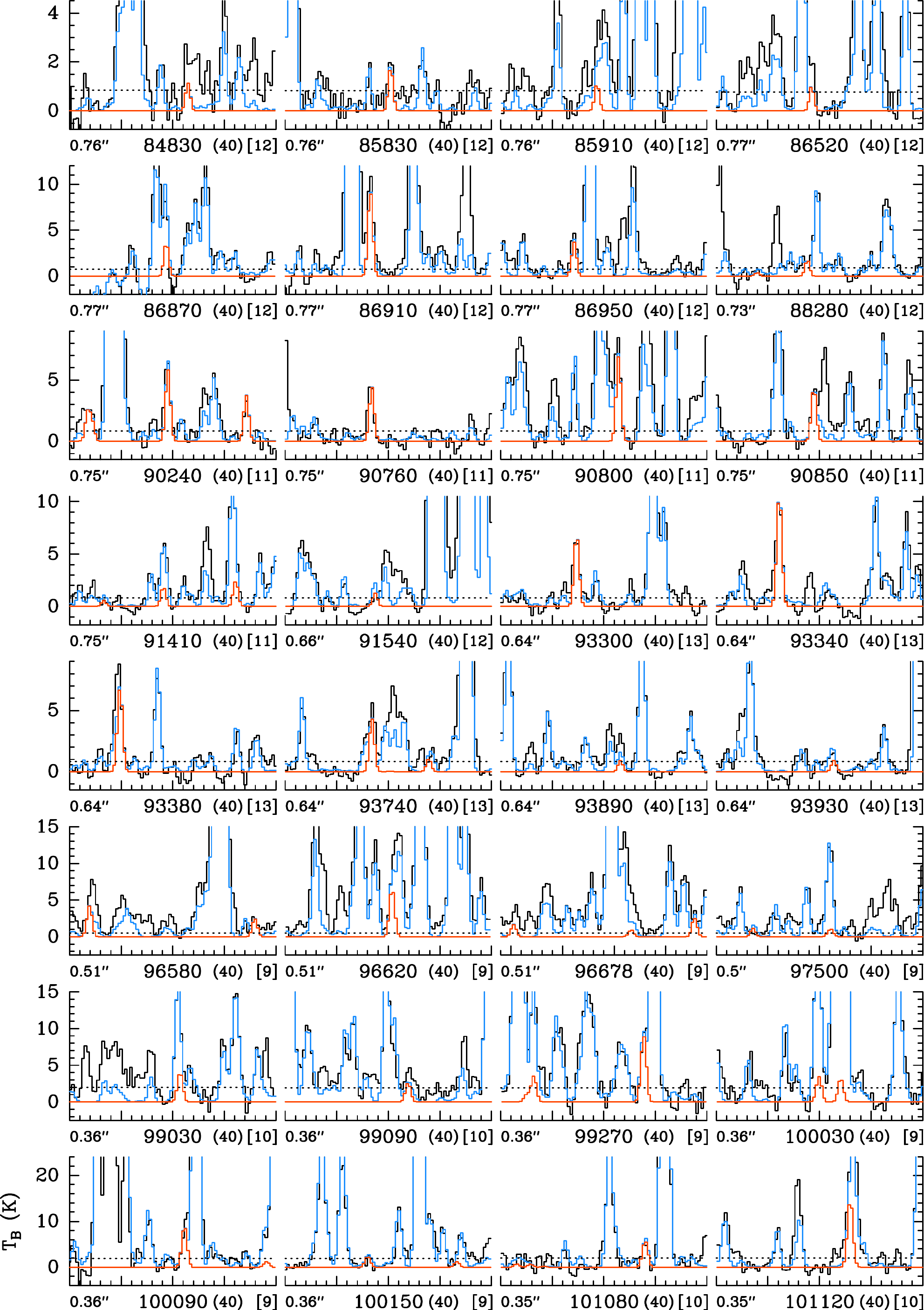}}}
\caption{Same as Fig.~\ref{f:remoca_ch3och3_ve0_n2b}, but for CH$_3$OCH$_3$, $\upsilon_{15} = 1$.
}
\label{f:remoca_ch3och3_v15e1_n2b}
\end{figure*}

\begin{figure*}
\addtocounter{figure}{-1}
\centerline{\resizebox{0.88\hsize}{!}{\includegraphics[angle=0]{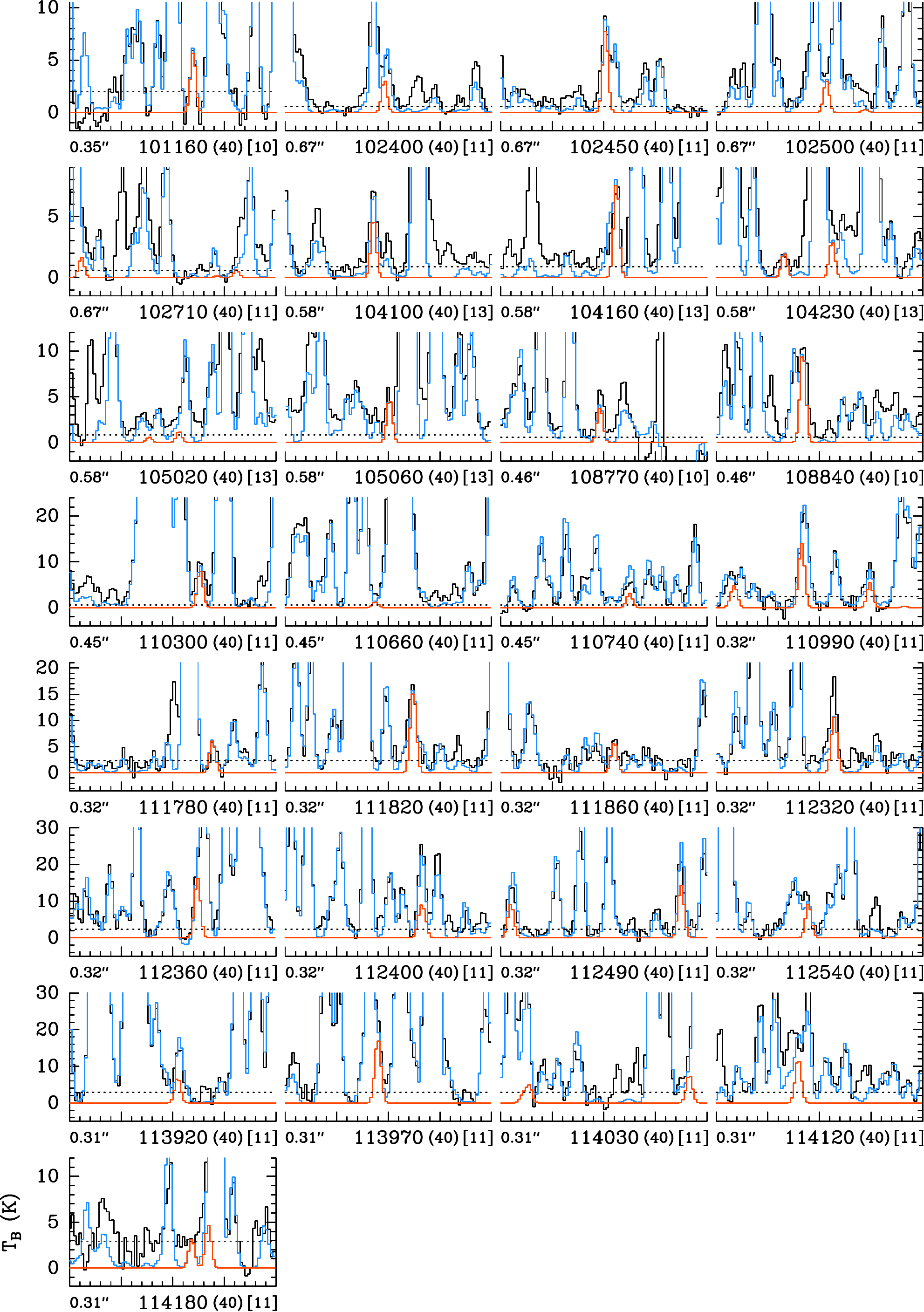}}}
\caption{continued.}
\end{figure*}

\begin{figure*}
\centerline{\resizebox{0.88\hsize}{!}{\includegraphics[angle=0]{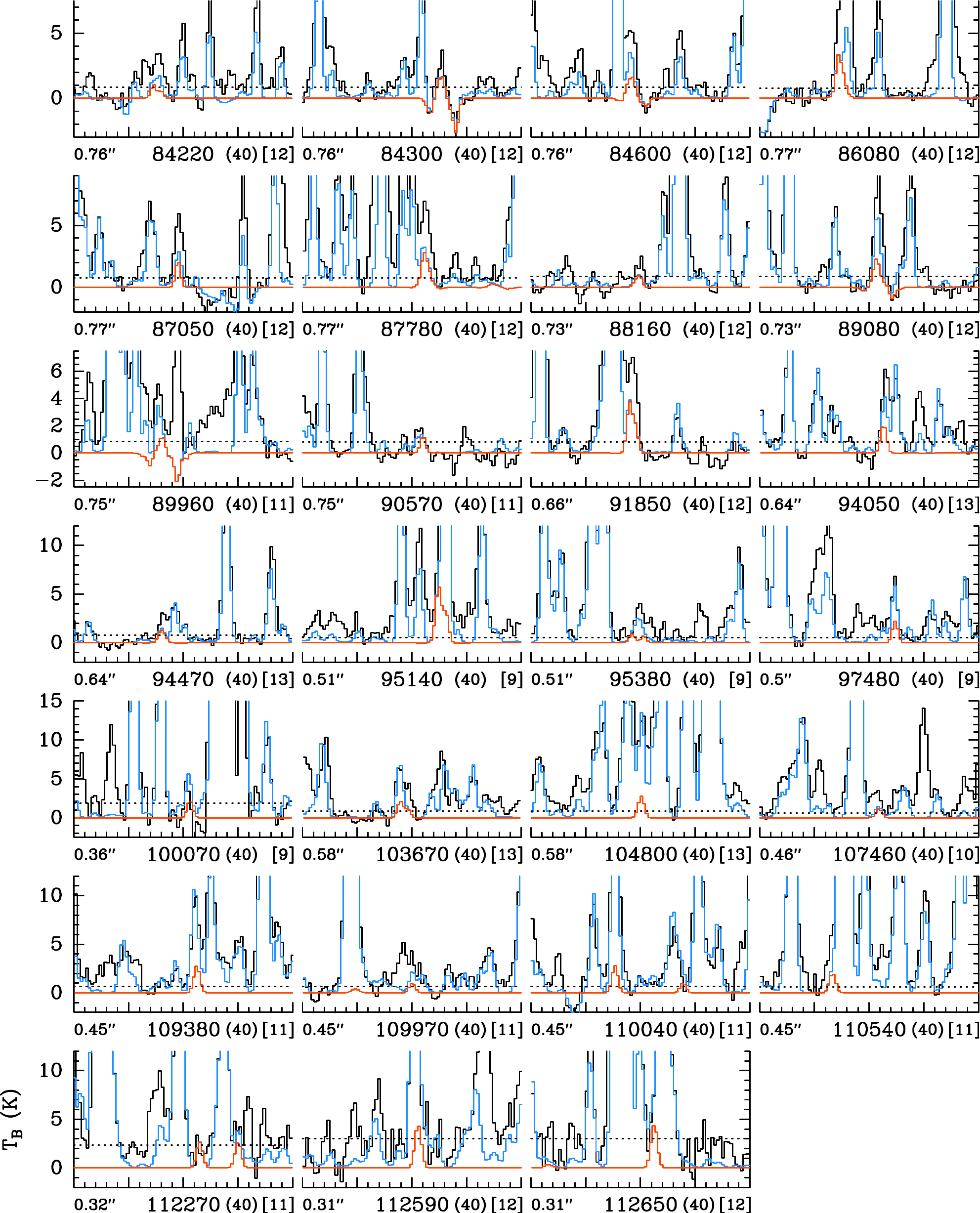}}}
\caption{Same as Fig.~\ref{f:remoca_ch3och3_ve0_n2b}, but for CH$_3$NH$_2$, $\upsilon = 0$.
}
\label{f:remoca_ch3nh2_ve0_n2b}
\end{figure*}

\end{appendix}

\bsp	
\label{lastpage}
\end{document}

%% file: tab_ch3nhch3_popfit_n1s.tex
\begin{table}
 \begin{center}
 \caption{
 Rotational temperature of dimethyl ether derived from its population diagram toward Sgr~B2(N1S).
}
 \label{t:popfit_n1s}
 \vspace*{0.0ex}
 \begin{tabular}{lll}
 \hline
 \multicolumn{1}{c}{Molecule} & \multicolumn{1}{c}{States$^{a}$} & \multicolumn{1}{c}{$T_{\rm fit}$$^{b}$} \\ 
  & & \multicolumn{1}{c}{\small (K)} \\ 
 \hline
CH$_3$OCH$_3$ & $\upsilon=0$, $\upsilon_{11}=1$, $\upsilon_{15}=1$ &   174 ( 7) \\ 
\hline 
 \end{tabular}
 \end{center}
 \vspace*{-2.5ex}
$^{a}$ Vibrational states that were taken into account to fit the population diagram.\\
$^{b}$ The standard deviation of the fit is given in parentheses. As explained in Sect.~3 of \citet{Belloche16} and in Sect.~4.4 of \citet{Belloche19}, this uncertainty is purely statistical and should be viewed with caution. It may be underestimated.
 \end{table}

%% file: tab_ch3nhch3_popfit_n2b.tex
\begin{table}
 \begin{center}
 \caption{
 Rotational temperature of dimethyl ether derived from its population diagram toward Sgr~B2(N2b).
}
 \label{t:popfit_n2b}
 \vspace*{0.0ex}
 \begin{tabular}{lll}
 \hline\hline
 \multicolumn{1}{c}{Molecule} & \multicolumn{1}{c}{States$^{a}$} & \multicolumn{1}{c}{$T_{\rm fit}$$^{b}$} \\ 
  & & \multicolumn{1}{c}{\small (K)} \\ 
 \hline
CH$_3$OCH$_3$ & $\upsilon=0$, $\upsilon_{11}=1$, $\upsilon_{15}=1$ & 127.2 (1.1) \\ 
\hline 
 \end{tabular}
 \end{center}
 \vspace*{-2.5ex}
 $^{a}$ Vibrational states that were taken into account to fit the population diagram.\\
 $^{b}$ The standard deviation of the fit is given in parentheses. As explained in Sect.~3 of \citet{Belloche16} and in Sect.~4.4 of \citet{Belloche19}, this uncertainty is purely statistical and should be viewed with caution. It may be underestimated.
 \end{table}

%% file: tab_ch3nhch3_weedsmodel_n1s.tex
\begin{table*}
 \begin{center}
 \caption{
 Parameters of our best-fit LTE model of methanol, dimethyl ether, and methylamine toward Sgr~B2(N1S), and upper limit for dimethylamine.
}
 \label{t:coldens_n1s}
 \vspace*{-1.2ex}
 \begin{tabular}{lcrccccccr}
 \hline\hline
 \multicolumn{1}{c}{Molecule} & \multicolumn{1}{c}{Status$^{a}$} & \multicolumn{1}{c}{$N_{\rm det}$$^{b}$} & \multicolumn{1}{c}{Size$^{c}$} & \multicolumn{1}{c}{$T_{\rm rot}$$^{d}$} & \multicolumn{1}{c}{$N$$^{e}$} & \multicolumn{1}{c}{$F_{\rm vib}$$^{f}$} & \multicolumn{1}{c}{$\Delta V$$^{g}$} & \multicolumn{1}{c}{$V_{\rm off}$$^{h}$} & \multicolumn{1}{c}{$\frac{N_{\rm ref}}{N}$$^{i}$} \\ 
  & & & \multicolumn{1}{c}{\small ($''$)} & \multicolumn{1}{c}{\small (K)} & \multicolumn{1}{c}{\small (cm$^{-2}$)} & & \multicolumn{1}{c}{\small (km~s$^{-1}$)} & \multicolumn{1}{c}{\small (km~s$^{-1}$)} & \\ 
 \hline
 CH$_3$OH$^{j,\star}$ & d & 57 &  2.0 &  230 &  2.0 (19) & 1.00 & 5.0 & 0.2 &       1 \\ 
\hline 
 CH$_3$OCH$_3$, $\upsilon=0$ & d & 32 &  2.0 &  170 &  1.4 (18) & 1.02 & 5.0 & 0.0 &      14 \\ 
 $\hspace*{11.0ex}\upsilon_{\rm 11}=1$ & d & 3 &  2.0 &  170 &  1.4 (18) & 1.02 & 5.0 & 0.0 &      14 \\ 
 $\hspace*{11.0ex}\upsilon_{\rm 15}=1$ & d & 3 &  2.0 &  170 &  1.4 (18) & 1.02 & 5.0 & 0.0 &      14 \\ 
\hline 
 CH$_3$NH$_2$$^{k}$ & d & 15 &  2.0 &  230 &  1.4 (18) & 1.25 & 5.0 & 0.0 &      15 \\ 
\hline 
 CH$_3$NHCH$_3$, $\upsilon=0$ & n & 0 &  2.0 &  170 & $<$  9.8 (16) & 1.40 & 5.0 & 0.0 & $>$     204 \\ 
\hline 
 \end{tabular}
 \end{center}
 \vspace*{-2.5ex}
 $^{a}$ d: detection, n: nondetection. 
 $^{b}$ Number of detected lines \citep[conservative estimate, see Sect.~3 of][]{Belloche16}. One line of a given species may mean a group of transitions of that species that are blended together. 
 $^{c}$ Emission size (\textit{FWHM}). 
 $^{d}$ Rotational temperature. 
 $^{e}$ Total column density of the molecule. $x$ ($y$) means $x \times 10^y$. 
 $^{f}$ Correction factor that was applied to the column density to account for the contribution of vibrationally excited states, in the cases where this contribution was not included in the partition function of the spectroscopic predictions. 
 $^{g}$ Linewidth (\textit{FWHM}). 
 $^{h}$ Velocity offset with respect to the assumed systemic velocity of Sgr~B2(N1S), $V_{\mathrm{sys}} = 62$~km~s$^{-1}$. 
 $^{i}$ Column density ratio, with $N_{\rm ref}$ the column density of the previous reference species marked with a $\star$.
 $^{j}$ The parameters were derived from the ReMoCA survey by \citet{Motiyenko20}.
 $^{k}$ The parameters were derived from the ReMoCA survey by \citet{Gyawali23}. 
 \end{table*}

%% file: tab_ch3nhch3_weedsmodel_n2b.tex
\begin{table*}
 \begin{center}
 \caption{
 Parameters of our best-fit LTE model of methanol, dimethyl ether, and methylamine toward Sgr~B2(N2b), and upper limit for dimethylamine.
}
 \label{t:coldens_n2b}
 \vspace*{-1.2ex}
 \begin{tabular}{lcrccccccr}
 \hline\hline
 \multicolumn{1}{c}{Molecule} & \multicolumn{1}{c}{Status$^{a}$} & \multicolumn{1}{c}{$N_{\rm det}$$^{b}$} & \multicolumn{1}{c}{Size$^{c}$} & \multicolumn{1}{c}{$T_{\rm rot}$$^{d}$} & \multicolumn{1}{c}{$N^{e}$} & \multicolumn{1}{c}{$F_{\rm vib}$$^{f}$} & \multicolumn{1}{c}{$\Delta V$$^{g}$} & \multicolumn{1}{c}{$V_{\rm off}$$^{h}$} & \multicolumn{1}{c}{$\frac{N_{\rm ref}}{N}$$^{i}$} \\ 
  & & & \multicolumn{1}{c}{\small ($''$)} & \multicolumn{1}{c}{\small (K)} & \multicolumn{1}{c}{\small (cm$^{-2}$)} & & \multicolumn{1}{c}{\small (km~s$^{-1}$)} & \multicolumn{1}{c}{\small (km~s$^{-1}$)} & \\ 
 \hline
 CH$_3$OH$^{j,\star}$ & d & 69 &  0.5 &  140 &  8.0 (19) & 1.00 & 3.5 & 0.0 &       1 \\ 
\hline 
 CH$_3$OCH$_3$, $\upsilon=0$ & d & 93 &  0.8 &  130 &  3.1 (18) & 1.00 & 3.5 & -0.5 &      26 \\ 
 $\hspace*{11.0ex}\upsilon_{\rm 11}=1$ & d & 24 &  0.8 &  130 &  3.1 (18) & 1.00 & 3.5 & -0.5 &      26 \\ 
 $\hspace*{11.0ex}\upsilon_{\rm 15}=1$ & d & 26 &  0.8 &  130 &  3.1 (18) & 1.00 & 3.5 & -0.5 &      26 \\ 
\hline 
 CH$_3$NH$_2$, $\upsilon=0$ & t & 2 &  0.5 &  140 &  2.7 (17) & 1.07 & 3.5 & 0.0 &     299 \\ 
\hline 
 CH$_3$NHCH$_3$, $\upsilon=0$ & n & 0 &  0.8 &  130 & $<$  5.9 (16) & 1.18 & 3.5 & 0.0 & $>$    1360 \\ 
\hline 
 \end{tabular}
 \end{center}
 \vspace*{-2.5ex}
 $^{a}$ d: detection, t: tentative detection, n: nondetection. 
 $^{b}$ Number of detected lines \citep[conservative estimate, see Sect.~3 of][]{Belloche16}. One line of a given species may mean a group of transitions of that species that are blended together. 
 $^{c}$ Emission size (\textit{FWHM}). 
 $^{d}$ Rotational temperature. 
 $^{e}$ Total column density of the molecule. $x$ ($y$) means $x \times 10^y$. 
 $^{f}$ Correction factor that was applied to the column density to account for the contribution of vibrationally excited states, in the cases where this contribution was not included in the partition function of the spectroscopic predictions. 
 $^{g}$ Linewidth (\textit{FWHM}). 
 $^{h}$ Velocity offset with respect to the assumed systemic velocity of Sgr~B2(N2b), $V_{\mathrm{sys}} = 74.2$~km~s$^{-1}$. 
 $^{i}$ Column density ratio, with $N_{\rm ref}$ the column density of the previous reference species marked with a $\star$. 
 $^{j}$ The parameters were derived from the ReMoCA survey by \citet{Belloche22}.
\end{table*}